\documentclass[12pt]{article}
\newlength{\abstractwidth}
\input epsf

\renewcommand{\thefootnote}{\fnsymbol{footnote}}
\renewcommand{\thanks}[1]{\footnote{#1}} 
\newcommand{\starttext}{
\setcounter{footnote}{0}
\renewcommand{\thefootnote}{\arabic{footnote}}}

\newcommand{\be}{\begin{equation}}
\newcommand{\bea}{\begin{eqnarray}}
\newcommand{\eea}{\end{eqnarray}}
\newcommand{\beq}{\begin{equation}}
\newcommand{\ee}{\end{equation}}
\newcommand{\eeq}{\end{equation}}

\renewcommand{\>}{\rangle}

\def\ba{\begin{eqnarray}}
\def\ea{\end{eqnarray}}

\def\N{{\cal N}}
\def\O{{\cal O}}
\def\G{{\cal G}}
\def\L{{\cal L}}
\def\R{{\cal R}}
\def\F{{\cal F}}
\def\p{{\partial}}
\def\D{{\cal D}}

\def\Im{{\rm Im}}
\def\tr{{\rm tr}}
\def\ch{{\rm ch}}
\def\12{{1 \over 2}}
\def\32{{3 \over 2}}
\def\72{{7 \over 2}}
\def\92{{9 \over 2}}
\def\d{{d \over 2}}

\def\P{\Phi}
\def\a{\alpha}
\def\b{\beta}

\def\d{\delta}
\def\l{\lambda}
\def\F{{\cal F}}
\def\half{{1\over 2}}

\begin{document}
\baselineskip=16pt

\begin{titlepage}
\bigskip
\hskip 3.7in\vbox{\baselineskip12pt
\hbox{UCLA/99/TEP/28}
\hbox{Columbia/Math/99}}

\bigskip\bigskip\bigskip\bigskip

\centerline{\Large \bf Lectures on Supersymmetric Yang-Mills Theory}
\medskip
\centerline{\Large \bf and Integrable Systems}

\bigskip\bigskip
\bigskip\bigskip

\centerline{\bf  Eric D'Hoker$^{a}$ and D.H.Phong$^{b}$} 

\bigskip
\bigskip

\centerline{$^a$ \it Department of Physics}
\centerline{ \it University of California, Los Angeles, CA 90095}
\bigskip
\centerline{$^b$ \it Department of Mathematics}
\centerline{ \it Columbia University, New York, NY 10027}

\bigskip\bigskip

\begin{abstract}
\noindent
We present a series of four self-contained lectures on the following
topics; 
 
(I) 
An introduction to 4-dimensional $1\leq \N \leq 4$ supersymmetric
Yang-Mills theory, including particle and field contents, $\N=1$
and $\N=2$ superfield
methods and the construction of general invariant Lagrangians; 

(II) 
A review of holomorphicity and duality
in $\N=2$ super-Yang-Mills,
of Seiberg-Witten theory and its formulation in terms of
Riemann surfaces; 

(III)
An introduction to mechanical Hamiltonian integrable systems, such as the Toda
and Calogero-Moser systems associated with general Lie algebras; a review of
the recently constructed Lax pairs with spectral parameter for twisted and
untwisted elliptic Calogero-Moser systems; 

(IV) 
A review of recent solutions of the Seiberg-Witten theory for general gauge
algebra  and adjoint
hypermultiplet content in terms of the elliptic Calogero-Moser integrable
systems. 
 
\end{abstract}
\end{titlepage}

\vfill\eject

\tableofcontents

\vfill\eject

\listoftables

\vfill\eject

\starttext
\baselineskip=15pt
\setcounter{equation}{0}
\setcounter{footnote}{0}

\section{Introduction}

Supersymmetry (often abbreviated as susy) maps particles and fields of integer
spin (bosons) into particles and fields of half odd integer spin (fermions)
and vice versa.  Supersymmetry was introduced in quantum field theory in
\cite{golf} (for textbooks, see \cite{susy}). Supersymmetry is a continuous
space-time symmetry, generated by a {\it supercharge} $Q_\alpha$, which is a
fermionic space-time spinor operator of spin 1/2. As such,
$Q_\alpha$ commutes with the usual local and global internal symmetries, such
as color and flavor symmetries,
\be
[Q_\alpha , {\rm internal \ charges}]=0\, .
\ee 
As a result, in a supersymmetric theory, the spectrum is arranged in pairs
(boson,\ fermion), where the boson and fermion have the same internal quantum
numbers and the same mass. 

In the present
lectures, we shall concentrate on Yang-Mills theories, which are built out of
particles and fields of spins at most equal to 1. Thus, we shall not deal with
theories of gravity which contain also particles and fields of spins 
larger than 1. Before reviewing the structure 
of the spectrum and of
the interactions of supersymmetric theories, we shall briefly explain why
supersymmetry is an important and integral part of modern particle physics
already today and is expected to become so even more in the near future. 

\subsection{Supersymmetry and the Standard Model} 

All the particle physics that we know of today is described to stunning
accuracy by the particle contents and by the interactions of the {\it Standard
Model}. The particle contents is given in table \ref{table:1}; all particles
have been observed, except for the Higgs. 

\begin{table}[b]
\begin{center}
\begin{tabular}{|c| c c|} \hline
Spin 0      & physical Higgs & not yet observed               \\ \hline \hline
            & leptons        & 
$ \left ( \matrix{\nu _e \cr e^-} \right )$ \ 
$ \left ( \matrix{\nu _\mu \cr \mu ^-} \right )$  \
$ \left ( \matrix{\nu _\tau \cr \tau^-}\right )$ \\   
Spin $\12$    &               &               \\
            & quarks         & 
$ \left (\matrix{u \cr d} \right )$ \ \
$ \left (\matrix{c \cr s} \right )$ \ \
$ \left (\matrix{t \cr b} \right )$\\ \hline \hline
            &  8 gluons $g$& strong interactions \\ 
Spin 1      &    photon $\gamma$     & electro-magnetic interactions \\ 
            & $W^\pm, \ Z$   & weak interactions \\ \hline
\end{tabular}
\end{center}
\caption{Particle Contents of the {\it Standard Model}}
\label{table:1}
\end{table}

The interactions between the spin 1 particles and quarks and leptons are
dictated by the gauge invariance of the standard model, 
$ SU(3)_c \times SU(2)_L \times U(1)_Y$.
To each of the factors of the gauge group, there is an independent
coupling constant, denoted by $g_3$, $g_2$ and $g_1$. The fine structure
constant $\alpha$, familiar from electromagnetism, and the weak mixing angle
$\theta _W$ are combinations of the couplings $g_2$ and $g_1$.

By inspection of the particle spectrum of the Standard Model, it is immediate
that there is not a single candidate for a supersymmetric pair of particles.
Thus, it would appear at first sight that supersymmetry has no useful place in
particle physics, since the most basic characteristic of a supersymmetric 
theory is not realized in the Standard Model. 

However, in Nature, symmetries are often {\it spontaneously broken}. This means
that the laws of physics are invariant under a symmetry, 
but the solution of the
equations (such as the ground state of the theory) is not invariant under this
symmetry. This situation is familiar from the Standard Model itself and occurs
in all three sectors of the theory. In the $SU(3)_c$ or Quantum Chromodynamics
(QCD) sector, the spontaneous breaking of approximate chiral symmetry
for the up and down quarks is responsible for the comparatively small masses of
the pions, as well as for the comparatively large masses of the
nucleons. (This effect is discussed in detail in the Banff lectures of
Frank Wilczek \cite{wilc}.) 
In the $SU(2)_L \times U(1)_Y$ sector, the spontaneous breaking
of $SU(2)$ is responsible for the masses of quarks, leptons, $W^\pm$ and $Z$
particles, as well as for the emergence of unbroken electromagnetic
$U(1)_{\rm em}$.

Thus, by analogy with mechanisms familiar from the Standard Model, we should
expect that also supersymmetry may be spontaneously broken if realized at all
in Nature. In the generic mechanisms of supersymmetry breaking, the spectrum
continues to be organized in pairs of bosons and fermions with equal quantum
numbers, but the mass of the fermion is different from the mass of the boson
of the pair.

\begin{table}[b]
\begin{center}
\begin{tabular}{|cc c |c|} \hline
Non-Susy Sector      && Susy Partners & Spin Susy Partners  
\\ \hline \hline
2 Higgs (required)      && Higgsinos     & 1/2 \\ \hline
Leptons $(\nu _e\ e^-) \ \cdots$ && 
sLeptons $(\tilde \nu _e \ \tilde e^-)\ \cdots $ & 0 \\  
Quarks  $ (u \ d )\ \cdots$  && sQuarks $ (\tilde u \ \tilde d )\ \cdots$ & 0 
\\ \hline 
Gluons    $g$        &&      Gluinos $\tilde g$ & \\ 
Photon $\gamma$   &&    photino   $\tilde \gamma$    & 1/2 \\ 
$W^\pm, \ Z$   &&  Winos $\tilde W^\pm$,  Zino $\tilde Z$\\ \hline
\end{tabular}
\end{center}
\caption{Particle Contents of the {\it Minimal Supersymmetric Standard Model}}
\label{table:2}
\end{table}

Considering now pairs of fermions and bosons, 
but allowing for unequal masses,  
as is typical of spectra with spontaneously broken supersymmetry,  we see that
the Standard Model still exhibits no candidates for such pairs, and thus
displays no room for supersymmetry to operate, even if spontaneously broken. To
have any chance at all of realizing supersymmetry, even spontaneously broken,
the particle spectrum of the Standard Model must be extended. 
The minimal way of
doing this, via the introduction of the smallest number of supplementary
particles, is called the {\it Minimal Supersymmetric Standard Model}. 
Basically,
for each currently known or {\it non-supersymmetric} particle, we supply a
hypothetical {\it supersymmetric partner}. 
There is one exception : supersymmetry
requires at least two Higgs fields. The spectrum is schematically exhibited
in table \ref{table:2}.

\medskip

Actually, the existence of super partners in supersymmetric theories is one of
the most dramatic and most general predictions that follows from the assumption
that supersymmetry has anything to do with Nature at all. Bold as the
assumption of this extension of the particle spectrum may be, this is not
the first time in the history of particle physics that the number of
degrees of freedom associated with an elementary particle is doubled up, as
schematically shown in the diagram below for the case of the electron

\medskip

\centerline{Thompson electron \ 
$\Longrightarrow$ \ + spin   
$\Longrightarrow$ \ + positron   
$\Longrightarrow$ \ + susy partner\, .} 

\medskip

The precise mechanism by which supersymmetry is to be broken is not well
understood at present. By analogy with QCD, the mechanism may be inherently
non-perturbative, but we do not know for a fact that this is the case.

\subsection{Supersymmetry and Unification of Forces}

The supersymmetric extension of the Standard Model, in which a hypothetical
supersymmetric partner is included for each known particle, has appeared in a
rather ad-hoc fashion. The full significance of supersymmetry really emerges
only when the principle of supersymmetry is considered in conjunction
with the principle of the unification of the strong, electro-magnetic and
weak forces (and most likely also gravity).

\medskip

Remarkably, the Standard Model automatically contains a new scale 
at energies of
order $10^{15} \ GeV$. It is surprising to find such a high energy scale in the
model, since the masses of the quarks, leptons and gauge bosons are on the
order of or smaller than the electro-weak scale of about $250 \ GeV$. 
To see how
this new scale emerges, one must take into account the effect of 
renormalization of the
three couplings $g_i$, $i=1,2,3$. (See for example \cite{peskin}.)
As the typical energy or mass scale $\mu$ of a
physical process is altered, the couplings $g_i$ vary according to the
renormalization group equations, whose lowest order solutions are given by
\be
{4 \pi \over g_i ^2 (\mu) } = {b_i \over 2 \pi } \ln {\mu \over \Lambda _i}
\qquad \qquad
i=1,2,3\, .
\ee
Here, $b_i$ are numerical constants, dependent on the gauge group and on the 
matter contents, and $\Lambda _i$ are integration constants; we shall not need
their explicit values here. The couplings have been measured to very high
accuracy at the scale of the mass of the $Z$ particle
$\mu = M_Z$, where \cite{part}
\bea 
g_1(M_Z) \sim 0.46 <
g_2(M_Z) \sim 0.64 <
g_3(M_Z) \sim 1.22
\eea 
These values determine the integration constants $\Lambda _i$ for given $b_i$.

The couplings $g_2$ and $g_3$ are asymptotically free, i.e. their values
decrease as the energy scale is increased because $b_2, b_3 >0$. Now, at low
energies, the strong coupling $g_3$ is larger than $g_2$, and since
$b_3 > b_2$, the coupling $g_3$ will decrease faster than $g_2$. Also, the
coupling $g_1$ is not asymptotically free and increases as the energy scale is
increased because $b_1 <0$. From these general considerations, it is not
surprising that the couplings $g_i$ will mutually intersect as the energy scale
is raised.

\medskip

Remarkably however, all three couplings $g_i(\mu)$ approximately intersect
in one point around $M_U \sim 10^{15} \ GeV$. Even more remarkably, this scale
is only a few orders of magnitude below perhaps the most primary scale in
physics, the Planck scale $M_P \sim 10^{19} \ GeV$ at which quantum gravity
is supposed to become important. Adopting the philosophy that such a remarkable
coincidence cannot be merely an accident and must have a fundamental physical
meaning, we are led to interpret the scale $M_U$, where the three couplings
approximately meet, as a scale where the three different forces are unified
into a single one. A unified theory will be governed by a simple gauge group.
The smallest such group which contains the gauge group of the Standard Model 
as a subgroup is $SU(5)$, but the inclusion of a massive neutrino can be
realized only in the larger $SO(10)$, and string theory might favor a
sequence all the way up to the maximal exceptional  group $E_8$,
\be 
SU(3) _c \times SU(2)_L \times U(1)_Y \subset SU(5) \subset SO(10) \subset
E_6 \subset E_8\, .
\ee   
Starting at the highest energy scales, just below $M_P$, and running down to
lower energy scales, there will initially be only a single force, governed by
the unified simple gauge group. The strong, electromagnetic and weak
interactions will emerge as the result of a phase transition at the
unification scale $M_U$, where the unified gauge symmetry is broken down to the
$SU(3)_c\times SU(2)_L \times U(1)_Y$ gauge group. If the
breaking of the gauge symmetry occurs in successive stages, various 
intermediate breaking scales will also emerge. 

\bigskip

The unification scheme is attractive for a number of additional reasons. 

\medskip

\noindent
$\oplus$ \ The Abelian gauge interactions $U(1)_Y$ and $U(1)_{\rm em}$
are not asymptotically free and the gauge couplings $g_1$ and $\alpha$ grow
indefinitely as the energy scale is increased. It is usually argued that
this behavior leads to a Landau ghost, thus to a failure of unitarity at high
energies and ultimately to an inconsistency of the theory. In the unification
schemes where the $U(1)$ gauge group arises from the breaking of an
asymptotically free gauge theory with simple gauge group, this problem is
automatically avoided, because the dynamics of the asymptotically simple group
will take over at sufficiently large energies.

\medskip

\noindent
$\oplus$ \ The particle spectrum of the standard model without right
handed neutrino fits economically in three family copies of the $\bar {\bf 5}
\oplus {\bf 10}$ of $SU(5)$, while the spectrum including the right handed
neutrino fits precisely in three family copies of the {\bf 16} spinor
representation of $SO(10)$. In view of the recent experimental results on
neutrino masses and oscillations, the $SO(10)$ scheme is definitely
the preferred one.

\bigskip

However, the unification scheme also has some fundamental problems.

\medskip

\noindent
$\ominus$ \ The additional gauge bosons can mediate baryon number
violating transitions, and the {\it proton life-time} comes out around $10^{29}
$ years, many orders of magnitude shorter than the present experimental bounds
of around $10 ^{33}$ years. (See for example \cite{proton}.)

\medskip

\noindent
$\ominus$ \ As the fundamental scale in the unification models is now
large and on the order of $10 ^{15} \ GeV$, it becomes an issue as to how the
comparatively small mass scales of the electro-weak gauge bosons and the quarks
and leptons come about. This is called the {\it hierarchy problem}. The
principles of quantum field theory guarantee that these masses are free
parameters, so in principle they could be adjusted by hand to be small.
This requires an absurd degree of {\it fine tuning} of the various couplings
in the theory, and is usually viewed as unnatural.

\medskip

\noindent
$\ominus$ \ Finally, the three gauge couplings $g_1$, $g_2$ and $g_3$
meet only approximately in a single point. With the high precision data made
available by the study of the $Z$-particle and its decays, we now know that
strictly within the scenario of the Standard Model and its extrapolation to
high energies, the three couplings will not intersect at one point. 

\bigskip

The Supersymmetric Extension of the Standard Model, combined with the principle
of unification of strong, electro-magnetic and weak forces supplies radical
solutions to the above problems. 

\medskip

\noindent
$\oplus$ \ The supersymmetric extension of the $SO(10)$ unified theory
provides perfect intersection of the couplings $g_1$, $g_2$ and $g_3$ at
$M_{SU}\sim 10^{16} \ GeV$. The running of the gauge couplings depends
only very slightly upon the scale and mechanism of supersymmetry breaking.
The result quoted here is obtained generically when the susy breaking scale is
on the order of the electro-weak scale of $100 \ GeV$.
 
\medskip

\noindent
$\oplus$ \ The proton life-time for the $SU(5)$ or $SO(10)$
supersymmetric unified theories is now much longer (in part because the
unification scale is an order of magnitude larger in the supersymmetric
theories) and estimated to be on the order  of $3 \times 10^{33}$ years. This
result is tantalizingly close to the present day bounds for this number, and if
supersymmetric unified theories are viable, we may expect to see proton decay
within our life-time.

\medskip

\noindent
$\oplus$ \ The hierarchy and fine-tuning problems are dramatically
reduced. The smallness of fermion masses is naturally protected by chiral
symmetries. In generic non-supersymmetric theories, the smallness of boson
masses is not protected and contributions to the boson mass quadratic in the
cutoff scale tend to drive boson masses towards the largest scale in the
theory. In exactly supersymmetric theories, bosons and fermions are degenerate,
so supersymmetry together with chiral symmetry will also protect the smallness
of boson masses. In more realistic theories, where supersymmetry is broken
(explicitly or spontaneously) at a scale of $M_B$, the smallness of boson
masses will be naturally protected up to the supersymmetry breaking scale. The
idea being that above this breaking scale, supersymmetry is exact for all
practical purposes, and boson and fermion masses above this scale will be
degenerate. 

\medskip

\noindent
$\oplus$ \ Finally, the argument of the previous point may be turned
around and used to provide a rough estimate of the supersymmetry breaking
scale. If supersymmetry is to protect boson masses up to the electro-weak
scale, then the supersymmetry breaking scale cannot be too much larger than
the electro-weak breaking scale, otherwise supersymmetry will cease to resolve
the hierarchy and fine-tuning problems. Including lower bounds from present
day experiments on supersymmetric particles, we obtain the challenging range 
\be
100 \ GeV \ \leq \ {\rm susy \ breaking \ scale} \ \leq \ 1000 \ GeV\, .
\ee
As experiments confront energy scales larger than the susy breaking scale, one
may naturally expect to find a wealth of supersymmetric partners. Remarkably,
if the above scheme is borne out in Nature, the experimental discovery of
supersymmetry may not be far off in the future !

\subsection{Supersymmetric Yang-Mills dynamics}

In view of the remarkable confluence of circumstancial evidence
pointing towards a supersymmetric extension of the Standard Model,
it has become a priority in particle physics to understand 
better the perturbative and non-perturbative dynamics of 
supersymmetric gauge theories and their spontaneous breaking.
Over the past few years, substantial progress has been made
in the understanding of the non-perturbative dynamics of Yang-Mills theory
with extended supersymmetry, namely $\N=2$ and $\N=4$. On the one 
hand, by generalizing Montonen-Olive \cite{montonen} duality, Seiberg
and Witten \cite{seiberg} obtained the exact low energy 
effective action as well as the exact spectrum of BPS states
of $\N=2$ super Yang-Mills theory (with SU(2) gauge group) 
and established that the 
dynamics is always that of a Coulomb phase. On the other hand,
by analyzing $D3$-brane configurations \cite{polchinski} in certain
limits, Maldacena \cite{maldacena} arrived at an equivalence
conjecture
between Type IIB superstring theory on $AdS_5 \times S^5$ and $\N=4$
superconformal Yang-Mills theory. Assuming this equivalence,
further conjectures, purely on $\N=4$ super Yang-Mills, have emerged,
suggesting an extensive degree of non-renormalization \cite{lee}
perhaps beyond that required by the $\N=4$
supersymmetry alone \cite{howewest}. While it is unlikely that
$\N=4$ super Yang-Mills is completely integrable in the 
sense of 2-dimensional models, both developments mentioned above
suggest that the theory may be integrable in restricted sense.
While these developments are impressive, the physically most
pressing questions such as the dynamics of supersymmetry
breaking remain only partially under control to this date.

The purpose of the present lectures is to present an introduction
to 4-dimensional supersymmetric gauge theories, including those
with extended supersymmetry, as well as to the Seiberg-Witten 
non-perturbative solutions. Remarkably, these solutions possess
a natural and extremely useful mapping onto integrable systems in
classical mechanics, a subject to which we also present a review.
Finally, the map onto integrable systems, in particular onto Toda and
Calogero-Moser systems, is used to give a Seiberg-Witten construction
for general gauge algebras and a hypermultiplet in the adjoint
representation of the gauge algebra (a theory obtained from $\N=4$
super Yang-Mills by adding a mass term to the hypermultiplet).

\vfill\eject

\setcounter{equation}{0}
\section{Supersymmetric Yang-Mills in 4 Dimensions}

We begin by reviewing the particle and field contents and the construction of
Lagrangians invariant under supersymmetry for spins less or equal to 1. For
textbooks see \cite{susy}. 

\subsection{Supersymmetry Algebra}

Poincar{\'e} symmetry is generated by the translations ${\bf R}^4$ and Lorentz
transformations $SO(3,1)$, with generators $P_\mu$ and $L_{\mu \nu}$
respectively. (Here and below, $\mu, \nu =0,1,2,3$.) The complexified Lorentz
group is isomorphic to $SU(2) \times SU(2)$, and its  finite-dimensional
representations are usually labeled by two positive (or zero) half integers
$(s_+, s_-)$, $s_\pm \in {\bf Z}/2$. Scalar, 4-vector, and rank 2 symmetric
tensors transform under $(0,0)$, $(\12, \12)$ and $(1,1)$ respectively, 
left and right chirality fermions transform under $(\12,0)$ and $(0,\12)$
respectively, and selfdual and anti-selfdual rank 2 anti-symmetric tensors
transform under $(1,0)$ and $(0,1)$ respectively.

Supersymmetry enlarges the Poincar{\'e} algebra by including spinor {\it
supercharges},
\be
I=1,\cdots , \N \qquad 
\cases{Q_\alpha ^I  \quad  \ \ \alpha =1,2 & {\rm left \ Weyl \ spinor} \cr
& \cr
\bar Q_{\dot \alpha I}=(Q_\alpha ^I)^\dagger & {\rm right \ Weyl \ spinor} \cr}
\ee 
Here, $\alpha$ is a Weyl spinor label, and $\N$ is the number of independent
supersymmetries of the algebra. The supercharges transform as Weyl spinors 
of $SO(3,1)$, and are translation invariant, so that $[P_\mu , Q_\alpha ^I]=0$.
The remaining super-Lie algebra structure relations are 
\bea \label{susy}
\{ Q_\alpha ^I, \bar Q _{\dot \beta J}\} 
& = & 2 \sigma ^\mu _{\alpha \dot \beta} P_\mu \delta ^I _J  \\
\{ Q_\alpha ^I,  Q _\beta  ^J \} 
& = & 2 \epsilon _{\alpha  \beta} Z^{IJ} 
\eea
Here, we have used 2-component spinor notation, which is related to 4-component
Dirac spinor notation by
\be 
\gamma ^\mu = \left ( \matrix{0 & \sigma ^\mu \cr \bar \sigma ^\mu & 0 \cr}
\right )
\qquad
Q^I = \left ( \matrix{Q^I _\alpha \cr \bar Q_I^{\dot \alpha}} \right )
\ee 
By construction, the generators $Z^{IJ}$ are anti-symmetric in the indices
$I$ and $J$, and commute with all generators of the supersymmetry algebra. For
the last reason, the $Z^{IJ}$ are usually referred to as {\it central charges},
and we have
\be
Z^{IJ} = - Z^{JI} 
\qquad \quad
[Z^{IJ}, {\rm anything} ] =0\, .
\ee
Note that for $\N=1$, the anti-symmetry of $Z$ implies that $Z=0$.

The supersymmetry algebra is left invariant under a global phase rotation of 
all supercharges $Q_\alpha ^I$, forming a group $U(1)_R$. In addition, when
$\N >1$, the different supercharges may be rotated into one another under a
unitary transformation, belonging to $SU(\N)_R$. 
These (automorphism) symmetries
of the supersymmetry algebra are called {\it R-symmetries}. In quantum field
theories, part or all of these R-symmetries may be broken by anomaly effects.  

\subsection{Massless Particle Representations}

To study massless representations, we choose a Lorentz frame in which the
momentum takes the form $P^\mu = (E,0,0,E)$, $E>0$. The susy algebra
relation (\ref{susy}) then reduces to
\be \label{massless}
\{ Q_\alpha ^I, (Q _\beta ^J)^\dagger \} 
= 2 (\sigma ^\mu P_\mu)_{\alpha \dot \beta} \delta _J ^I 
= \left ( \matrix{4E & 0 \cr 0 & 0 \cr } \right ) _{\alpha \dot \beta} \delta
_J ^I\, .  
\ee
We consider only unitary particle representations, in which the operators
$Q_\alpha ^I$ act in a positive definite Hilbert space. The relation for
$\alpha = \dot \beta =2$ and $I=J$ implies
\be \label{vanishing} 
\{ Q_2 ^I , (Q_2 ^I)^\dagger \} = 0 
\quad \Longrightarrow \quad
Q_2 ^I =0, \quad Z^{IJ}=0\, .
\ee
The relation $Q_2 ^I=0$ follows because the left hand side of 
(\ref{vanishing}) 
implies that the norm of $Q_2 ^I |\psi\rangle$ vanishes for any state
$|\psi\rangle$ in the Hilbert space. The relation $Z^{IJ}=0$ then follows from
(\ref{susy}) for $\alpha =2$ and $\dot \beta =1$. The remaining supercharge
operators are 

\begin{itemize}

\item $Q_1 ^I$ \ which lowers helicity by $1/2$;

\item $\bar Q_{\dot 1} ^I = (Q_1 ^I )^\dagger$ \ which raises 
helicity by $1/2$.

\end{itemize}

\noindent
Together, $Q_1 ^I$ and $(Q_1 ^I)^\dagger$, with $I=1,\cdots , \N$ form a
representation of dimension $2^\N$ of the Clifford algebra associated with the
Lie algebra $SO(2\N)$. All the states in the representation may be obtained by
starting from the highest helicity state $|h \rangle$ and applying products of
$Q_1 ^I$ operators for all possible values of $I$.

Here, we shall only be interested in CPT invariant theories, such as quantum
field theories and string theories, for which the particle spectrum must be
symmetric under a sign change in helicity. If the particle spectrum obtained
as a Clifford representation in the above fashion is not already CPT
self-conjugate, then we shall take instead the direct sum with its CPT
conjugate. For helicity $\leq 1$, the spectra are listed in table 3.
The $\N=3$ and $\N=4$ particle spectra then coincide, and the quantum field
theories are identical. 

\begin{table}[b]
\begin{center}
\begin{tabular}{|c|c| c|c|c|c|c|} \hline 
Helicity  & $\N=1$ & $\N=1$ & $\N=2$ & $\N=2$ & $\N=3$ & $\N=4$ \\
$\leq 1$  & gauge  & chiral & gauge  & hyper  & gauge  & gauge  \\ 
\hline \hline
1         & 1      & 0      & 1      & 0      & 1      & 1      \\ \hline
$1/2$     & 1      & 1      & 2      & 2      & 3+1    & 4      \\ \hline
0         & 0      & 1+1    & 1+1    & 4      & 3+3    & 6      \\ \hline
$-1/2$    & 1      & 1      & 2      & 2      & 1+3    & 4      \\ \hline
$-1$      & 1      & 0      & 1      & 0      & 1      & 1      \\ \hline
\hline Total $\#$ & $2 \times 2$ & $2 \times 2$ & $ 2 \times 4$ & 8 & $ 2
\times 8$ & 16  \\ \hline
\end{tabular}
\end{center}
\caption{Numbers of Massless States as a function of $\N$ and helicity}
\label{table:3}
\end{table}

\subsection{Massive Particle Representations}

For massive particle representations, we choose the rest frame with $P^ \mu =
(M,0,0,0)$, so that the first set of susy algebra structure relations takes the
form
\be
\{ Q_\alpha ^I, (Q_\beta ^J)^\dagger \} = 2M \delta _\alpha ^\beta \delta _J
^I\, .
\ee
To deal with the full susy algebra, it is convenient to make use of the
$SU(\N)$ $R$-symmetry to diagonalize in blocks of $2\times 2$ 
the anti-symmetric matrix
$Z^{IJ} = - Z^{JI}$. To do so, we split the label $I$ into two labels :
$I=(a,i)$ where $a=1,2$ and $i=1,\cdots ,r$, where $\N=2r$ for $\N$ even (and
we append a further single label when $\N$ is odd). We then have
\be
Z = {\rm diag} (\epsilon Z_1, \cdots , \epsilon Z_r, \# )
\qquad 
\epsilon ^{12} = - \epsilon ^{21} =1\, ,
\ee
where $\# $ equals 0 for $\N=2r+1$ and $\#$ is absent for $\N=2r$. The $Z_i$,
$i=1,\cdots ,r$ are real {\it central charges}. In terms of linear
combinations $ {\cal Q} ^i _{\alpha \pm} \equiv
\12 \bigl ( Q_\alpha ^{1i} \pm \sigma ^0 _{\alpha \dot \beta} (Q_\beta
^{2i})^\dagger \bigr ), $
the only non-vanishing susy structure relation left is (the $\pm$ signs below 
are correlated)
\be \label{BPS} 
\{ {\cal Q}^i _{\alpha \pm} , ( {\cal Q}^j _{\beta \pm} )^\dagger \}
=
\delta _j ^i \delta _\alpha ^\beta (M \pm Z_i)\, .
\ee 
In any {\it unitary particle representation},  the operator on the left hand
side of (\ref{BPS}) must be positive, and thus we obtain the {\it BPS bound}
(for Bogomolnyi-Prasad-Sommerfield)
\be 
M \geq |Z_i| \qquad i=1,\cdots , r = \bigg [ {\N \over 2} \bigg ]\, .
\ee
Whenever one of the values $|Z_i|$ equals $M$, the BPS bound is (partially)
saturated and either the supercharge ${\cal Q}^i_{\alpha +}$ or ${\cal
Q}^i_{\alpha -}$ must vanish. The supersymmetry representation then suffers
{\it multiplet shortening}. More precisely, if we have $M = |Z_i|$ for
$i=1,\cdots, r_o$, and $M>|Z_i|$ for all other values of $i$, the susy algebra
is effectively a Clifford algebra associated with $SO(4\N -2r_o)$, the
corresponding representation is said to be $1/2^{r_o}$ BPS, and has dimension 
$2^{2\N-r_o}$. 

\begin{table}[b]
\begin{center}
\begin{tabular}{|c|c| c|c|c|c|c|} \hline 
Spin      & $\N=1$ & $\N=1$ & $\N=2$ & $\N=2$     & $\N=2$     & $\N=4$    \\
$\leq 1$  & gauge  & chiral & gauge  & BPS gauge  & BPS hyper  & BPS gauge \\
\hline \hline 
1         & 1      & 0      & 1      & 1          & 0          & 1    \\ \hline
$1/2$     & 2      & 1      & 4      & 2          & 2          & 4    \\ \hline
0         & 1      & 2      & 5      & 1          & 4          & 6    \\ \hline
\hline 
Total $\#$& 8      & 4      & 16     & 8          & 8          & 16   \\ \hline
\end{tabular}
\end{center}
\caption{Numbers of Massive States as a function of $\N$ and spin}
\label{table:4}
\end{table}

\subsection{Field Contents of Supersymmetric Field Theories}

The analysis of the preceding two subsections has revealed that the
supersymmetry particle representations for $1 \leq \N \leq 4$, with spin less
or equal to 1, simply consist of customary spin 1 vector particles, spin 1/2 
fermions and spin 0 scalars. Correspondingly, {\it the fields in 
supersymmetric theories with spin less or equal to 1 are customary spin 1 gauge
fields,\footnote{By the principles of unitary quantum field
theory, spin 1 vector particles are described by gauge fields.} spin 1/2 Weyl
fermion fields and spin 0 scalar fields, but these fields are restricted to
enter in multiplets of the relevant supersymmetry algebras.} Here, we shall
need only the $\N=2$ BPS gauge field, instead of the full $\N=2$ gauge
multiplet, and henceforth we shall restrict to this case.

\medskip

Let $\G$ denote the gauge algebra, associated with a compact Lie group $G$.
Then, two kinds of multiplets occur. For any $1\leq \N \leq 4$, we have a
gauge multiplet, which transforms under the adjoint representation of $\G$.
For $\N=3,4$, this is the only possible multiplet. Furthermore, the quantum
field theories with $\N=3$ supersymmetry turn out to coincide with those with
$\N=4$ supersymmetry in view of CPT invariance. Thus, we shall limit our
discussion to the $\N=4$ theories, as is customarily done.  For $\N=1$ and
$\N=2$, we also have {\it matter multiplets} : for $\N=1$, this is the {\it
chiral multiplet}, and for $\N=2$ this is the {\it hypermultiplet}, both of
which may transform under an arbitrary (unitary, and possibly reducible)
representation ${\cal R}$ of $\G$.

\medskip

We shall now briefly discuss each of these field multiplets in terms of their
components, which are the customary gauge field $A_\mu$, left Weyl fermions
$\psi _\alpha$ and $\lambda _\alpha$ and scalar fields $\phi$ and $H$. (The
notation is conventional.) A Dirac fermion $\psi _D$ is the
direct sum of a left and right Weyl fermion $\psi _D = (\psi \ \bar \lambda)$,
while a {\it Majorana} (or {\it real}) fermion $\psi _M$ 
is a Dirac fermion with
$\lambda = \psi$ : $\psi _M = (\psi \ \bar \psi)$. 
In 4 space-time dimensions, a
Majorana fermion is equivalent to a Weyl fermion.

\begin{itemize}

\item $\N=1$ {\it Gauge Multiplet} \ $(A_\mu \ \lambda_\alpha)$, where $\lambda
_\alpha$ is the gaugino Majorana fermion;

\item $\N=1$ {\it Chiral Multiplet} \ $(\psi _\alpha \ \phi)$, where $\psi
_\alpha$ is a left Weyl fermion and $\phi$ a complex scalar, in the
representation ${\cal R}$ of $\G$.

\item $\N=2$ {\it Gauge Multiplet} \ $(A_\mu \ \lambda _{\alpha \pm} \ \phi)$, 
where $\lambda _{\alpha \pm}$ form a Dirac fermion, and $\phi$ is the complex
{\it gauge scalar}. Under the $SU(2)_R$ symmetry of $\N=2$ supersymmetry, the
fields $A_\mu$ and $\phi$ are singlets, while the fields $\lambda _+$ and
$\lambda _-$ transform as a doublet.

\item $\N=2$ {\it Hypermultiplet} \ 
$(\psi_{\alpha +}\ H_\pm\ \psi_{\alpha +})$,
where $\psi _{\alpha \pm}$ form a Dirac spinor and $H_\pm$ are complex scalars,
transforming under the representation ${\cal R}$ of $\G$. Under the $SU(2)_R$
symmetry, $\psi _\pm$ are singlets, while $H_+$ and $H_-$ transform as a
doublet.

\item $\N=4$ {\it Gauge Multiplet} \ $(A_\mu \ \lambda _\alpha ^A \ \phi ^I)$,
where $\lambda _\alpha ^A$, $A=1,2,3,4$ are left Weyl fermions (equivalent to
two Dirac fermions) and $\phi ^I$, $I=1,\cdots , 6$ are real scalars
(equivalent to three complex scalars). Under the $SU(4)_R$ symmetry of $\N=4$
supersymmetry, the gauge field $A_\mu$ is a singlet, the fermions $\lambda
_\alpha ^A$ transform in the fundamental {\bf 4}, the scalars $\phi ^I$
transform in the rank 2 anti-symmetric {\bf 6}. 

\end{itemize}

It is very useful to record the transformation properties and representation
contents under supersymmetry {\it subalgebras} of the $\N=2$ and $\N=4$
supersymmetry algebras,

\begin{itemize}

\item ($\N=2$ gauge) = ($\N=1$ gauge) $\oplus$ ($\N=1$ chiral in
adjoint rep. of $\G$);

\item ($\N=2$ hyper) = ($\N=1$ chiral) $\oplus$ ($\N=1$ chiral in complex
conjugate rep.);

\item ($\N=4$ gauge) = ($\N=2$ gauge) $\oplus$ ($\N=2$ hyper in adjoint
rep. of $\G$).

\end{itemize}  

\noindent
We conclude by remarking that all of the above multiplets are parity
self-conjugate, except for the $\N=1$ chiral multiplet, since it is built out
of a single left Weyl fermion.

\subsection{N=1 Supersymmetric Lagrangians}

Lagrangians invariant under supersymmetry are just customary Lagrangians of
gauge, spin 1/2 fermion and scalar fields, (these fields are arranged in
multiplets of the supersymmetry algebra, as established previously) with 
certain special relations amongst the coupling constants and masses. We begin
with the construction of Lagrangians invariant under the smallest degree of
supersymmetry, namely $\N=1$.
For our purposes, the Lagrangians of interest are of two restricted kinds

\begin{itemize}

\item (1) Renormalizable $\N=1$ supersymmetric gauge theories;

\item (2) More general {\it low energy effective $\N=1$ supersymmetric
theories}, with the property that any monomial term in the Lagrangian has a
total of no more than two derivatives on all boson fields and no more than one
derivative on all fermion fields. Such restricted Lagrangian may be viewed as
describing phenomena in the limit of low energy and momenta, and are well
familiar from soft pion physics.

\end{itemize}

\noindent
In 4 space-time dimensions, all Lagrangians in group (1) automatically belong
in group (2). Thus, we seek to construct all Lagrangians in (2).

\medskip

We consider first the case of only the $\N=1$ gauge multiplet $(A_\mu \ \lambda
_\alpha)$, and proceed by writing down all possible gauge invariant polynomial
terms of dimension 4 using minimal coupling. One finds
\be \label{lag}
\L = 
- {1 \over 2 g^2} \tr F_{\mu \nu} F^{\mu \nu}
+{\theta \over 8 \pi ^2} \tr F_{\mu \nu} \tilde F^{\mu \nu} 
-{i \over 2} \tr \bar \lambda \bar \sigma ^\mu D_\mu \lambda\, ,
\ee
where $g$ is the gauge coupling, $\theta$ is the instanton angle, the field
strength is $F_{\mu \nu} 
= \partial _\mu A_\nu - \partial _\nu A_\mu + i [A_\mu , A_\nu]$, 
$\tilde F_{\mu\nu}=\12 \epsilon_{\mu\nu\rho\kappa}F^{\rho\kappa}$ 
is the Poincar{\'e} dual of $F$, and $D_\mu = \partial _\mu
\lambda + i [A_\mu , \lambda]$. Remarkably, $\L$ is automatically invariant
under the $\N=1$ supersymmetry transformations 
\bea \label{xi}
\delta _\xi A_\mu 
& = & 
i \bar \xi \bar \sigma _\mu \lambda - i \bar \lambda \bar \sigma _\mu \xi 
\nonumber \\
\delta _\xi \lambda 
& = &
\sigma ^{\mu \nu } F_{\mu \nu } \xi 
\eea
where $\xi$ is a spin 1/2 valued infinitesimal supersymmetry parameter. Note
that the addition in (\ref{lag}) of a Majorana mass term $m\lambda
\lambda$ for $\lambda$ would spoil supersymmetry.

\medskip

Unfortunately, as soon as scalar fields are to be included, such as is the case
when dealing with chiral multiplets, it is no longer so easy to guess
supersymmetry invariant Lagrangians. For example, for a single chiral multiplet
$(\psi \ \phi)$ already, the restricted case (1) of just renormalizable
Lagrangians is much more involved and much less intuitive. We quote here the
result and shall leave the derivation of this case as well as of the more 
general ones for the next subsections. We have
\be 
\L = - \partial _\mu \phi ^* \partial ^\mu \phi 
- i \bar \psi \bar \sigma ^\mu \partial _\mu \psi 
- \bigg | {\partial U \over \partial \phi } \bigg | ^2
- {\rm Re} \bigg ( \psi \psi {\partial ^2 U \over \partial \phi ^2} \biggr )
\ee
where $U(\phi)$ is forced by $\N=1$ supersymmetry to be a complex analytic
(holomorphic) scalar function of $\phi$, called the {\it superpotential}.
Renormalizability furthermore restricts $U$ to be a polynomial of degree no
larger than 3 : $U(\phi)= U_0 + U_1 \phi + U_2 \phi ^2 + U_3 \phi ^3$. Here,
$U_i$ are complex coupling constants, with $U_3$ specifying both the scalar
quartic self-coupling and Yukawa coupling, and $U_2$ specifying both the
scalar and fermion masses.   

\subsection{N=1 Superfield Methods}

Generally, the construction of invariants is highly facilitated by the use of
tensor-type methods, in which all fields which transform linearly into one
another are assembled into a single multiplet. Such methods are familiar from
the construction of Lorentz invariants or gauge invariants.

\medskip

The construction of field multiplets containing all fields that transform
linearly into one another under supersymmetry presents a new challenge,
because such multiplets must hold simultaneously bosonic and fermionic fields.
To achieve such superpositions of commuting Bose fields and anti-commuting
Fermi fields consistently with Lorentz invariance, requires the introduction of
new anti-commuting spin 1/2 parameters or coordinates. For $\N=1$
supersymmetry, we introduce a (constant) left Weyl spinor coordinate $\theta
_\alpha$ and its complex conjugate $\bar \theta ^{\dot \alpha} = (\theta
_\alpha)^\dagger$, satisfying
\be
[x^\mu, \theta _\alpha] = [x^\mu , \bar \theta ^{\dot \alpha}]
= \{ \theta _\alpha , \theta _\beta \} = \{ \theta _\alpha, \bar \theta ^{\dot
\beta}\} = \{ \bar \theta ^{\dot \alpha} , \bar \theta ^{\dot \beta} \}=0\, .
\ee
Often, $\theta$ and $\bar \theta$ are regarded as additional space-time
coordinates on the same footing as the customary coordinate $x^\mu$, which
together then parametrize {\it superspace}.

\medskip

Superderivatives are defined by
\be 
D_\alpha \equiv {\partial \over \partial \theta ^\alpha} + i \sigma ^\mu
_{\alpha \dot \alpha} \bar \theta ^{\dot \alpha} \partial _\mu
\qquad \qquad 
\bar D_{\dot \alpha} \equiv -{\p \over \p \bar \theta ^{\dot \alpha}} 
- i \theta ^{ \alpha} \sigma ^\mu _{\alpha \dot \alpha}  \partial _\mu
\ee
where differentiation and integration of $\theta$ coordinates are defined by
\be
{\partial \over \partial \theta ^\alpha} 
(1,\theta ^\beta, \bar \theta ^{\dot \beta}) 
\equiv  
\int \! d\theta ^\alpha (1,\theta ^\beta, \bar \theta ^{\dot \beta})
\equiv 
(0,\delta _\alpha {} ^\beta, 0)\, .
\ee
We also define Lorentz scalar differentials by 
\be
d^2 \theta \equiv {1 \over 4} d\theta _\alpha d \theta ^\alpha, 
\qquad 
d^2 \bar \theta \equiv {1 \over 4} d \bar
\theta ^{\dot \alpha} d \bar \theta _{\dot \alpha},
\qquad 
d^4 \theta \equiv d^2 \theta d^2 \bar \theta\, ,
\ee
so that
\be
\int d^2 \theta \ \theta \theta  =
\int d^2 \bar \theta \ \bar \theta \bar \theta =
\int d^4 \theta \ \theta \theta \bar \theta \bar \theta =1
\ee
For general notations and conventions for spinors and their contractions, see
Appendix A.

\medskip

A {\it superfield} is defined as a general function of the superspace
coordinates $x^\mu, \theta _\alpha, \bar \theta ^{\dot \alpha}$. Since the
square of each $\theta ^\alpha$ or of each $\bar \theta ^{\dot \alpha}$
vanishes, superfields admit finite Taylor expansions in powers of $\theta$ and
$\bar \theta$. Thus, the most general superfield $S (x,\theta, \bar \theta)$
yields the following {\it component expansion}
\bea \label{sfield}
S (x,\theta, \bar \theta)
& = &
\phi (x) + \theta \psi(x) + \bar \theta \bar \chi (x) + \bar \theta \bar
\sigma ^\mu \theta A_\mu (x) + \theta \theta f(x) + \bar \theta \bar \theta
 g^* (x) \nonumber \\
&& + i \theta \theta \bar \theta \bar \lambda (x) - i \bar \theta \bar \theta 
\theta \rho (x) + \12 \theta \theta \bar \theta \bar \theta D(x)\, .
\eea
Naturally, in quantum field theory, one restricts to superfields that are
either bosonic or fermionic, so that the superfield has definite (anti-)
commutation relations with $\theta$ and $\bar \theta$,
\bea
{\rm bosonic \ superfield} & \qquad & [S, \theta ^\alpha]=[S, \bar \theta
_{\dot \alpha}]=0 \nonumber \\
{\rm fermionic \ superfield} & \qquad & \{S, \theta ^\alpha\}=\{S, \bar
\theta _{\dot \alpha}\}=0\, . 
\eea
Thus, if $S$ is bosonic, the component fields $\phi$, $A_\mu$, $f$, $g$ and
$D$ are bosonic as well, while the fields $\psi$, $\chi$, $\lambda$ and $\rho$
are fermionic. On the other hand, if $S$ is fermionic, the component fields
$\phi$, $A_\mu$, $f$, $g$ and $D$ are fermionic as well, while the fields
$\psi$, $\chi$, $\lambda$ and $\rho$ are bosonic.  Mathematically, the
superfields belong to a ${\bf Z}_2$ graded algebra of functions on superspace,
with the even grading associated with bosonic fields and the odd grading
asssociated with fermionic fields. We shall denote the grading by
$g(S)$. 

\medskip

Superderivatives on superfields satisfy the following graded differentiation
rule
\bea
D_\alpha (S _1 S _2) 
&=& (D_\alpha S _1) S _2 
+ (-)^{g(S _1)  g(S _2)} S _1 (D_\alpha S _2)
\nonumber \\  
D_{\dot \alpha} (S _1 S _2) 
&=& (D_{\dot \alpha} S _1) S _2 + (-)^{g(S _1) g(S _2)} S _1
(D_{\dot \alpha} S _2)\, ,  
\eea
where $g(S _i)$ is the grading of the field $S _i$.

\medskip

The only {\it elementary} superfields we shall need here are bosonic
and they have the property
that the zero-th order term, the field $\phi (x)$, is a Lorentz scalar.
We shall also encounter a composite fermionic superfield, whose first
component is a spinor, and which is associated with the gauge field
strength. If
$S$ is a bosonic scalar superfield, then $\phi$, $f$, $g$, $D$ are scalars
while $\psi$, $\chi$ $\lambda$, $\rho$ are left-handed Weyl spinors, and
$A_\mu$ is a gauge field.
    
\medskip

On superfields, supersymmetry transformations are naturally realized in a
linear way via super-differential operators (just as on ordinary fields,
translations and Lorentz transformations are realized in a linear way via
differential operators). The infinitesimal supersymmetry parameter is still a
constant left Weyl spinor $\xi$, as in (\ref{xi}) and we have
\be \label{tfon}
\delta _\xi S = (\xi Q + \bar \xi \bar Q) S 
\ee
with the supercharges defiend by
\be
Q_\alpha = {\partial \over \partial \theta ^\alpha} - i \sigma ^\mu
_{\alpha \dot \alpha} \bar \theta ^{\dot \alpha} \partial _\mu
\qquad \qquad
\bar Q_{\dot \alpha} = -{\p \over \p \bar \theta ^{\dot \alpha}} 
+ i \theta ^{ \alpha} \sigma ^\mu
_{\alpha \dot \alpha}  \partial _\mu
\ee
The super-differential operators $D_\alpha$ and $Q_\alpha$ 
differ only by a sign
change, and generate left and right actions of supersymmetry respectively.
Their relevant structure relations are
\be
\{ Q_\alpha, \bar Q_{\dot \beta}\}= 2 \sigma ^\mu _{\alpha \dot \beta} P_\mu
\qquad \qquad
\{ D_\alpha, \bar D_{\dot \beta}\}= - 2 \sigma ^\mu _{\alpha \dot \beta} P_\mu
\ee
where $P_\mu = i \p _\mu$. 
Since left and right actions mutually commute, all 4 components of $D$
anti-commute with all 4 components of $Q$ : $\{ Q_\alpha , D_\beta\}=
\{Q_\alpha , \bar D^{\dot \beta}\}=0$, and their complex conjugate relations.
Further useful relations are that the product of any three $D$'s or any three
$Q$'s vanishes,
\be  \label{useful}
D_\alpha D_\beta D_\gamma = Q_\alpha Q_\beta Q_\gamma =0
\ee
as well as their complex conjugate relations.

\subsection{Irreducible Superfields of N=1}

The type of superfield introduced above is in general highly reducible, and the
irreducible components may be found by imposing supersymmetric conditions on
the superfield. 

\medskip

\noindent
(a) The {\it Chiral Superfield} $\Phi$ is obtained by imposing the condition 
\be \label{dphi} 
\bar D_{\dot \alpha} \Phi =0\, .
\ee
The {\it anti-chiral superfield} $\Phi ^\dagger$ is obtained by imposing
$D_\alpha \Phi ^\dagger =0$. These conditions are invariant under the
supersymmetry transformations  of (\ref{tfon}) since $D$ or $\bar D$ and $Q$ or
$\bar Q$ anti-commute.  Equation (\ref{dphi}) may be solved in terms of the
composite coordinates
\be
x^\mu _\pm = x^\mu \pm i \theta \sigma ^\mu \bar \theta,
\ee
which satisfy
\be
\bar D_{\dot \alpha} x^\mu _+=0,
\quad\quad
D_\alpha x_-^\mu =0,
\ee 
and we have (a factor of $\sqrt 2$ has been inserted multiplying $\psi$ to give
this field standard normalization)
\bea \label{chiral}
\Phi (x,\theta, \bar \theta) 
&=&
\phi (x_+) + \sqrt 2 \theta \psi (x_+) + \theta \theta F(x_+) 
\nonumber \\
\Phi ^\dagger (x,\theta , \bar \theta)
&=&
\phi ^* (x_-) + \sqrt 2 \bar \theta \bar \psi (x_-) 
+ \bar \theta \bar \theta F^*(x_-)
\eea 
The component fields $\phi$ and $\psi$ are the scalar and left Weyl spinor
fields of the chiral multiplet respectively, as discussed previously. The field
$F$ has not appeared previously. 
The field equation for $F$ is always algebraic,
so that $F$ is a non-dynamical or {\it auxiliary field} of the chiral
multiplet. 

\medskip

\noindent
(b) The {\it Vector Superfield} is obtained by imposing the condition
\be
V=V^\dagger
\ee 
on a general superfield of the type (\ref{sfield}). This
condition sets $\chi = \psi$, $g=f$ and $\rho=\lambda$ in (\ref{sfield}), and
requires that the fields $\phi$, $A_\mu$ and $D$ be real. It is conventional to
use a specific notation for vector superfields and it is convenient to define
its expansion by
\bea
V(x,\theta, \bar \theta)
& = & 
v(x) 
+ \theta \chi (x) +  \bar \theta \bar \chi (x) 
+ \theta \theta f(x) + \bar \theta \bar \theta f^*(x) 
+ \bar \theta \bar \sigma ^\mu \theta A_\mu (x) 
\nonumber \\ 
&& + i \theta \theta \bar \theta \bigl (\bar \lambda (x) +\12 \bar \sigma ^\mu
\partial _\mu \chi (x) \bigr ) 
-i \bar \theta \bar \theta \theta \bigl ( \lambda (x)  +\12 \sigma ^\mu
\partial _\mu \bar \chi (x) \bigl ) 
\nonumber \\
&& + \12 \theta \theta \bar \theta \bar \theta \bigl (D(x) 
+ \12 \partial _\mu \partial ^\mu v(x) \bigr )
\eea
The advantage of the shifts by derivatives 
in the fields will become clear shortly.

\bigskip

The {\it gauge superfield} is a special case of a vector superfield. On a
single (Abelian) vector superfield $V$, 
the reality condition $V^\dagger =V$ is preserved
upon addition of a chiral superfield $\Lambda$ and its complex conjugate
$\Lambda ^\dagger$, as follows,
\be \label{gauge}
V \longrightarrow V' = V + i\Lambda -i \Lambda ^\dagger\, .
\ee
(Below, we shall assume that $\Lambda $ has the same component field 
decomposition as $\Phi$
in (\ref{chiral}), in order to save on some notation.) 
Under this transformation, the component fields $\lambda$ and $D$ of $V$ are
unchanged, $v$, $\chi$ and $f$ transform in a purely algebraic way,
\bea \label{algebraic}
v \longrightarrow v' &=& v+i \phi -i \phi ^* \nonumber \\
\chi \longrightarrow \chi ' &=& \chi + i \sqrt 2 \psi \nonumber \\
f \longrightarrow f' &=& f + iF 
\eea
while the field $A_\mu$ transforms as an Abelian gauge field
\be
A_\mu \longrightarrow A_\mu ' = A_\mu + \partial _\mu (\phi + \phi ^*)\, .
\ee 
Thus, it is natural to view (\ref{gauge}) as the superfield generalization of
a gauge transformation on an Abelian gauge superfield $V$. 

\bigskip

The non-Abelian generalization of the gauge field is such that $V$ takes values
in the Lie algebra $\G$ of the gauge group $G$ 
(or equivalently transforms under
the adjoint representation of $\G$) and that the transformation 
(\ref{gauge}) is
replaced by the following non-linear gauge transformation law,
\be \label{nonabgauge}
e^ V \longrightarrow e^ {V'} = e^{-i\Lambda ^\dagger} e^V e^{i \Lambda} \,.
\ee 
which again preserves the reality condition $V^\dagger = V$, assuming that
$\Lambda$ is a chiral superfield transforming under the adjoint representation
of the gauge algebra $\G$.

\medskip
As is clear from (\ref{algebraic}), (and an analogous result holds for the
non-Abelian case), the component fields $v$, $\chi$ and $f$ may be gauged away
in an algebraic way, without implying any dynamical constraints. The gauge in
which this is achieved is called the {\it Wess-Zumino gauge}, and is almost
always imposed when performing practical calculations in 
the superfield formulation. What remains is
the gauge superfield in {\it Wess-Zumino gauge}, given by
\be \label{vector}
V(x,\theta \bar \theta)
=
\bar \theta \bar \sigma ^\mu \theta A_\mu (x)
+ i \theta \theta \bar \theta \bar \lambda (x) - i \bar \theta \bar \theta 
\theta \lambda (x) + \12 \theta \theta \bar \theta \bar \theta D(x)
\ee
The component fields $A_\mu$ and $\lambda$ are the gauge and gaugino fields of
the gauge multiplet respectively, as discussed previously. The field $D$ has
not appeared previously and is an {\it auxiliary field}, just as $F$ was an
auxiliary field for the chiral multiplet. 

\medskip

The role of the auxiliary fields $F$ and $D$ in the superfield formalism is to
provide a linearization of the supersymmetry transformations, as well as to
allow for an off-shell realization on the fields of the supersymmetry
algebra, as given in (\ref{tfon}).

\medskip

Working out the supersymmetry transformation (\ref{tfon}) on  chiral and
vector superfields in terms of components, we see that the only
contribution to the auxiliary fields is from the $\theta \partial$ term of
$Q$ and thus takes the form of a total derivative. However, because the form
(\ref{vector}) was restricted to a Wess-Zumino gauge, $F$ and $D$
transform by a total derivative only if $F$ and $D$ are themselves gauge
singlets, in which case we have
\bea
\delta _\xi F &=& i \sqrt 2 \partial _\mu (\bar \xi \bar \sigma ^\mu \psi) \\
\delta _\xi D &=& \partial _\mu (i\bar \xi \bar \sigma ^\mu \lambda
-i \bar \lambda \bar \sigma ^\mu \xi)\, .
\eea
(For non-gauge singlet auxiliary fields, the transformation laws will involve
in addition non-total-derivative terms; we shall not need their expressions
here.) These transformation properties guarantee that the $F$ and
$D$ auxiliary fields yield supersymmetric invariant Lagrangian terms. Thus, we
have two different ways of building terms for an invariant Lagrangian,
\bea
{\rm F-terms} & \qquad & \L_F = F = \int d^2 \theta \ \Phi \\
{\rm D-terms} & \qquad & \L_D = \12 D = \int d^4 \theta \  V\, .
\eea      
The first starts from a chiral superfield $\Phi$ and the second from a vector
superfield $V$.

\subsection{General N=1 Susy Lagrangians via Superfields}

The $F$ and $D$ terms used to construct invariants in the previous subsection
need not be elementary fields, and may be gauge invariant composites of
elementary fields instead. Allowing for this possibility, we may now derive the
most general possible $\N=1$ invariant Lagrangian in terms of superfields.
To do so, we need the following ingredients.

\medskip

(1) \ Any complex analytic function $U$ depending only on left chiral
superfields $\Phi ^i$ (but not on their complex conjugates) is itself a left
chiral superfield,
\be
\bar D _{\dot \alpha} \Phi ^i =0 
\quad \Longrightarrow  \quad
\bar D_{\dot \alpha} U(\Phi ^i)=0\, .
\ee
Thus, for any complex analytic function $U$, called the {\it superpotential},
we may construct an invariant contribution to the 
Lagrangian by forming an $F$-term
\be
\L _U = \int d^2 \theta \ U(\Phi ^i) + {\rm complex \ conjugate}\, .
\ee
Using the component expansion of (\ref{chiral}), this Lagrangian takes the
component form
\be
\L _U = \sum _i F^i {\partial U \over \partial \phi ^i}
-\12 \sum _{i,j} \psi ^i \psi ^j {\partial ^2 U \over \partial \phi ^i \partial
\phi ^j} + {\rm complex \ conjugate}
\ee

\medskip

(2) \ Actually, the gauge field strength is a fermionic left chiral (spinor)
superfield $W_\alpha$, which is constructed out of the gauge superfield $V$ by
\be
W_\alpha = - {1 \over 4} \bar D \bar D \big (e^{-V} D_\alpha e^{+V} \big )\, .
\ee
In view of (\ref{useful}), we automatically have $\bar D _{\dot \beta}
W_\alpha =0$, so that $W_\alpha$ is chiral. Decomposing $W_\alpha$ in
components, one finds
\be \label{fieldstrength}
W_\alpha (x, \theta , \bar \theta ) = 
-i \lambda _\alpha (x_+) + \theta _\alpha D (x_+) 
- {i \over 2} (\sigma ^\mu \bar \sigma ^\nu)_\alpha {}^\beta \theta _\beta 
F_{\mu \nu} (x_+)
+ \theta \theta \sigma ^\mu _{\alpha \dot \beta} D_\mu \bar \lambda ^{\dot
\beta} (x_+)
\ee
The gauge field strength may be used as a chiral superfield along with
elementary (scalar) chiral superfields to build up $\N=1$ supersymmetric
Lagrangians via $F$-terms. In view of our restriction to Lagrangians
with no more than two derivatives on Bose fields, $W$ can enter at most
quadratically. Denoting by $W^a _\alpha$ the components of the gauge multiplet,
with the index $a$ running over the adjoint representation, we shall be
interested in bilinears of the form
\bea
W^{\alpha  a} W_\alpha ^b
=
-\lambda ^a \lambda ^b -i \theta \lambda ^a D^b - i \theta \lambda ^b D^a
-\12 \theta (\sigma ^\mu \bar \sigma ^\nu) (\lambda ^a F^b _{\mu \nu} 
+ \lambda
^b F^a _{\mu \nu}) \qquad \quad
\nonumber \\  
 - \theta \theta \biggl ( i \lambda ^a \sigma ^\mu \partial _\mu \bar
\lambda ^b +i \lambda ^b \sigma ^\mu \partial _\mu \bar \lambda ^a
+{1 \over 4} (F^{\mu \nu  a} + i \tilde F^{\mu \nu  a})
(F^b _{\mu \nu} + i \tilde F^b _{\mu \nu}) - D^a D^b \biggr )
\eea
in Wess-Zumino gauge. It is understood that all fields on the right hand side
depend upon the composite coordinate $x_+ ^\mu = x^\mu + i \theta \sigma ^\mu
\bar \theta$, just as was the case in (\ref{fieldstrength}).

\medskip

Elementary (scalar) chiral superfields may enter in any functional way without
generating more than 2 derivatives. Thus, the most general gauge kinetic and
self-interaction term is from the $F$-term of the gauge field strength
$W_\alpha$ and the elementary (scalar) chiral superfields $\Phi ^i$ as follows,
\be
\L _G = \int d^2 \theta \ \tau _{ab}(\Phi ^i) W^a W^b + {\rm complex \
conjugate}\, .
\ee
Here, $a$ and $b$ stand for the gauge index running over the adjoint
representation of $\G$. The functions $\tau _{ab}(\Phi ^i)$ are again
required to be complex analytic. It is assumed that the function $\tau
_{ab}(\Phi^i)$ under suitable gauge transformations of $\Phi^i$ will transform
under the symmetrized square of the adjoint representation of $\G$.

\medskip

The gauge kinetic Lagrangian may also be worked out in components, and is given
by (index summation over indices $i,j$ and $a,b$ are suppressed for the sake of
brevity)
\bea
\L _G & = &
- \lambda ^a \lambda ^b 
\biggl ( F^i {\partial \tau _{ab} \over \partial \phi ^i}
-\12 \psi ^i \psi ^j  {\partial ^2 \tau _{ab} \over \partial \phi ^i \partial
\phi ^j} \biggr )
\nonumber \\
&&-{1 \over 2 \sqrt 2} {\partial \tau _{ab} \over \partial \phi ^i}
\psi ^i \biggl ( -i \lambda ^a D^b - i \lambda ^b D^a
-\12  (\sigma ^\mu \bar \sigma ^\nu) (\lambda ^a F^b _{\mu \nu} + \lambda
^b F^a _{\mu \nu}) \biggr )
\nonumber \\
&&- \tau _{ab} \biggl ( i \lambda ^a \sigma ^\mu \partial _\mu \bar
\lambda ^b +i \lambda ^b \sigma ^\mu \partial _\mu \bar \lambda ^a
+{1 \over 4} (F^{\mu \nu  a} + i \tilde F^{\mu \nu  a})
(F^b _{\mu \nu} + i \tilde F^b _{\mu \nu}) - D^a D^b \biggr )
\qquad
\eea 

\medskip

(3) \ The left and right chiral superfields 
$\Phi ^i $ and $(\Phi ^i) ^\dagger$,
as well as the gauge superfield $V$, may be combined into a gauge invariant
vector superfield $K(e^V \Phi ^i , (\Phi ^i) ^\dagger)$, provided the gauge
algebra is realized linearly on the fields $\Phi ^i$. (For the general case,
including when part or all of the gauge group is realized non-linearly, see
Wess and Bagger \cite{susy}; we shall not need this case here.) 
The function $K$ is
called the {\it K{\"a}hler potential}. Assuming that the gauge transformations
$\Lambda$ act on $V$ by (\ref{nonabgauge}), 
the chiral superfields $\Phi$ transform
as
\be
\Phi \longrightarrow \Phi ' = e^{-i \Lambda } \Phi\, ,
\ee
so that $e^V \Phi$ transforms as $\Phi$. An invariant Lagrangian may
be constructed via a $D$-term,
\be 
\L _K = \int d^4 \theta  K(e^V \Phi ^i, (\Phi ^i) ^\dagger)\, .
\ee
Upon expanding $\L_K$ in components, one sees immediately that the leading
terms are 
\be 
\L _K \sim - D_\mu \phi ^* D^\mu \phi -i \bar \psi \bar \sigma ^\mu D_\mu
\psi\, , 
\ee
and thus already generates an action with two derivatives on boson fields.
As a result, $K$ must be a function only of the superfields
$\Phi ^i$ and $(\Phi ^ i) ^\dagger$ and $V$, but not of their derivatives.
The expansion of the full K{\"a}hler part Lagrangian is
\bea
\L_K
&=&
-g_{ii^*} D_\mu \phi ^i D^\mu \phi ^{i^*}
-i g_{ii^*} \bar \psi ^{i^*} \bar \sigma ^\mu D_\mu \psi ^i
+{1 \over 4} R_{ik^*jl^*} \psi ^i \psi ^j \bar \psi ^{k^*} \bar \psi ^{l^*}
\nonumber \\
&&+g_{ii^*} (F^i -\12 \Gamma _{jk}^i \psi ^j \psi ^k)
(F^{i^*} -\12 \Gamma _{{j^*}{k^*}}^{i^*} \psi ^{j^*} \psi ^{k^*})
\nonumber \\
&&-{i \over 2} D^a  (T^a)^j{}_i \phi ^i {\partial K \over \partial \phi ^j}
+ \sqrt 2 g_{ii^*} (T^a)^i{}_k \phi ^k \bar \psi ^{i^*} +{\rm complex \
conjugate}\, .
\eea
Here, complex conjugation is to be added only on the last line. The matrices
$T^a$ are the representation matrices of $\G$ in the representation under which
the $\Phi ^i$ transform. The covariant derivatives are given by
\bea
D_\mu \phi ^i & = & \partial _\mu \phi ^i  - A_\mu ^a (T^a)^i{}_j \phi ^j
\nonumber \\
D_\mu \psi ^i & = & \partial _\mu \psi ^i - A_\mu ^a (T^a)^i{}_j \psi ^j
+ \Gamma _{jk}^i D_\mu \phi ^j \psi ^k
\eea
and $\Gamma$ is the usual Levi-Civita connection for the K{\"a}hler metric 
\be
g_{ii^*} \equiv {\partial ^2 K \over \partial \phi ^i \partial \phi ^{i^*}}
\ee
and $R_{ik^*jl^*}$ is its Riemann curvature tensor. 
 
\medskip

Putting toghether contributions from $\L_K$, $\L_U$ and $\L_G$, we have the
most general $\N=1$ supersymmetric Lagrangian with the restrictions of above.

\subsection{Renormalizable N=2,4 Susy Lagrangians}

We shall now derive the
Lagrangians invariant under the larger supersymmetry algebras associated with
$\N=2$ and $\N=4$ supersymmetry. In this subsection, we shall restrict attention
to renormalizable Lagrangians, leaving the general case for the next subsection.
Thus, we seek an
$\N=2$ microscopic renormalizable super-Yang-Mills theory with gauge algebra
$\G$, and treat $\N=4$ as a special case thereof. Following our general
classification of supersymmetry representations,  two fields are possible. The
{\it gauge multiplet} with component fields
$(A_\mu \
\lambda _\pm \ \phi)$ transforms under the adjoint representation of $\G$,
which we simply denote by $\G$, and the {\it hypermultiplet} with component
fields $(\psi _+ \ H_\pm \ \psi _-)$ transforms under a representation $\R$ of
$\G$.

\medskip

To construct the most general renormalizable Lagrangian for these fields,
we use our results on invariant $\N=1$ supersymmetric Lagrangians. To do
so, we decompose the $\N=2$ gauge and hypermultiplets under an $\N=1$
supersymmetry,
\bea
\N=2 \ {\rm gauge} \ ({\rm rep} \ \G) & : & \ \ V \oplus \Phi
\qquad \ \ \ V \sim (A_\mu \ \lambda _+) 
\qquad \ \ \Phi \sim (\lambda _- \ \phi) \\
\N=2 \ {\rm hyper} \ ({\rm rep} \ \R) & : & H_1 \oplus H_2
\qquad H_1 \sim (H_+ \ \psi _+)
\qquad H_2 \sim (H_- \ \psi _-)
\eea    
and promote the $\N=1$ supermultiplets $V$, $\Phi$, $H_1$ and $H_2$ to $\N=1$
superfields by including their respective auxiliary fields. Thus, $V$ may be
viewed now as an $\N=1$ vector (or gauge) superfields, while $\Phi$ and $H_f$,
$f=1,2$ may be viewed as $\N=1$ chiral superfields. 

\medskip

Following the results of \S
2.8, we may immediately write down the most general $\N=1$ and $\G$-gauge
invariant Lagrangian for these fields,
\be \label{micro}
\L = {\rm Re} \int d^2 \theta \biggl ( \tau  W^aW^a + U(\Phi, H_f) \biggr )
+\int d^4 \theta \biggl ( \Phi ^\dagger e^{V_\G} \Phi + \sum _f H_f ^\dagger
e^{V_\R} H_f \biggr )
\ee
Here, $V_\G$ and $V_\R$ stand for the field $V$ in the representations $\G$ and
$\R$ respectively. The complex analytic function $\tau _{ab} (\Phi, H_f)$ that
was allowed for a general $\N=1$ supersymmetric Lagrangian must be constant for
renormalizability, and equal to $\tau\delta _{ab}$ by gauge invariance, for
a simple algebra $\G$. The constant $\tau$ is related to the gauge
coupling $g$ and the instanton angle $\theta $ by
\be\label{tau}
\tau = {\theta \over 2\pi} + {4\pi i \over g^2}\, .
\ee
The reality of $g$ guarantees that ${\rm Im} \tau >0$. 
The superpotential $U(\Phi, H_f)$ would be a general $\G$-invariant function
for $\N=1$ supersymmetry, but for $\N=2$ is restricted to be of the form
\be
U(\Phi, H_a) = H_1 ^T \Phi H_2 + H_1 m H_2
\ee
where $m$ is the hypermultiplet mass matrix.

\subsection{N=2 Superfield Methods : unconstrained superspace}

In order to realize $\N=2$ susy transformations linearly on complete 
multiplets,
we seek a formulation in terms of superfields \cite{gsw}, just as for $\N=1$.  
The coordinates of $\N=2$ {\it unconstrained superspace} are  collectively
denoted $z_M$, and given by
\be
z_M = \{ x^\mu, \ \theta _{\alpha i}, \ \bar \theta ^{\dot \alpha i} , \qquad
i=1,2\}\, ,
\ee
with $[x,\theta]=\{ \theta , \theta\}=0$ as usual.
The internal $SU(2)_R$ index $i$ is raised or lowered with the help of the
anti-symmetric  $\epsilon _{ij}$ or $\epsilon ^{ij}$ tensor, where $\epsilon
^{12}=-\epsilon _{12}=1$. Unconstrained superspace is real in the sense that
it is invariant under the following complex conjugation
\be
(x^\mu )^\dagger = x^\mu \qquad 
(\theta _{\alpha i})^\dagger = \bar \theta ^{\dot \alpha i}\, .
\ee  
Superderivatives (in the absence of central charges) are defined by
\bea 
D_\alpha ^i 
= {\partial \over \partial \theta ^\alpha _i} + i \sigma ^\mu
_{\alpha \dot \alpha} \bar \theta ^{\dot \alpha i} \partial _\mu
\qquad \qquad 
\bar D_{\dot \alpha i} 
=  -{\partial \over \partial \bar \theta ^{ \dot \alpha i}} 
 - i \theta ^ \alpha _i \sigma ^\mu _{\alpha \dot \alpha} \partial _\mu \, ,
\eea
while supercharges (in the absence of central charges) are given by
\bea 
Q_\alpha ^i 
=  {\partial \over \partial \theta ^\alpha _i} - i \sigma ^\mu
_{\alpha \dot \alpha} \bar \theta ^{\dot \alpha i} \partial _\mu
\qquad \qquad 
\bar Q_{\dot \alpha i} = 
-{\partial \over \partial \bar \theta ^{ \dot \alpha i} } 
+ i \theta ^ \alpha _i \sigma ^\mu _{\alpha \dot \alpha}   \partial _\mu \, .
\eea
Superderivatives and supercharges generate commuting left and right
actions of susy, so that
 $\{ Q_\alpha ^i , D_\beta ^j \}=\{Q_\alpha ^i , \bar D^{\dot \beta j}\}=0$,
together with complex conjugate relations. 

\medskip

An {\it unconstrained superfield} is defined as a general function of  $z_M$,
with a Taylor expansion of the form
\bea 
S (z_M)
& = &
\phi (x) + \theta _i \psi ^i (x) 
+ \bar \theta ^j \bar \chi _j (x) 
+ \bar \theta ^i \bar \sigma ^\mu \theta _j A_{\mu i}^j (x) 
+ \theta _{\alpha i} \theta _{j \beta}  f^{\alpha \beta ij} (x) 
\nonumber \\ && 
+ \bar \theta ^{\dot \alpha i}  \bar \theta ^{\dot \beta j} 
g_{\dot \alpha \dot \beta ij} ^* (x)
+ \cdots 
+ \theta _i \theta _j \theta _k \theta _l \bar \theta ^m \bar \theta ^n
\bar \theta ^o \bar \theta ^p D^{ijkl} _{mnop} (x)\, .
\eea
Supersymmetry transformations are linearly realized on unconstrained 
superfields
via the super-differential operators $Q$ and $\bar Q$ by 
\be 
\delta _\xi S = (\xi ^i Q_i + \bar \xi _i \bar Q^i) S \,  ,
\ee
where $\xi ^i$ and $\bar \xi _i$ are infinitesimal spinor parameters.
The unconstrained superfield $S$, just as a general $\N=1$ superfield, is
reducible. Thus, we seek suitable sets of $\N=2$ supersymmetric
constraints to decompose $S$ into irreducible components. Since $Q$
and $D$ mutually anticommute, we may impose on $S$ that some
of its superderivatives vanishes. 

\medskip

We define a {\it $\N=2$ chiral superfield} by imposing the constraints
\be \label{chirs}
\bar D_{\dot \alpha i} \Phi =0 \qquad \qquad i=1,2\, .
\ee
Notice that in order to preserve $SU(2)_R$ symmetry, we are forced to
 impose the
above constraint for both values of $i$. The chirality constraints 
(\ref{chirs})
may be solved by introducing the combinations
\bea 
x^\mu _\pm = x^\mu \pm i \theta _i  \sigma ^\mu \bar \theta ^i 
\eea
which satisfy
\bea
\bar D_{\dot \alpha i} x^\mu _+=0, \qquad \qquad D ^{\alpha i} x_- ^\mu=0\, .
\eea
Thus, the superfield $\Phi$ depends only upon $x_+ ^ \mu$ and $\theta _{\alpha
i}$ and has the component expansion
\bea
\Phi (x_+, \theta) &=& \phi (x_+) + \theta ^{\alpha i} \psi _{\alpha i} (x_+)
+ \theta ^{\alpha i} \theta ^{\beta j} f _{\alpha \beta ij} (x_+)
\nonumber \\
&& + \theta ^{\alpha i} \theta ^{ j} \theta ^{ k} \chi _{\alpha ijk}  (x_+)
+ \theta  \theta  \theta  \theta  D (x_+)\, . \ 
\eea
When $\Phi$ is a scalar superfield, for example, the components $\phi$,
$f^{[\alpha \beta](ij)}$ and $D$ are scalars, $\psi$ and $\chi$ are
 spin 1/2 and
$f^{(\alpha \beta)[ij]}$ is a self-dual rank 2 antisymmetric tensor
field. 
(This 
chiral superfield will naturally incorporate the field
 strength of the $\N=2$
gauge multiplet, as we shall see later).
Any complex analytic function $\F (\Phi)$ of an $\N=2$ chiral superfield is
automatically also an $\N=2$ chiral superfield.

\medskip

The term with the highest power of $\theta$ of a (gauge invariant) $\N=2$
chiral superfield transforms under $\N=2$ susy by a total space-time
derivative, so that it may be used to construct $\N=2$ invariant Lagrangians.
Introducing the chiral measure $d^4 \theta \equiv d^2\theta ^1 d^2 \theta ^2$, 
the combination 
\be
\L _D = D =  \int d^4 \theta \ \Phi
\ee
is an invariant Lagrangian. 

\medskip

The $\N=2$ chiral superfield is reducible, unlike its $\N=1$ counterpart. To
achieve an Abelian irreducible superfield, we may additionally impose the
constraints
\bea
D^{\alpha i} D_{\alpha } ^j W = \bar D _{\dot \alpha} ^i \bar D^{\dot \alpha j}
\bar W \, .
\eea 
As a result we have $f_{[\alpha \beta](ij)}=\chi _{\alpha ijk}=0$, 
while $\phi$, $\psi$, $f_{(\alpha \beta)[ij]}$ and the auxiliary field  $D$ are
unaffected. It is customary to represent the solution to these constraints
in the form of an $\N=1$ superfield expansion, 
\be
W(x_+,\theta) = \Phi (x_+, \theta ^1) + \sqrt 2 \theta ^{\alpha 2} W_\alpha
(x_+, \theta ^1) + \theta ^2 \theta ^2 G(x_+,\theta ^1)\, .
\ee
Here, $\Phi$ is the $\N=1$ chiral superfield contained in the $\N=2$ gauge 
multiplet, while we shall denote by $V$ the $\N=1$ gauge superfield contained
in the $\N=2$ gauge multiplet, and denote the field strength superfield
associated with $V$ by $W_\alpha$. The auxiliary field $G$ is then a field
that may be expressed in terms of $\Phi $ and $V$ as follows,
\be
G(x_+,\theta ^1) 
= - \12 \int d^2 \bar \theta ^1 
\Phi (x_+-i \theta_1 \sigma \bar \theta ^1, \theta ^1, \bar \theta ^1) ^\dagger
e^{-2V(x_+-i\theta_1 \sigma \bar \theta^1 , \theta ^1, \bar \theta ^1)}\, .
\ee
The component field contents of $W$ is that of an Abelian  
$\N=2$ gauge multiplet.

\medskip

The $\N=2$ gauge superfield for a non-Abelian gauge algebra $\G$ is obtained
by introducing the $\G$-valued supergauge fields $A_{ \alpha i}$ 
and $\bar A_{\dot
\alpha i}$, and the associated gauge covariant superderivatives
\be
{\cal D} _{\alpha i} = D_{\alpha i} + i A_{\alpha i}
\qquad \qquad
\bar {\cal D} _{\dot \alpha i} = \bar D_{\dot \alpha i} + i \bar A_{\dot \alpha
i}\, .
\ee 
The $\N=2$ gauge superfield is now governed by the Poincar{\'e} algebra, the
gauge algebra $\G$ and the $SU(2)_R$-symmetry covariant constraints
\bea \label{constr}
\bar {\cal D}_{\dot \alpha i} W =0 \qquad \qquad
{\cal D}^{\alpha i} {\cal D}_{\alpha } ^j W 
= \bar {\cal D} _{\dot \alpha} ^i \bar {\cal D}^{\dot \alpha j} \bar W \, .
\eea 
The component field contents is that of an $\N=2$
multiplet for gauge algebra $\G$.

\medskip

Invariant actions for the $\N=2$ gauge supermultiplet may be obtained by
integrating a chiral gauge invariant function of $W$. An $\N=2$
 chiral superfield
is obtained for any complex analytic function $\tr \F(W)$, called the {\it
prepotential}. The most general gauge invariant,
$\N=2$ supersymmetric Lagrangian is given by
\be
S _\F = \int dz_M \ \tr \F(W) + \ {\rm complex \ conjugate}
\ee 
where the superspace measure is defined by  $dz_M \equiv
d^4x_+ d^4 \theta$. Upon integrating out $\theta ^2$ and $\bar \theta ^2$ from
$dz_M$, we readily recover the $\N=1$ superfield form for the action $S_\F$,
\be
S_\F = - \12 \int d^2 \theta \biggl ( {\p \F \over \p \Phi ^a} G^a \biggr )
- \int d^2 \theta {\p ^2 \F \over \p \Phi ^a \p \Phi ^b} W^{\alpha a} W^b
_\alpha
\ee
where the index $a$ labels the components of the adjoint representation.

\subsection{N=2 Superfield Methods : harmonic/analytic superspaces}

The above construction of the gauge superfield on $\N=2$ unconstrained
 superspace
does not appear to extend to the off-shell hypermultiplet. The obstacle arises
from the fact that strong constraints have to be imposed on the superfield in
order to retain only dynamical fields with spin less or equal to 1/2. Any such
set of constraints turns out to be so strong that it entails the field 
equations
as well. Harmonic and analytic superspace methods were introduced to 
remedy this
situation. We discuss here only the case of $\N=2$. 

\medskip

In {\it harmonic superspace} \cite{gikos}  new bosonic coordinates are
introduced,  in addition
to the space-time coordinates $z_M$. From a supergroup viewpoint, additional
bosonic coordinates appear naturally for any extended supersymmetry.
 For example,
the conformal supergroups $SU(2,2|\N)$ have maximal bosonic subgroups $SU(2,2)
\times SU(\N)_R$, where the internal $SU(\N)_R$ is associated
with new bosonic coordinates in superspace.

\medskip

$\N=2$ harmonic superspace is parametrized by $x^\mu$, $\theta _{\alpha i}$,
$\bar \theta ^{\dot \alpha i}$ and the new bosonic coordinates $u^\pm _i$,
$i=1,2$, with $u^{\pm i} = \epsilon ^{ij} u^\pm _j$ and
\be
\left ( \matrix{u_1 ^- & u_1 ^+ \cr u_2 ^- & u_2 ^+} \right ) \in SU(2)_R
\qquad 
\left \{ \matrix{
u^{+i} u^- _i =1\cr
u^{\pm i} u^\pm _i =0} \right . 
\ee
The variables $u_i ^\pm$ form a complex orhonormal frame, transforming
 under the
fundamental representation of $SU(2)_R$  in its index $i$,
and under a $U(1)$ in its $\pm$ indices. Harmonic superspace is real in the
sense that it is invariant under the following complex conjugation
\be
(x^\mu )^\dagger = x^\mu \qquad (\theta _{\alpha i})^\dagger = \bar \theta
^{\dot \alpha i} \qquad (u^{\pm i})^\dagger = u^\mp _i
\ee

\medskip

Supersymmetry transformations on harmonic superspace are just the supersymmetry
transformations on unconstrained superspace, implemented with the invariance of
$u^\pm _i$,
\bea
\delta x^\mu  =  i \xi ^i \sigma ^\mu \bar \theta _i -i \theta ^i \sigma
^\mu \bar \xi _i
\qquad \qquad 
\delta \theta _{\alpha i}  &=&  \xi _{\alpha i}
\nonumber \\
\delta u^\pm _i   =   0 
\hskip 1.8in
\delta \bar \theta _{\dot \alpha} ^i  &=&  \bar \xi _{\dot \alpha } ^i
\eea
An {\it analytic basis for harmonic superspace} with parameters
 $x_A^\mu$, $\theta
^\pm _\alpha$, $\bar \theta ^\pm _{\dot \alpha}$, $u^\pm _i$ is obtained  via
the following invertible change of variables
\bea
x^\mu _A  \equiv  x^\mu -i \theta ^+ \sigma ^\mu \bar \theta ^-
-i \theta ^- \sigma ^\mu \bar \theta ^+
\qquad 
\theta ^\pm _\alpha  & \equiv &  \theta ^i _\alpha u^\pm _i
\nonumber \\
\bar \theta _{\dot \alpha} ^\pm & \equiv & \bar \theta _{\dot \alpha} ^i
u^\pm _i 
\eea 
In the analytic basis, the susy transformation laws take the form
\bea
\delta x^\mu _A & = & -2i( \xi ^i \sigma ^\mu \bar \theta ^+  + \theta ^+
\sigma ^\mu \bar \xi _i) u_i ^-
\qquad \qquad 
\delta \theta _{\alpha } ^\pm  =  \xi _{\alpha } ^i u_i ^\pm
\nonumber \\
\delta u^\pm _i  & = &  0
\hskip 2.3in
\delta \bar \theta _{\dot \alpha} ^\pm   =  \bar \xi _{\dot \alpha } ^i u_i
 ^\pm \, .
\eea
Besides the space-time derivative $D_\mu = \p /\p x_A ^\mu$, we have $SU(2)_R$
covariant superderivatives  
\bea
D^+ _\alpha & = & {\p \over \p \theta ^{\alpha -}}
\qquad \qquad
D^- _\alpha  =  - {\p \over \p \theta ^{\alpha +}} + 2i \sigma ^\mu _{\alpha
\dot \alpha} \bar \theta ^{\dot \alpha -} D_\mu
\nonumber \\
\bar D^+ _{\dot \alpha} & = & {\p \over \p \bar \theta ^{\dot \alpha -}}
\qquad \qquad
\bar D^- _{\dot \alpha}  =  - {\p \over \p \bar \theta ^{\dot \alpha +}} - 2i
\theta ^{ \alpha -} \sigma ^\mu _{\alpha \dot \alpha}  D_\mu
\eea
and covariant derivatives with respect to the internal coordinates $u_i ^\pm$, 
\bea
D^{++} & = & u^{i+} {\p \over \p u^{i-}} 
-2i \theta ^+ \sigma ^\mu  \bar \theta ^+ D_\mu 
+ \theta ^{\alpha +} {\p \over \p \theta ^{\alpha -}}
+ \bar \theta ^{\dot \alpha +} {\p \over \p \bar \theta ^{\dot \alpha -}}\, , 
\eea
as well as $D^{--}= (D^{++})^\dagger $ and $D^0 = 1/2 [D^{++}, D^{--}]$.
 
\medskip

{\it Analytic superspace} is defined as the subspace of harmonic superspace
parametrized by the coordinates 
\be
z_A = \{ x^\mu _A, \ \theta ^+ _\alpha, \ \bar \theta ^+ _{\dot \alpha} \}
\quad {\rm  and} \quad  u_i^\pm
\ee 
only (i.e. it is the subspace in which $\theta ^-
_\alpha = \bar \theta _{\dot \alpha} ^-=0$). All $\N=2$ susy transformations
leave this subspace invariant. Analytic superspace is real in the sense that
it has a natural complex conjugation which will be denoted by ${}^c$, and
acts by
\be
(x_A ^\mu)^c = x_A ^\mu,
\qquad \quad
(\theta _\alpha ^\pm)^c = \bar \theta _{\dot \alpha} ^\pm
\qquad \quad
(\bar \theta _{\dot \alpha} ^\pm)^c = -  \theta _{ \alpha} ^\pm
\qquad \quad
(u^{i \pm} )^c = - u^\pm _i\, .
\ee
The square of the operation ${}^c$ equals plus/minus the identity on
bosonic/fermionic coordinates. (The complex conjugation ${}^\dagger$ introduced
previously induces a complex conjugation in the analytic basis of harmonic
superspace as well $(x^\mu _A)^\dagger = x^\mu _A$, $(\theta _\alpha
^\pm)^\dagger = \bar \theta ^\mp _{\dot \alpha}$, $(u^{i\pm})^\dagger = \mp u_i
^\mp$ which does not leave analytic superspace invariant.)

\medskip

Functions on analytic superspace naturally generalize the
notion of chiral superfield for $\N=1$. Viewed as functions on harmonic
superspace, they satisfy the constraints
\be
D_\alpha ^+ \Phi \ =  \ \bar D _{\dot \alpha} ^+ \Phi \ = 0
\ee
and are thus functions of the form $\Phi(z_A, u)$ with
the following decomposition
\bea
\Phi (z_A, u)
& = & 
\phi (x_A,u) 
+ \theta ^+ \psi (x_A,u) +  \bar \theta ^+ \bar \chi (x_A,u) 
+ \theta ^+ \theta ^+ f(x_A,u) + \bar \theta ^+ \bar \theta ^+ g(x_A,u)
\nonumber \\ 
&& + \bar \theta ^+ \bar \sigma ^\mu \theta ^+ A_\mu (x_A,u)  
+ i \theta ^+ \theta ^+ \bar \theta ^+  \bar \lambda (x_A,u) 
-i \bar \theta ^+ \bar \theta ^+ \theta ^+ \nu (x_A,u)  
\nonumber \\
&& + \12 \theta ^+ \theta ^+ \bar \theta ^+ \bar \theta ^+ D(x_A,u) 
\eea
If $\Phi$ has $U(1)$ charge $q$, then the component field $\phi$
 has charge $q$,
the fields $\psi$ and $\chi$ have charge $q-1$, $f$, $g$ and 
$A_\mu$ have charge
$q-2$, $\lambda$ and $\nu$ have charge $q-3$ and $D$ has charge $q-4$. For $q$
even, one may impose the reality condition $( \Phi ^{(q)})^c = \Phi ^{(q)}$.

The dependence on the internal coordinates $u_i ^\pm$ produces
an infinite number of component fields which are functions of $z_A$. Indeed, an
analytic superfield with $U(1)$ charge $q>0$ has a
decomposition of the form
\be
\Phi ^{(q)} (z_A,u) = \sum _{n=0} ^\infty \phi ^{(i_1 \cdots i_{n+q} j_1 \cdots
j_n)}(z_A) u^+ _{(i_1} \cdots u^+_{i_{n+q}} u^- _{j_1} \cdots u^- _{j_n)}\, .
\ee
Both the $\N=2$ gauge multiplet and the hypermultiplet may be treated
as
 analytic
superfields, with an off-shell $\N=2$ Lagrangian, which contains precisely the
appropriate number of fields on-shell. 

\medskip

An invariant integration on analytic superspace may be defined in terms of 
the measure $dz_A \cdot du$, which may be expressed in terms of the 
``space-time
measure" $dz_A\equiv d^4 x_A d^2 \theta ^+ d^2 \bar \theta ^+$ and the 
``internal
measure" $du$. The latter is defined by
\be
\int du \ u^+ _{(i_1} \cdots u^+_{i_{m}} u^- _{j_1} \cdots u^- _{j_n)} 
 =\delta
_{m+n,0}\, .
\ee
In order to get a $U(1)$-invariant Lagrangian, the integration against
 the measure
$dz_A$ of $U(1)$-charge $-4$ requires an analytic superfield of $U(1)$
charge 4. 

\medskip

The {\it hypermultiplet analytic superfield} has $U(1)$-charge 1, and is
denoted by $\Phi ^+$. A good field equation for this superfield will
be such that
the infinite tower of auxiliary fields associated with the $u$-dependence of
$\Phi ^+$ collapses to just the correct number of physical fields. Thus, the
field equations cannot just be built up from the superderivatives
 $D^-$ and $\bar
D^-$, but must involve derivation with respect to $u$. The natural candidate is
the operator $D^{++}$, and the correct hypermultiplet free field equation is
$D^{++} \Phi ^+ =0$.
The most general $U(1)$-invariant action may be easily constructed along
these lines, and one finds \cite{gikos},
\bea \label{hyperaction}
S[\Phi ^+]& = &\int dz_A \ du \ \bigl \{ (\Phi ^+)^c D^{++} \Phi ^+ 
+ U(\Phi ^+, (\Phi ^+)^c) \big \}
 \\
U(\Phi ^+, ( \Phi ^+)^c) &=& a (\Phi ^+)^4 + b ( \Phi ^+)^c (\Phi ^+)^3
 + c
((\Phi ^+)^c)^2 (\Phi ^+)^2 + b^* (( \Phi ^+)^c)^3 \Phi ^+ 
+ a^* ((\Phi ^+)^c)^4 
\nonumber
\eea
The couplings all have dimensions of mass${}^{-2}$ and thus are
non-renormalizable.

\medskip

The $\N=2$ {\it gauge analytic superfield} has $U(1)$-charge 2, and is
denoted $V^{++}$. It is real under the complex conjugation
 $ {}^c$ that leaves
analytic superspace invariant, $( V^{++})^c = V^{++}$. Gauge transformations
are generated by analytic superfields $\Lambda$ of $U(1)$ charge 0, since they
must act consistently on the analytic superfields $V^{++}$ and  on
analytic hypermultiplets $\Phi ^+$. The appropriate transformation laws are
\bea
(V^{++})'  =  e^{i \Lambda} \ (V^{++} - i D^{++}) \ e^{-i \Lambda} 
\qquad \qquad
(\Phi ^+) '  =  e^{i \Lambda } \Phi ^+\, .
\eea 
Since $\Lambda$ is analytic, the superderivatives $\D _\alpha ^+
=D_\alpha ^+$ and
$\bar \D _{\dot \alpha} ^+ = \bar D _{\dot \alpha} ^+$ are automatically
covariant, while the remaining covariant derivatives are constructed by
\bea
\D ^{++}       =  D^{++} +i V^{++} 
\qquad \qquad
\D _\alpha ^-   =  D_\alpha ^- + i A_\alpha ^- 
\qquad \quad
\D _{\dot \alpha} ^-  =  D_{\dot \alpha} ^- + i A_{\dot \alpha } ^-
\eea 
The superconnections $A_\alpha ^-$ and $A_{\dot \alpha }^-$ may be 
expressed in
terms of $V^{++}$, but we shall not need their explicit form here. 

\medskip

The gauge field strength $W$ is an analytic superfield and was
constructed in terms of the superfield $V^{++}$ in \cite{gikos}, 
\be
W= -{i \over 4} e^{i v} \biggl \{ \bar D ^+ _{\dot \alpha}
 \bar D^{\dot \alpha +}
\biggl ( e^{-iv } D^{--} e^{iv} \biggr ) \biggr \} e^{-i v}\, .
\ee 
Here, $v$ is a general superfield (i.e. neither analytic nor chiral),
 defined by
\be
\bigl ( D^{++} +i V^{++} \bigr ) \ e^{iv} =0\, .
\ee
An explicit expression for $W$ in terms of $V^{++}$ was obtained in \cite{zup}
and is given by
\be
W(z,u) = {i \over 4} \bar D^+ _{\dot \alpha} \bar D^{\dot \alpha +}
\sum _{n=1} ^\infty (-i)^n \int du_1 \cdots du_n \ {V^{++}(z,u_1) \cdots
V^{++}(z,u_n) \over (u u_1)(u_1 u_2) \cdots (u_n u) }
\ee
where we have defined $(u_a u_b) = u_a ^{+i} u_{bi} ^+$. Remarkably, the
superfield $W$ constructed in the analytic superspace approach turns out to be
independent of $u$ altogether, and its kinematics thus reduces
to that discussed in the preceding section on unconstrained superspace. 

The most general $\N=2$ supersymmetric and $\G$-gauge invariant action is thus
\be \label{fulllag}
S = {\rm Re} \int dz_M \ du \ \tr \F (W; \tau)  
+ \int dz_A \ du \ \bigl \{ ( \Phi ^+)^c D^{++} \Phi ^+ 
+ U(\Phi ^+, (\Phi ^+)^c) \big \}   
\ee
where the potential  $U$ was given in (\ref{hyperaction}). The prepotential
function $\F$ is an arbitrary complex analytic function of its
arguments.  In view of arguments similar to the ones given for $\N=1$, $\F$ is
also complex  analytic in its coupling constants $\tau$. 
The renormalizable case
corresponds to $\F(W;\tau) = i\tau W^2$ and $U=0$.

\vfill\eject

\setcounter{equation}{0}
\section{Seiberg-Witten Theory}

Seiberg-Witten theory deals with the construction of the non-perturbative
dynamics of $\N=2$ super-Yang-Mills theory in the limit of low energy and
momemnta. The prime motivation is the exploration of the fully
non-perturbative dynamics of gauge theories, a problem of quantum field theory 
that has been outstanding for several decades now. The restriction to $\N=2$
theories is made because the stronger constraints of extended supersymmetry
lead to simplified -- yet highly non-trivial -- dynamics of super-Yang-Mills
theory. The dynamics of the even more restricted $\N=4$ theory is even
simpler than that of the $\N=2$ theory, but the low energy approximation,
obtained in Seiberg-Witten theory, becomes rather trivial in this theory.

\subsection{Wilson Effective Couplings and Actions}

Despite the large degree of supersymmetry, it is rather unlikely that the
quantum field theory defined by the Lagrangian (\ref{micro}) will be exactly
solvable, in the sense that all its correlation functions or even all its
$S$-matrix elements could be obtained in explicit form. Fortunately, many of
the physically most interesting questions, that have to do with the
non-perturbative dynamics of the theory, are related to the properties of the
vacuum and the lowest energy (and momentum) excitations above the vacuum.
Thus, there already is great interest in obtaining information on the
non-perturbative dynamics of these theories in the limit of low energy only.
The Wilson renormalization group \cite{kw} provides the ideal tool for deriving
such low energy effective theories.

\begin{itemize}

\item The starting point is the {\it bare} quantum field theory with a
(energy-momentum) cutoff $\Lambda$, in which all masses, energies and momenta 
of all fields are restricted to be less than $\Lambda$. (It is usually safest
to consider such cutoffs in the Euclidean version of the theory, where they
can be imposed in a Lorentz invariant way.) The bare coupling is denoted
$g(\Lambda)$. 

\item It is assumed that the masses, energies and momenta of physical relevance
are much smaller that $\Lambda$.

\item One introduces an arbitrary {\it renormalization scale} $\mu$, with 
\be 
{\rm physical\ masses,\ energies,\ momenta} \ \ll \mu \ll \Lambda\, .
\ee
One considers the {\it Wilson effective theory}, 
in which only the particles and
fields of masses, energies and momenta less than $\mu$ (i.e. the {\it light
fields}) are retained as quantum degrees of freedom from the original bare
theory. All particles and fields with masses, energies and momenta between the
renormalization scale $\mu$ and the cutoff scale $\Lambda$ (i.e. the {\it
heavy fields}) are {\it integrated out}. (The separation between light and
heavy modes is not usually gauge invariant. One may either work with gauge
invariant degrees of freedom directly, which is always hard, or, in
perturbation theory, one may replace gauge invariance with BRST invariance.
Since BRST invariance is a global symmetry, separation into light and heavy
modes may be achieved as usual.) The effect of this integration over the heavy
fields is summarized in terms of (an infinite series of) 
additional terms in the
Lagrangian. The resulting object is usually referred to as the {\it Wilson
effective Lagrangian or effective action}. 

\end{itemize}

For example, in a theory of only a gauge field $A$, with gauge field strength
$F$ and a single coupling $g$, the bare theory (ignoring gauge fixing and ghost
terms) involves the bare field strength $F_{(\Lambda)}$ and the bare coupling
$g(\Lambda)$. The Wilson effective theory involves the field strength
$F_{(\mu)}$ and the effective gauge coupling $g(\mu)$, as well as an infinite
series of higher dimension operators with numerical coefficients which are
functions of $\mu$ and $\Lambda$,
\bea
\L _{\rm bare} & = & {1 \over 2 g(\Lambda )^2} \tr F^2_{(\Lambda}) 
\nonumber \\
\L _{\rm eff}  & = & {1 \over 2 g(\mu )^2}  \tr F^2 _{(\mu )} 
+ {1 \over \Lambda
^2} f(g, {\mu \over \Lambda}) \tr F^3 _{( \mu )} + \cdots 
\eea
Here, it is understood that the fields $F_{( \Lambda )}$ and $F_{( \mu )}$
contain only modes of masses and energies less than $\Lambda $ and less than
$\mu$ respectively.

\medskip

The Wilson effective Lagrangian at scale $\mu$ essentially amounts to a Taylor
series expansion in the {\it light fields} obtained by integrating out the
heavy fields. Such a Taylor expansion is analytic by construction, and does not
exhibit any infrared singularities by construction. (This may be contrasted
with the effective Lagrangian defined as generating functional of connected
one-particle-irreducible Feynman graphs, which does exhibit infrared
singularities whenever massless particles are around.)

\medskip

However, there is a generic situation in which the Wilson effective Lagrangian
does exhibit singularities. This occurs when one considers not a single theory,
but rather a family of theories obtained for example by varying one or several
of the masses in the theory. Let $m$ be such as mass. Assume that we begin by
analyzing a theory in which the mass $m$ is larger than the renormalization
scale $\mu$. Thus, by the rules for the construction of the Wilson effective
theory, we integrate out completely all the modes associated with this mass $m$,
and obtain a certain effective Lagrangian $\L _m$. No infrared singularity
occurs since all the modes that were integrated out are massive. Now, consider
varying $m$, and rendering $m$ smaller than $\mu$. The effective Lagrangian
$\L_m$ still makes sense. However, as soon as $m <\mu$, $\L _m$ strictly
speaking is no longer the Wilson effective action, since particles with mass
$m<\mu$ have been completely integrated out. Nonetheless, $\L _m$ is a useful
effective action, especially because it is analytic in $m$, while the true
Wilson effective action will have discontinuities at $m=\mu$. As $m \to 0$, the
effective action $\L _m$ will in general develop singularities, associated with
the fact that massless modes will now have been integated out along with
massive ones. It is a fact that the only singularities in the Wilson effective
action arise due to particles and fields becoming massless. Thus, the
singularity structure of the effective action corresponds to the appearance of
massless particles in the spectrum.

\subsection{Holomorphicity and Non-Renormalization}  

Supersymmetric theories invariably have {\it holomorphic data}, such as the
superpotential $U(\Phi)$ or the {\it gauge coupling} $\tau _{ab}(\Phi)$.
Under the Wilson renormalization group procedure, 
which preserves supersymmetry,
these holomorphic data are modified in a holomorphic way, and become
the effective superpotential $U^{\rm eff}(\Phi)$ and the effective gauge
coupling $\tau ^{\rm eff} _{ab}(\Phi)$. The analyticity of the Wilson
renormalization group procedure, pointed out in the preceding subsection,
guarantees that complex analyticity will also be preserved \cite{shifman,
seiberg88}. This property of supersymmetric data places severe restrictions on
the renormalization group flow in supersymmetric theories, and provides one of
the key ingredients for Seiberg-Witten theory.

\medskip

The couplings entering these holomorphic data, such as the mass $m$ and Yukawa
coupling $\lambda$ in $U(\Phi)$, or the gauge coupling $\tau$ in $\tau
_{ab}(\Phi)$, are naturally complex numbers, and the dependence of $U$ or
$\tau_{ab}$ on them may be viewed as holomorphic as well. Remarkably, this
holomorphicity property is also preserved under the Wilson renormalization
procedure. The reason is that the complex parameters $m$ and $\lambda$ and
$\tau$ may be viewed as vacuum expectation values of chiral superfields. The
holomorphic dependence on these new superfields is preserved under the Wilson
renormalization procedure, just as that of the dynamical chiral superfields
was. The net result is that there will be highly restricted renormalization
effects (or none at all) on such complex couplings. To render this discussion
more concrete, we present two important examples.

\medskip

\noindent
{\it (1) Absence of mass and Yukawa coupling renormalization}

Consider a renormalizable theory of a single chiral scalar superfield $\Phi$,
with superpotential $U(\Phi;\lambda ,m)= m\Phi ^2 + \lambda \Phi ^3$.
Holomorphicity considerations \cite{seiberg88}, discussed above, imply that the
effective superpotential $U^{\rm eff}(\Phi; \lambda,m)$ will be a holomorphic
function of the superfield $\Phi$ and of the couplings $\lambda$ and $m$. Now,
as
$\lambda$ and $m$ have both been promoted to superfields, the superpotential
$U(\Phi;\lambda,m)$ has two $U(1)$ symmetries. The first $U(1)$ is just phase
rotations of $\Phi$, $\lambda$ and $m$ in such a way that $U$ is invariant.
The second $U(1)_R$ is the $R$-symmetry generator, under which the
superpotential must have charge 2. The charge assignments are listed in table
\ref{table:5}. Both symmetries remain after quantization and are preserved
under the Wilson renormalization group flow, so that $U^{\rm eff}$ has the same
quantum numbers as $U$. From the table \ref{table:5}, it is clear that the
only quantity invariant under both $U(1)$ and $U(1)_R$ is $\lambda\Phi /m$.
Since the Wilson effective Lagrangian is analytic in the light fields, $U^{\rm
eff}$ gets only contributions of the form
\be
m\Phi ^2 \biggl ( {\lambda \Phi \over m} \biggr )^n\, .
\ee
But, these are all tree level contributions and can arise in the effective
superpotential $U^{\rm eff}$ only if they were already part of $U$. (One loop
and higher order contributions would include non-local terms in the effective
action, as well as contributions that are not analytic in the fields, such as
is the case in the Coleman-Weinberg effective potential, which has terms of the
type $\phi ^4 \ln \phi ^2$.) Thus,
$\lambda $ and $m$ are not renormalized, a property that holds for all values
of the couplings, perturbative or non-perturbative. These conclusions are borne
out by explicit (super-) Feynman graph arguments in perturbation theory.

\begin{table}[t]
\begin{center}
\begin{tabular}{|c||c| c|c|c|c|} \hline 
Field   & $\Phi$ & $m$ & $\lambda$ & $U \ {\rm and} \ U^{\rm eff}$ &
$\lambda \Phi/m$ \\
\hline \hline 
$ U(1)  $  & 1 & -2  & -3 & 0 & 0  \\ \hline
$ U(1)_R$  & 1 & 0   & -1 & 2 & 0   \\ \hline
\hline 
\end{tabular}
\end{center}
\caption{$U(1)\times U(1)_R$ quantum numbers of chiral superfields}
\label{table:5}
\end{table}

\medskip

\noindent
{\it (2) Gauge Coupling Renormalization : $\beta$-functions}

The complex gauge coupling $\tau$ enters in a holomorphic way in the gauge
kinetic and self-coupling terms, and we wish to derive the effects of the 
Wilson renormalization group procedure on this term as well \cite{shifman}.
\be
\int \! d^2 \theta \ \tau \tr (WW) 
\qquad \Longrightarrow \qquad   
\int \! d^2 \theta \ \tau _{\rm eff} \tr (WW)\, .
\ee
We may again use $U(1)_R$ symmetry, but now, in the presence of gauge
interactions, this symmetry suffers a triangle anomaly. In view of 
Adler-Bardeen type arguments, the perturbative contribution to the $U(1)_R$
anomaly is purely 1-loop and may be evaluated directly from the triangle
graph, with no need for the inclusion of radiative corrections. The result
is summarized by the {\rm Wilson $\beta$-function} for the coupling $g$
(ignoring temporarily $\theta$), 
\be
\beta _W (g) = {\partial g(\mu) \over \partial \ln \mu} \bigg |_{\Lambda ,
g(\Lambda)}
= \bigl ( -3 C_2(\G) + T_2 (\R) \bigr ) {g^3 \over 16 \pi ^2}\, .
\ee 
Here, we use the standard notations for the quadratic Casimir and Dynkin index
in an arbitrary representation $\R$ of $\G$
\be 
T_\R ^a T_\R ^a = C_2(\R) I_\R
\qquad \qquad
\tr (T_\R ^a T_\R ^b) = T_2(\R) \delta ^{ab}\, .
\ee
Integrating the equation for the coupling flow, and restoring the $\theta$
dependence in terms of a complex valued generalization of the renormalization
scale $\mu$, we have
\be\label{beta}
\tau (\mu) - \tau (\mu')
= {1 \over 2 \pi i} \bigl ( -3 C_2(\G) + T_2(\R) \bigr ) \ln {\mu ' \over \mu}
\ee
For $3C_2(\G) > T_2(\R)$, the theory is asymptotically free, while for
$3C_2(\G) = T_2(\G)$, the theory has vanishing $\beta$-function and will be
conformally invariant provided no explicit mass parameters arise.
For $3C_2(\G) < T_2(\R)$, the theory becomes strongly coupled at short
distances : the microscopic dynamics of such theories is not well formulated in
terms of these degrees of freedom and we shall not consider this case any
further here.

\medskip

\noindent
{\it (3) Applications to $\N=2$ and $\N=4$ theories} 

Let us now consider an $\N=2$ super-Yang-Mills theory, where the gauge algebra
is $\G$ and the representation of the hypermultiplet is $\R_H$. Then, the
repesentation in terms of $\N=1$ chiral multiplets is $\R= \G \oplus \R_H
\oplus \R_H ^*$, and we have $T_2(\R)= C_2(\G) + 2 T_2(\R_H)$, so that
\be
\tau (\mu) - \tau (\mu')
= {1 \over  \pi i} \bigl ( - C_2(\G) + T_2(\R_H) \bigr ) \ln {\mu ' \over \mu}
\ee
Moving onto the $\N=4$ theory, for which $\R_H = \G$, we see that the $\beta$
function vanishes identically and the theory is UV finite. If the vacuum
expectation value of the gauge scalars vanishes, 
this theory is exactly (super-)
conformally invariant, while if the vacuum expectation value is non-zero,
superconformal invariance is spontaneously broken. 

\subsection{Low Energy Dynamics of N=2 super-Yang-Mills}

The gauge multiplet part of the renormalizable $\N=2$ supersymmetric Lagrangian
is derived from expanding the superfields Lagrangian of (\ref{micro}) in terms
of component fields and dropping the contribution of the hypermultiplet,
\bea
\L & = & 
-{1 \over 2 g^2} \tr F F - {\theta \over 16 \pi ^2} \tr F \tilde F 
- {1 \over 2 g^2} \tr D_\mu \phi ^\dagger D^\mu \phi 
+ {1 \over 2 g^2} \tr [\phi ^\dagger, \phi ]^2
\nonumber \\ 
&& -{i \over 2g^2} \tr \bar \lambda \bar \sigma ^\mu D_\mu \lambda 
-{i \over 2g^2} \tr \bar \psi \bar \sigma ^\mu D_\mu \psi
+i {\sqrt 2 \over g^2} \tr ([\phi ^\dagger , \psi] \lambda) 
-i {\sqrt 2 \over g^2} \tr ([\bar \lambda , \phi]\bar \psi) \, .
\eea
We note the symmetry in $\lambda\leftrightarrow\psi$,
which is a manifestation of R-symmetry.
To derive the low energy dynamics of this theory, we begin by identifying the
ground state(s). Since $\L$ is supersymmetric, its Hamiltonian is positive, and
the energy of any state, including the ground state is positive or zero. If we
can find a state with exactly zero energy, then this state is one of the
possible ground states or vacua of the theory, and supersymmetry in
this vacuum  will be unbroken.

\medskip

At the semi-classical level, a solution of vanishing energy will require that
the following quantities vanish simultaneously,
\be \label{vac}
F_{\mu \nu}=0 \qquad \quad D_\mu \phi =0 \qquad \quad \tr ([\phi ^\dagger,
\phi]^2)=0
\ee
so that the gauge field $A_\mu$ is pure gauge. By a gauge choice, we may pick
$A_\mu=0$, so that $\phi$ must be constant, and only the last equation of 
(\ref{vac}) remains to be solved. Since $\phi \in \G$, 
which is the Lie algebra
of a compact gauge group, $\tr ([\phi ^\dagger, \phi ]^2)=0$ implies $[\phi
^\dagger, \phi ]=0$. Now, $\phi$ is a complex field, in the adjoint
representation of $\G$, and may thus be decomposed into its hermitian and
anti-hermitian parts $\phi = \phi _1 + i \phi _2$, with $\phi _{1,2}
^\dagger=\phi _{1,2}$ and $[\phi _1, \phi _2]=0$. Thus, $\phi _1$ and $\phi_2$
may be simultaneously diagonalized, each with real eigenvalues. This implies
that the original $\phi$ may be diagonalized with complex eigenvalues. 
Thus, we
find not just a single zero energy state, but rather a family of zero energy
states, each of which is a candidate vacuum. It is customary to denote these
ground states by $|0\>$, but it is understood that this symbol really is
parametrized by the vacuum expectation values of the gauge scalar $\phi$. 
While
these results were derived here within the semi-classical approximation, they
hold true in the full quantum theory as well.

\medskip

A convenient basis for the diagonal matrices is provided by the Cartan
generators $h_j$ of $\G$, with $j=1,\cdots , n={\rm rank} (\G)$, which obey
$[h_i,h_j]=0$. The vacuum expectation value of the gauge scalar is then
\be
\langle 0 | \phi | 0 \rangle = \sum _{j=1} ^n a_j h_j
\ee
The complex parameters $a_j$, $j=1,\cdots, n$ are called the {\it quantum
moduli} or {\it quantum order parameters} of the vacuum. The semiclassical
limits of these parameters are denoted by $\bar a_j$, $j=1,\cdots ,n$ and are
called the {\it classical moduli} or {\it classical order parameters}.

\medskip

For generic values of the moduli $a_j$, the vacuum expectation value 
$\langle 0| \phi | 0 \rangle$ will have all distinct eigenvalues. Since the
field transforms under the adjoint representation of $\G$, the standard Higgs
mechanism will break the gauge symmetry to that of the Cartan subalgebra
$U(1)^n$. Actually, the role of the different Cartan generators may be
permuted under the set of residual (discrete) gauge transformations which form
the Weyl($\G$) group. Thus, the detailed symmetry breaking is given by
\cite{seiberg}
\be
\G \longrightarrow U(1)^n /{\rm Weyl}(\G)\, .
\ee
In other words, the theory after symmetry breakdown reduces to $n$ copies of
(supersymmetric) electro-magnetism, and for this reason is often said to be
in the {\it Coulomb phase}.

\medskip

At the semi-classical level, non-generic values of $a_j$ may yield a larger
symmetry group. It will be one of the fundamental results of Seiberg-Witten
theory that in the full quantum theory, such larger residual symmetry groups do
not survive quantization, so that the theory is always in the Coulomb phase.
For sufficiently large gauge algebra or for sufficiently large hypermultiplet
contents, novel phenomena, such as the coexistence of massless electrically
charged particles with massless magnetic monopoles is possible at so-called
Argyres-Douglas points or curves \cite{ad}.

\subsection{Particle and Field Contents}

 Since we are left with $n={\rm rank}(\G)$ factors of $U(1)$, we have
$n$ different massless $U(1)$ gauge $\N=2$ supersymmetry multiplets. The fields
in these multiplets may be read off from our general classification,
\be
\bigl ( A_{j \mu}, \ \lambda _{j \pm}, \ \phi _j \bigr )
\qquad \qquad 
\langle 0 | \phi _j | 0 \rangle = a_j
\qquad \qquad 
j=1,\cdots ,n\, .
\ee  

$\bullet$ For each of the ${\rm dim} (\G) - {\rm rank}(\G)$ roots $\alpha$ of
$\G$, we have a massive vector boson multiplet appearing in the spectrum as a
single particle state, described by the field multiplet
\be
\bigl ( V_{\alpha \mu}, \ \lambda _{\alpha \pm}, \ \phi _\alpha \bigr )
\qquad \qquad
\alpha = {\rm root \ of } \ \G\, .
\ee 
Counting the number of fields for each spin, and comparing with the $\N=2$
entries of table \ref{table:4}, we see that this massive vector boson multiplet
must be a BPS multiplet ! As a result, we have the BPS formula for the masses
of this multiplet in terms of the central charges, which are given by $\alpha
\cdot a$,
\be
M^W _\alpha = |\vec{\alpha} \cdot \vec{a}|
\qquad \qquad 
\vec{a} = (a_1 , \cdots , a_n)\, .
\ee

$\bullet$ As the simple gauge algebra $\G$ is broken to a subalgebra which
includes $U(1)$ factors, general arguments show that `t Hooft-Polyakov magnetic
monopoles will arise \cite{thp}. In fact, precisely one magnetic monopoles
arises for each possible embedding of $SU(2)$ into $\G$, i.e. for each root
$\beta$ of $\G$. 

It is instructive to recall the set-up of the basic `t Hooft-Polyakov
solution in the Bogomolnyi-Prasad-Sommerfield (BPS) limit \cite{bps}. 
The limit consists in ignoring the quartic Higgs field potential, but retaining
the boundary conditions on the Higgs field that it should tend towards a
constant value at spatial $\infty$. (This is automatically the set-up
we have  in
$\N=2$ super-Yang-Mills when the gauge symmetry is broken to a $U(1)$ factor.)
Now, the magnetic monopole is a stable, static soliton solution, magnetically
charged under the unbroken $U(1)$. In the  BPS limit, the energy is given by
\bea \label{monopole}
E 
& = & 
{1 \over g^2} \int d^3x \biggl ( \tr (D_i \phi)^2 + \tr B_i^2 \biggr ) 
\nonumber\\ & = &
{1 \over g^2} \int d^3 x \tr \bigl (D_i \phi \pm  B_i \bigr )^2
\mp {2 \over g^2} \int d^3 x \partial _i \bigl ( \tr \phi B_i \bigr )
\eea 
where the $\pm$ signs are correlated. The second term on the rhs of
(\ref{monopole}) is proportional to the magnetic charge. The first term on the
rhs of (\ref{monopole}) is always positive or zero, and this fact guarantees
that the energy is bounded from below by the magnetic charge of the monopole.
This so-called {\it BPS bound} is really nothing else than the BPS bounds we
have encountered in the study of extended supersymmetry. A BPS solution (which
preserves $\N=1$ supersymmetry) is obtained by saturating the BPS bound upon
setting the first term on the rhs of (\ref{monopole}) to zero. 

This construction will give us a BPS formula (at the semi-classical level) for
the mass of the `t Hooft-Polyakov magnetic monopole in terms of the vacuum
expectation value of the gauge scalar. Appling this construction for any
$SU(2)$ subgroup of $\G$, specified uniquely by a root $\vec{\beta}$
of $\G$, and using the fact that $\phi = \vec{a} \cdot \vec{a}$, we find the
semi-classical magnetic monopole BPS mass formula
\be
M^M _\beta = \bigg |\ \vec{\beta} \cdot  \tau \vec{a} \ \bigg |
\qquad \qquad 
\tau =  {\theta \over 2 \pi} + {4 \pi i \over g^2} 
\ee   
Here, we have also included the effect of the $\theta$ angle, which is to
produce an electric charge on the magnetic monopole and make it into a {\it
dyon} \cite{wittendyon}. 

$\bullet$ In fact, a general dyon (single particle state) may be specified by
two roots : one root $\vec{\alpha}$ for its electric charge as in the
construction of vector boson masses; and one (co)root $\vec{\beta}$ for its
magnetic charge as in the construction of the magnetic monopole masses. The
full quantum BPS mass formula for such a dyon is \cite{montonen}, \cite{dyon}
\be
M^D _{(\vec{\alpha}, \vec{\beta})} = \big |\vec{\alpha} \cdot \vec{a}
+ \vec{\beta} \cdot \vec{a}_D \big |\, .
\ee  
In the semi-classical limit, the quantity $\vec{a}_D$ is given by $\vec{a}_D =
\tau \vec{a}$, but at the full quantum level the expression for $\vec{a}_D$ in
terms of $\vec{a}$ will be modified. One of the goals of Seiberg-Witten theory
will be to predict precisely what the relation between these two quantities is.
Given this information, the masses of all BPS states (including vector bosons,
magnetic monopoles and dyons) will be known exactly in terms of the vacuum
moduli parameters $a_j$. In the more constrained $\N=4$ theory, the relation
$a_D = \tau a$ will not be renormalized at the full quantum level.

\subsection{Form of the N=2 Low Energy Effective Lagrangian}

With the preceding subsection, we know precisely what the particle spectrum of
the $\N=2$ super-Yang-Mills theory will be. Upon restriction to the low energy
part of the spectrum, the effects of the massive vector bosons and magnetic
monoples and dyons will decouple. Only the $n$ massless $\N=2$ Abelian $U(1)$
gauge supermultiplets will remain.\footnote{We note 
here though that for special
values of the vacuum moduli, one or several of the dyons may also becomes
massless. This phenomenon and its symptoms in the Wilson effective action 
are central features of the theory. They will
be investigated extensively in the next subsections.} Thus, we are left with
the problem of determining {\it the most general low energy effective action of
$n$ massless $U(1)$ gauge supermultiplets}, with fields $(A_\mu ^j, \ \lambda
_\pm ^j, \ \phi ^j)$, as function of the vacuum moduli parameters $a_j$,
which are the expectation values of the gauge scalars $\phi^j$.

\medskip

We have two methods available to do this : we could start directly from the
manifestly $\N=2$ supersymmetric formulation of \S 2.10; or start from an
$\N=1$ standard superfield formulation and attempt to impose the extra $\N=2$
susy by hand. It turns out to be very instructive to start with the latter. To
do this, we first construct the most general
$\N=1$ invariant Lagrangian with this field content, and we shall then enforce
the supplementary constraints of $\N=2$ supersymmetry \cite{[10]}.
In terms of $\N=1$
superfields, the $n$ multiplets decompose into $r$ Abelian $U(1)$ vector
superfields $V_j \sim (A_{j\mu}, \ \lambda _-^j=\lambda ^j)$ and $n$ chiral
superfields 
$\Phi ^j \sim
(\lambda _+^j =\psi ^j, \ \phi ^j)$, which are neutral under the $U(1)^n$ gauge
group. The field strength superfields are simply
\be
W_{\alpha}^j = - {1 \over 4} \bar D \bar D D_\alpha V^j\, .
\ee  
The full effective action is now gotten by putting together the superpotential
part $\L_U$, the gauge coupling part $\L_G$ and the K{\"a}hler part $\L_K$ of
subsection \S 2.8. Since the chiral superfields $\Phi^j$ are neutral, 
the K{\"a}hler
potential does not involve the gauge fields $V_j$. We have
\be \label{super}
\L = \int \! d^4 \theta\ K(\Phi ^j, (\Phi ^j) ^\dagger)
+ {\rm Re} \int \! d^2 \theta \biggl \{ U(\Phi ^j) 
+ \tau _{ij}(\Phi ^k) W^i W^j
\biggr \}\, .
\ee
Here, the superpotential $U(\Phi ^j)$ and the gauge coupling $\tau _{ij}
(\Phi^k)$ are complex analytic functions of the chiral superfields $\Phi ^j$.

The first step towards enforcing $\N=2$ supersymmetry is to expand the
superfields  of (\ref{super}) in terms of components and eliminate the
auxiliary fields. This step is necessary because the auxiliary fields of
the $\N=1$ superfields do not have a natural place in or good transformation
properties under $\N=2$ supersymmetry. The result is 
\bea
\L & = & - g_{i\bar j} \big ( D_\mu \phi ^i D^\mu \bar \phi ^{\bar j} + i \bar
\psi ^{\bar j} \bar \sigma ^\mu D_\mu \psi ^i \big )
- \12 g^{i\bar j} {\partial U \over \partial \phi ^i} {\partial \bar U \over
\partial \bar \phi ^{\bar j}} + {\rm Re} \biggl \{ \psi ^i \psi ^j {\partial ^2
U \over \phi ^i \partial \phi ^j} \biggr \}
\nonumber \\  
&& -{\rm Re} \bigg \{  
\tau _{ij} \big ( {i\over 2} F^i_{\mu \nu} F^{j\mu \nu} 
- \12 F^i_{\mu \nu} \tilde F^{j\mu \nu} +
\bar \lambda ^i \bar \sigma ^\mu D_\mu \lambda ^j \big ) \bigg \}\, .
\eea
Here, the K{\"a}hler metric $g_{i\bar j}$ is given in terms of the K{\"a}hler
potential $K$ by
\be
g_{i\bar j} (\phi, \bar \phi) = {\partial ^2 K(\phi, \bar \phi) \over \partial
\phi ^i \partial \bar \phi ^{\bar j}}\, .
\ee
The covariant derivative $D_\mu$ is taken with respect to the Levi-Civita 
connection $\Gamma ^i_{jk}$ of the metric $g_{i\bar j}$,
\be
D_\mu \phi ^i = \partial _\mu \phi ^i + \Gamma ^i _{jk} \phi ^j \partial _\mu
\phi ^k\, .
\ee  
No gauge field enters here, because the chiral multiplet in which $\phi ^i$ and
$\psi ^i$ belong are neutral under the gauge group.

\medskip

To enforce $\N=2$ supersymmetry directly is still not so easy. However, as part
of $\N=2$ supersymmetry, we know that the fields 
$A_\mu ^j$ and $\phi ^j$ of the
$\N=2$ gauge supermultiplets have to transform as singlets under the $SU(2)_R$
symmetry, while the fermions $\lambda ^i$ and $\psi ^i$ must combine into an
$SU(2)_R$ doublet. Thus, we may first enforce, as a necessary condition for
$\N=2$ supersymmetry, the requirement that $\psi^i$ and $\lambda ^i$ enter
symmetrically.

\medskip

Clearly, there is no Yukawa interaction for the field $\lambda ^i$, and thus,
the Yukawa interaction must also be absent for the $\psi$ fields. Thus,
$U(\phi^i)$ can be at most linear in $\phi^i$. In fact, if $U(\phi ^i)$ has any
dependence on $\phi ^i$ at all, then the $|\partial U|^2$ term will always be
strictly positive and supersymmetry will be broken spontaneously.  Discarding
this possibility, $U$ must be constant, and may hence be set to $U=0$ : the
superpotential is absent !

\medskip

Next, the kinetic terms of $\lambda ^i$ and $\psi^i$ must be equal. This
requires the relation
\be \label{cond}
{1 \over 2i} \bigl ( \tau _{ij}(\phi) - \overline{\tau _{ij}(\phi)} \bigr )
= {\partial ^2 K(\phi, \bar \phi) \over \partial \phi ^i \partial 
\bar\phi ^j}\, .
\ee
To solve (\ref{cond}), we make use of the holomorphicity of $\tau _{ij}$ and of
its symmetry under $i\leftrightarrow j$. Considering $\tau_{ij}$ as a given
holomorphic function, 
$K$ is determined by a linear differential equation, which
is trivially solved in terms of holomorphic functions $T_i(\phi)$, 
\be 
K(\phi, \bar \phi) = T_i(\phi ) \bar \phi ^i + \overline{T_i(\phi)} \phi ^i
\ee
with the relation
\be \label{grad}
{1 \over 2i} \tau _{ij}(\phi) = {\partial T_j (\phi) \over \partial \phi ^i} =
{\partial T_i(\phi) \over \partial \phi ^j}\, .
\ee 
The second equality in (\ref{grad}) holds in view of the fact that $\tau_{ij}$
is symmetric, and immediately implies that $T_i(\phi)$ is a pure gradient
of a single holomorphic function, called the {\it prepotential} and usually
denoted by $\F(\phi)$. We have then
\be
T_i(\phi) = {1 \over 2i} {\partial \F (\phi) \over \partial \phi ^i}
\ee

Remarkably, the entire low energy effective action $\L$ has now been completely
determined in terms of this single prepotential function $\F(\phi)$. In view
of the arguments presented earlier showing holomorphicity as a function of
the masses, gauge coupling  and Yukawa couplings, we know that $\F$ is also
holomorphic as a function of the gauge coupling $\tau$ and the hypermultiplet
masses $m$ : $\F(\phi; m_a, \tau)$. The $\N=2$ low energy effective action for
the $r$ $U(1)$ gauge multiplets is  thus
\be \label{invlag}
\L =  {\rm Im }(\tau _{ij}) \  F_{\mu \nu}^i F^{j\mu \nu}
+ {\rm Re} (\tau _{ij})\ F_{\mu \nu} ^i \tilde F^{j\mu \nu}
+ {\rm Im} \bigl ( \partial _\mu \bar \phi ^j \partial ^\mu \phi _{Dj}\bigr )
+ \ {\rm fermions}
\ee 
with 
\be
\phi _{Dj} =  {\partial \F (\phi) \over \partial \phi ^j}
\qquad
\qquad 
\tau _{ij}(\phi) = {\p ^2 \F (\phi) \over \p \phi ^i \p \phi ^j} \, .
\ee
Here, it is understood that the gauge scalar $\phi^j$ takes on a vacuum
expectation value $a^j$, which are arbitrary complex numbers. 

\medskip

That the Lagrangian (\ref{invlag}) is fully invariant under the $\N=2$ susy
follows almost trivially by obtaining it from the expansion of the manifestly
$\N=2$ supersymmetric Lagrangian for gauge group $U(1)^n$ of (\ref{fulllag})
and without hypermultiplets. 

\subsection{Physical Properties of the Prepotential}

Since $\F(\phi)$ only depends upon the field 
and not upon space-time derivatives
$\partial _\mu \phi$, $\F$ is uniquely determined by its values on the
expectation values $a_j$, so we consider now the complex analytic function
$\F(a_j;m_a,\tau)$. $\F$ is subject to two fundamental physical properties.

\begin{itemize}

\item $\F(a_j;m_a,\tau)$ is complex analytic by $\N=2$ supersymmetry;

\item $\tau_{ij}(a) = {\partial ^2 \F \over \partial a_i \partial a_j}$
obeys ${\rm Im} (\tau _{ij}) >0$ since this function plays the role of a metric
on the gauge kinetic terms. Unless ${\rm Im} (\tau _{ij}) >0$, some gauge fields
will propagate with negative probability.

\end{itemize}

As a result of the above two properties of $\F$, we find that $\F$ cannot be a
single valued function of the vacuum moduli $a_j$ (or of the gauge coupling
$\tau$).  Indeed, if $\F$ were single-valued, then ${\rm Im}(\tau_{ij})$ will
be a single-valued harmonic function, which by the second point above is
bounded from below. But, by the maximum pinciple, such a harmonic
function must in fact be constant. If it is not constant, then it can never be
bounded from below or from above. These properties form the basis for the
construction of $\F$ given by Seiberg and Witten \cite{seiberg}.

\medskip

Actually, we can easily explore the properties of the function $\F$ in the
regime of large vacuum expectation values, since the effects of renormalization
are then known from asymptotic freedom, and we find for say $\G=SU(N)$, 
\be \label{theta}
\tau _{ij} = {\theta _{ij} \over 2\pi} 
+ {4 \pi i \over g_{ij}^2} \sim { i \over
2\pi} \ln (a_i - a_j)^2/\mu ^2\, ,
\ee
as $a_j \to \infty$.
Integrating up to derive $\F$, we find
\be \label{eff}
\F \sim {i \over 8\pi} \sum _{i,j} (a_i - a_j)^2 \ln (a_i - a_j)^2/\mu ^2\, .
\ee
The expression for $\F$ in this limit clearly shows that $\F$ is neither
constant nor single-valued.

\medskip
 
In fact, the nature of the multiple-valuedness (also called the {\it
monodromy}) of $\F$ and of $\tau_{ij}$ is understood in part. A $2\pi$ rotation
of any
$a_i-a_j$ amounts to a shift of the $\theta _{ij}$ angle by $4\pi$. This
shift is physically immaterial since the instanton angles $\theta _{ij}$ are
periodic with period $2\pi$. However, this cannot be the source of all the
monodromy of $\F$. For, if it were, then the expression  (\ref{theta}) would
continue to be the source of this monodromy down to $a_i - a_j \to 0$. But, as
$a_i - a_j \to 0$, we would invariably led into a domain where ${\rm Im}
(\tau_{ij}) <0$, violating the positivity property ${\rm Im}(\tau_{ij})>0$
explained previously. 
The source of additional monodromy is much more subtle and will be explained
next.

\subsection{Electric-Magnetic Duality}

We begin by discussing first the case of $\G=SU(2)$ gauge group and no
hypermultiplets. The fields are then the single gauge scalar $\phi$, the gauge
field $A_\mu$ with field strength $F_{\mu \nu}$, and the prepotential is a
complex analytic function $\F(\phi)$ dependent on a single complex $\phi$ only.
The effective Lagrangian is then
\bea
\L = {\rm Im} \biggl \{ \tau (\phi) (F_{\mu \nu} 
&+& i \tilde F_{\mu \nu})^2 \biggr \}
+ \partial _\mu \! \left ( \matrix{\phi _D \cr \phi} \right ) ^\dagger J 
\partial ^\mu \! \left ( \matrix{\phi _D \cr \phi} \right ) + \ {\rm fermions}
\nonumber \\
 J&=&\left ( \matrix{0 & i \cr -i & 0 \cr} \right )
\eea
The kinetic term is manifestly invariant under the linear action of an
$M\in SL(2,{\bf R}) = Sp(2, {\bf R})$ on $\phi_D$ and $\phi$, with
\be \label{monod}
\left ( \matrix{\phi _D \cr \phi} \right )
\longrightarrow 
M \left ( \matrix{\phi _D \cr \phi} \right )
\qquad \qquad
M^\dagger J M = J
\ee
This group is generated by {\it translations}
$T_\beta$ and an inversion $S$, given by
\be \label{st}
T_\beta = \left ( \matrix{1 & \beta \cr 0 & 1 \cr} \right )
\qquad \qquad
S = \left ( \matrix{0 & -1 \cr 1 & 0 \cr} \right )\, .
\ee

The action of $T_\beta$ on $\phi$, $\phi _D$ and $\tau(\phi)$ is as follows
\bea
\phi _D & \longrightarrow & \phi _D + \beta \phi \nonumber \\
\phi    & \longrightarrow & \phi \nonumber \\
\tau(\phi) & \longrightarrow & \tau (\phi) + \beta\, .
\eea
This shift is precisely of the form of a shift in the angle $\theta$, and will
be immaterial only if $\beta $ is an integer multiple of 2. 
Thus, while the full $SL(2,{\bf R})$ leaves the kinetic term invariant,
only a subgroup $\Gamma _0(2) \subset SL(2,{\bf Z})$ will leave the quantum
mechanical physics of the problem unchanged.
The action of $S$ on $\phi$, $\phi _D$ and $\tau(\phi)$ is as follows
\bea
\phi _D & \longrightarrow & -\phi  \nonumber \\
\phi    & \longrightarrow & \phi _D \nonumber \\
\tau(\phi) & \longrightarrow & - 1/\tau (\phi) \, .
\eea  
Considering the special case where $\theta =0$, the transformation on the
remaining coupling $g$ in $\tau$ is by inversion : $ g \to 1/g$, and thus
exchanges small and large coupling regimes.

\medskip

Thus far, we have implemented the action of the transformation $S$ only on the
scalar kinetic term, and it remains to implement $S$ on the gauge field. But,
the gauge part of the action does not manifestly exhibit a symmetry in which
$g \to 1/g$. The key novel ingredient is the fact that upon $g \to 1/g$, the
gauge field $A_\mu$ must simultaneously be replaced by its dual field
$A_{D\mu}$. This transformation interchanges the roles of electric
and magnetic fields, as was considered by Montonen and Olive \cite{montonen}. 

\medskip

The fact that $g \to 1/g$ is related to an electric-magnetic duality
transformation may be seen intuitively from the Dirac quantization condition.
If we denote by $g$ the unit of electric charge and by $g_D$ the unit of
magnetic  charge, then the Dirac condition requires $g \cdot g_D \sim 1$ in
units where $\hbar =1$. Thus, as $g \to 1/g$, we have equivalently $g
\leftrightarrow g_D$, and electric and magnetic charge are being exchanged.

\medskip

Electric-magnetic duality in the free Maxwell's equations
$\partial _\mu \tilde F^{\mu \nu} = \partial _\mu F^{\mu \nu}=0$
is a manifest symmetry, realized by $F^{\mu \nu} \leftrightarrow \tilde F^{\mu
\nu}$. At the level of the Lagrangian, the symmetry is obscured by the fact
that the Bianchi identity $\partial _\mu \tilde F^{\mu \nu}=0$ has been
solved for in terms of the gauge potential, thus obscuring duality symmetry.

\medskip

To exhibit duality in the action, we must arrange to have both the field
equations $\partial _\mu F^{\mu \nu}=0$ and the Bianchi identities $\partial
_\mu \tilde F^{\mu \nu}=0$ emerge as field equations. This may be done by
formulating the action in terms of an independent field strength field
$F_{\mu \nu}$ and by enforcing the Bianchi identity by introducing a Lagrange
multiplier field $A_{D\mu}$. We shall carry out this duality transformation
only on the gauge field, but a full supersymmetric argument is possible as
well. We have
\be
S _G  =  \int {\rm Im} \tau (\phi) (F + i \tilde F)^2 + 2 \int A_{D\mu}
\partial _\nu \tilde F^{\mu \nu}\, .
\ee 
We consider the Lagrange multiplier $A_{D\mu}$ as a field in its own right,
and introduce its field strength $F_{D\mu \nu}=\partial _\mu A_{D\nu} -
\partial _\nu A_{D\mu}$. Upon integration by parts, the last term in $S_G$ is
then $\int F_{D\mu \nu} F^{\mu \nu}$, and we may complete the square in $F$ as
follows 
\be
S_G = \int {\rm Im} \biggl [
\tau (\phi) \big \{ F + i \tilde F +{1 \over \tau (\phi)} (F_D + i \tilde F_D)
\big \}^2 - {1 \over \tau (\phi)} (F_D + i \tilde F_D)^2 \biggr ]\, .
\ee
Integrating out the unconstrained field $F$ by Gaussian integration, we are
left with the same action $S_G$, but formulated entirely in terms of the dual
gauge field $A_{D\mu}$, and given by 
\be
S_G = \int {\rm Im} \biggl [
- {1 \over \tau (\phi)} (F_D + i \tilde F_D)^2 \biggr ]\, .
\ee
Thus, the $S$ transformation in which $\tau (\phi) \to -1/\tau(\phi)$ has been
realized on the gauge field as well. The precise monodromy group is generated
by the transformations $T_2$ and $S$ of (\ref{st}), and this group is a
subgroup $\Gamma _0(2)$ of the group $SL(2,{\bf Z})$, which is generated by
$T_1$ and $S$.

\subsection{Monodromy via Elliptic Curves for SU(2) Gauge Group}

We have now obtained a precise formulation of the monodromy problem  associated
with the Seiberg-Witten theory for $SU(2)$ gauge group.

\begin{itemize}

\item $\left ( \matrix{a _D \cr a} \right )$ transforms linearly under the
$\Gamma _0(2) \subset SL(2,{\bf Z})$ monodromy group;

\item $\tau (a) = \partial a_D /\partial a$ obeys the condition ${\rm
Im}\tau(a)>0$;

\item The monodromy $T_2$ is realized in the limit where $a\to \infty$.

\end{itemize}

Generally, such monodromy problems are very hard to solve. 
Here however, we have
one very important hint : a complex quantity $\tau$ with positive
definite imaginary part. Such quantities are also fundamental in the theory of
Riemann surfaces, where they describe the moduli of Riemann surfaces and the
positivity condition ensures regularity of the surface. In our
case, the relevant Riemann surface is an {\it elliptic curve} or equivalently a
{\it torus}. 

\medskip

The torus is conveniently described as an elliptic curve in ${\bf CP}^2$,
parametrized by two complex coordinates $(k,y)$, and given by
\be
y^2 = (k-\Lambda ^2) (k+\Lambda ^2) (k-u)\, .
\ee
Here $\Lambda $ is an overall scale, and $u$ parametrizes the
modulus $\tau$ of the elliptic curve. One may view the elliptic curve as
realized by a double cover of the complex plane 
(the two sheets corresponding to
the $\pm y$ branches) with branch points 
$\pm \Lambda ^2$, $u$ and $\infty$, and
branch cuts for example along 
$[- \Lambda ^2, + \Lambda ^2]$ and $[u, \infty]$.

\medskip

A canonical basis for the homology 1-cycles (in which the cycles $A$ and $B$
intersect precisely once) is provided by the following closed curves;
$A$ : circling the banch cut $[-\Lambda ^2, +\Lambda ^2]$ once; and $B$ : from
branch points $+\Lambda ^2$ to $u$ circling from the upper to the lower sheets
and back. This choice of homology cycles is not unique however, even when we
require single intersection. Any change of homology basis by a symplectic
transformation with integer coefficients will preserve canonical intersection
\be
\left ( \matrix{B \cr A} \right )
\longrightarrow 
M \left ( \matrix{B \cr A} \right )
\qquad \qquad
M^\dagger J M = J
\qquad \qquad 
M\in SL(2,{\bf Z})
\ee
Notice the similarity with the transformation properties of $\phi _D$ and
$\phi$ in  (\ref{monod}).

\medskip

There is a unique holomorphic 1-form (or {\it Abelian differential of the first
kind}), $dk/y$, whose Abelian integral is
\be
z(P) - z(Q) = \int _Q ^P {dk \over y}\, .
\ee 
This integral is multiple valued, and the periods of the torus are
defined by the following contour integrals
\be
\omega = \oint _A {dk\over y}
\qquad \qquad
\omega _D = \oint _B {dk \over y}
\ee
The value of the  matrix of periods $(\omega _D\ \omega)$ is subject to
monodromy as the canonical homology basis of $A$ and $B$ cycles is changed,
and transforms according to
\be
\left ( \matrix{\omega _D \cr \omega} \right )
\longrightarrow 
M \left ( \matrix{\omega_D \cr \omega} \right )
\qquad \qquad
M^\dagger J M = J
\qquad \qquad 
M\in SL(2,{\bf Z})
\ee
This transformation law is strongly suggestive in view of its similarity with
the transformation law for $\phi _D$ and $\phi$ in (\ref{monod}).
The modulus of the elliptic curve is defined by 
\be
\tau = {\omega_D \over \omega} 
\qquad \qquad 
{\rm Im }\tau >0\, ,
\ee
where the last condition guarantees that the curve is non-singular.

\medskip

We are now ready to identify the data of Seiberg-Witten theory and those of the
family of elliptic curves parametrized by the modulus $\tau$, or equivalently
by the complex parameter $u$. The modulus of the curve $\tau$ is identified
with the Seiberg-Witten gauge coupling $\tau (a)$, so that
\be
\tau = {\omega _D \over \omega} =
\tau (a) = {\partial a_D \over \partial a}
=\biggl ( {\partial a_D \over \partial u} \biggr ) 
\biggl ( {\partial a \over \partial u} \biggr ) ^{-1}
\ee
These equalities are readily solved by simply identifying
\be
{\partial a \over \partial u} = \omega = \oint _A {dk \over y}
\qquad \qquad
{\partial a_D \over \partial u} = \omega _D = \oint _B {dk \over y}\, .
\ee
Integration in $u$ may be carried out on both equations, and we obtain the
final form of the Seiberg-Witten solution,
\be
a = \oint _A d\lambda 
\qquad \qquad
a_D = \oint _B d\lambda
\ee
where the differential 1-form $d\lambda $, usually called the {\it
Seiberg-Witten differential} is such that its $u$-derivative is holomorphic.
Its form may be obtained explicitly by integration in $u$, and is unique up to
the addition of any exact differential,
\be
{\partial d\lambda \over \partial u}
= {dk \over y}  = {dk \over \sqrt{(k^2-\Lambda ^4)(k-u)}}
\qquad 
\Longrightarrow 
\qquad
d\lambda = {(k-u) dk \over y} + \ {\rm exact}
\ee
Notice that the Seiberg-Witten differential $d\lambda$ has a pole at
 $k=\infty$.

\medskip

The explicit solution is obtained by using the homology $A$ and $B$ cycles
defined above, and is given by the following parametric representation of
complete elliptic integrals
\be  \label{explicit}
a_D(u) = {\sqrt 2 \over \pi} \int _{\Lambda ^2} ^u dk {\sqrt{k-u} \over
\sqrt{k^2 - \Lambda ^4}}
\qquad  \qquad
a(u) = {\sqrt 2 \over \pi} \int _{\Lambda ^2} ^{\Lambda ^2} 
dk {\sqrt{k-u} \over
\sqrt{k^2 - \Lambda ^4}}\, .
\ee
The $T_2$ monodromy condition is readily verified in the limit $u\to\infty$
\bea
a_D(u)&= &{\sqrt{2u}\over\pi}\int_{\Lambda^2/u}^1dy{\sqrt{y-1}
\over \sqrt{y^2-{\Lambda^4\over u^2}}}
\sim -i{\sqrt{2u}\ln u\over\pi}\\
a(u)&\sim &
{\sqrt{2u}\over\pi}\int_{-\Lambda^2}^{\Lambda^2}
{dk\over\sqrt{\Lambda^4-k^2}}
=\sqrt{2u}
\eea
The function $a_D(a)$ and $\tau(a)$ are then obtained by eliminating $u$.

\subsection{Physical Interpretation of Singularities}

The elliptic curve $y^2 = (k-\Lambda ^2) (k+\Lambda ^2) (k-u)$ has precisely 3
singularities at $u=\pm \Lambda ^2, \ \infty$ where the curve degenerates. 
These
singularities translate into singular behavior of $a_D(a)$ and $\tau(a)$. For
all other $u\in {\bf C}$, these quantities are regular.

\medskip

Recall from the discussion of the Wilson effective action that singularities
will arise in $\tau(a)$ and $\F(a)$ if and only if the mass(es) of
some particle(s) that was integrated out in the Wilson remnormalization group
procedure actually becomes zero. Now, we know the masses of all massive vector
bosons, magnetic monopoles and dyons, because these particles belong to short
BPS multiplets of $\N=2$ supersymmetry and so their masses obey the BPS bound
\be
M= |n_2 a + n_m a_D|
\ee
where $n_e$ and $n_m$ are the electric and magnetic charges respectively.
Whenever $a_D /a$ becomes a rational number, some dyon masses will become zero.
In our case, there are only three candidate values for the vacuum moduli where
this can happen, and we now study those in turn, using the explicit
solution of (\ref{explicit}).

\begin{itemize}

\item $u\to \infty$ \hfill\break
corresponds to the large $a$ limit, where asymptotic freedom applies and we
find 
\bea
a(u) & \sim & \sqrt{u} \nonumber \\
a_D(u) & \sim & \sqrt{u} \ln {u \over \Lambda ^2} \sim a \ln a
\eea

\item $u \to + \Lambda ^2$ 
\bea
a(u) & \sim & {i \over \pi} a_D \ln {a_D \over \Lambda} \nonumber \\
a_D(u) & \sim & (u-\Lambda ^2)
\eea 
Using the BPS mass formula, we see that at this point any particle with
zero electric charge $n_e$ will be massless : these  are just the pure
magnetic monopoles, with $(n_e=0, \ n_m\not=0)$.

\item $u \to - \Lambda ^2$ 
\bea
(a - a_D)(u) & \sim & (u+\Lambda ^2) \nonumber \\ 
a(u) & \sim &  {i \over \pi} (a_D -a) \ln {a_D -a \over \Lambda}
\eea 
Using the BPS mass formula, we see that at this point any particle with
zero charge $n_e+n_m$ will be masseless : these are the dyons, with $(n_e, \
n_m=-n_e)$.

\end{itemize}

The monodromy matrices at each of these singularities can be derived from the
presence of the logarithm in each of the above formulas. One finds
\be \label{sol}
D_\infty = \left ( \matrix{1 & 2 \cr 0 & 1 \cr} \right )
\qquad 
D_{+\Lambda ^2} = \left ( \matrix{1 & 0 \cr -2 & 1 \cr} \right )
\qquad 
D_{-\Lambda ^2} = \left ( \matrix{-1 & 2 \cr -2 & 3 \cr} \right )\, .
\ee
As a consistency check that the above singularities are the only ones required
to satisfy the monodromy conditions in a consistent way, one may check that the
product of all three monodromies is proportional to the identity matrix
$D_\infty D_{+\Lambda ^2} D_{-\Lambda ^2}=-I$.

Physically, the remarkable new phenomenon that emerges from this study of the
Wilson effective action is that magnetic monopoles and dyons, 
which are usually
encountered as heavy semi-classical soliton solutions, may actually become
massless at strong coupling.

\subsection{Hypergeometric Function Representation}

In their original paper, Seiberg and Witten \cite{seiberg} were actually able
to identify the additional monodromies at $u=\pm\Lambda^2$,
without solving first for $(a,a_D)$. Their arguments
were as follows. Since ${\cal F}$ is not global, it has
at least 3 singularities. 
As explained earlier, the appearance of a singularity
in the Wilson effective action is due to a particle
becoming massless. 
The key physical input is the realization that the particle becoming
massless is a monopole. The magnetic coupling $\tau_D$ 
can then be evaluated in the
dual formulation in terms of the dual gauge field $A_D$,
giving the monodromy matrix $D_{+\Lambda^2}$. The remaining
monodromy matrix $D_{-\Lambda^2}$ follows from
the relation $D_\infty D_{+\Lambda ^2} D_{-\Lambda ^2}=-I$ for the homotopy
group of the sphere with 3 punctures. 
In this way, the three monodromy matrices of
(\ref{sol}) are recovered.
Here we indicate briefly how the
monodromy problem can be solved in terms of hypergeometric functions, if all
three monodromies $D_{\infty}$, $D_{+\Lambda^2}$, and $D_{-\Lambda^2}$
are already known and given by (\ref{sol}).

\medskip

We look for $(a_D,a)$ as solutions of a second order
differential equation with regular singular points.
For simplicity, set $\Lambda=1$.
Consider the equation
\be \label{hyper}
\bigl ( -{d^2\over dz^2}+V(z) \bigr ) \psi(z)=0
\ee
where $V(z)$ is a meromorphic function with at most double poles.
Then the equation admits two independent solutions $\psi_1,\psi_2$
which get transformed into a linear combination of
each other under analytic continuation around each of the
poles of $V(z)$. In the present case, there should be 3 singularities
at $z=-\Lambda^2,\Lambda^2,\infty$, and a candidate for $V(z)$ is
\be
V(z)
=
-{1\over 4}
\biggl ({1-\lambda_1^2\over (z+1)^2}
+
{1-\lambda_2^2\over (z-1)^2}
-
{1-\lambda_1^2-\lambda_2^2+\lambda_3^2
\over
(z+1)(z-1)} \biggr )
\ee
As $z\to\infty$, $V(z)\sim -{1\over 4}{1-\lambda_3^2\over z^2}$.
The solutions behave asymptotically like $z^{(1\pm\lambda_3)/2}$
if $\lambda_3\not=0$, and like $\sqrt z$ and $\sqrt z \ln \,z$
when $\lambda_3=0$. Comparing with the desired behavior for
$a$ and $a_D$, we set $\lambda_3=0$. Next, considering asymptotics
as $z\to \Lambda^2$, we are led to the choice $\lambda_1=\lambda_2=1$.

\medskip

Now the solutions $\psi$ of (\ref{hyper}) are related to
those of the standard hypergeometric equation
$x(1-x)f''+[c-(a+b+1)x]f'-ab f=0$ by
\be
\psi(z)=(z+1)^{(1-\lambda_1)/2}(z-1)^{(1-\lambda_2)/2}
f({z+1\over 2})\, .
\ee 
with $a=b=-{1\over 2}$, $c=0$.
Since a basis for the hypergeometric
equation is given by 
\bea 
f_1(x) & = & (1-x)^{c-a-b}F(c-a,c-b;c+1-a-b;1-x)\, , \nonumber \\
f_2(x) & = & (-x)^{-a}F(a,a+1-c;a+1-b;{1\over x})\, ,
\eea
the natural candidate for $(a,a_D)$ is
\bea
a_D(u)&=&i{u-1\over 2}F({1\over 2},{1\over 2};2;{1-u\over 2})\\
a(u) &= &\sqrt 2 (u+1)^{1\over 2}F(-{1\over 2},
{1\over 2};1;{2\over u+1})\, .
\eea
Recalling that the hypergeometric function admits an integral 
representation
\be 
F(a,b;c,z)
={\Gamma(c)\over\Gamma(b)\Gamma(c-b)}
\int_0^1dt\ t^{b-1}(1-t)^{c-b-1}(1-tz)^{-a}
\ee
we obtain the following integral representation for $(a,a_D)$
\bea
a_D(u) &=& 
i{u-1\over 2}{\Gamma(2)\over\Gamma({1\over 2})\Gamma({3\over 2})}
\int_0^1 dt\ t^{-{1\over 2}}(1-t)^{1\over 2}
\biggl (1-t{u-1\over 2} \biggr )^{-{1\over 2}}\\
a(u)&=&
\sqrt 2(u+1)^{1\over 2}{\Gamma(1)\over\Gamma({1\over 2})\Gamma({1\over 2})}
\int_0^1dt\ t^{-{1\over 2}}(1-t)^{-{1\over 2}}
\biggl (1-t{u+1\over 2u} \biggr )^{1\over 2}
\eea
A change of variables gives back the expressions (\ref{explicit}) proposed 
earlier for $(a,a_D)$.

\vfill\eject

\setcounter{equation}{0}
\section{More General Gauge Groups, Hypermultiplets}

In the previous section, we reviewed Seiberg-Witten theory for an $\N=2$
super-Yang-Mills theory with gauge group $SU(2)$ and no hypermultiplets
present. The key ingredients were

\begin{itemize}

\item The SW curve $\Gamma (u)$ given by $y^2 = (k-\Lambda ^2) (k+\Lambda ^2)
(k-u)$;

\item The SW differential $d\lambda = (k-u) y^{-1} dk$;

\item The relation between the periods of $d\lambda$, the quantum moduli $a$
and $a_D$ and the prepotential $\F$ by
\be
a= {1 \over 2 \pi i} \oint _A d\lambda 
\qquad \quad
a_D = {1 \over 2 \pi i} \oint _B d\lambda 
\qquad \qquad
a_D = {\partial F \over \partial a}
\ee

\end{itemize}

\noindent
(Henceforth, we shall include an extra factor of $2 \pi i$ in the 
normalization of $d \lambda$.)
In this section, we shall review the known generalizations of this
construction to

\begin{itemize} 

\item higher gauge algebras $SU(N)$, $SO(N)$, $Sp(N)$, and
exceptional algebras;

\item including hypermultiplets in various representations of the gauge
algebra, with the constraint that the theory remain asymptotically free or
have vanishing $\beta$-function.

\end{itemize}

A useful starting point is again the symmetries of the gauge scalar kinetic
term, which we shall rewrite as follows
\be
2 {\rm Im} \partial _\mu \phi ^i \partial ^\mu \phi _{Di}
=
i \partial _\mu \left ( \matrix{\phi_{Di} \cr \phi_i} \right ) ^\dagger
J_{ij} 
\left ( \matrix{\phi _{Di} \cr \phi_i} \right )
\qquad 
J = \left ( \matrix{0 & -I \cr I & 0} \right )
\ee
where the indices $i,j=1,\cdots , n={\rm rank}(\G)$. This part of the low
energy effective Lagrangian is invariant under the continuous symmetry group
$Sp(2n, {\bf R})$. The subgroup of Abelian transformations of the type
\be \label{beta1}
T_{\beta} \ : \ \left ( \matrix{\phi _{Di} \cr \phi_i} \right )
\longrightarrow 
\left ( \matrix{I & \beta \cr 0 & I} \right )
\left ( \matrix{\phi _{Di} \cr \phi_i} \right )
\ee
plays a special role. Its effect on the gauge coupling matrix $\tau _{ij}$ is
by shifts by $\beta $ : 
\be 
T_\beta \ : \ {\rm Re}(\tau_{ij}) 
\longrightarrow 
{\rm Re} (\tau _{ij}) + \beta _{ij}\, .
\ee
However, these transformations have the effect of shifting the instanton
angles $\theta _{ij}$ by $2\pi \beta _{ij}$, and thus the entries $\beta_{ij}$
must be integers. This restriction restricts the full invariance group of the
gauge scalar kinetic term to the subgroup $Sp(2n, {\bf Z})$. 

\medskip

The group $Sp(2n,{\bf Z})$ may be viewed as being generated by transformations
$T_\beta$, as in (\ref{beta1}), with integer entries, together with the
inversion $\tau \longrightarrow \tau ^{-1}$. This transformation exchanges weak
and strong coupling, and may be interpreted as a generalized electric-magnetic
duality transformation, extending the case of $SU(2)$ gauge group. 

\subsection{Model of Riemann Surfaces}

Consider a compact Riemann surface $\Gamma$ without boundary of genus $g$; such
a surface may be thought of as a sphere with $g$ handles attached to it. 
Let $\{
A_i, \ B_i, \ i=1,\cdots ,g\}$ be a basis  of closed 1-cycles for the homology
$H_1(\Gamma, {\bf Z})$ of $\Gamma$, so that the intersection form $\#$ is
canonical
\bea
\# (A_i,A_j) = \# (B_i, B_j) & = & 0 \nonumber \\
\# (A_i,B_j) = -\# (B_j, A_i) & = & \delta _{ij}\, . 
\eea
Dual to this canonical homology basis are the $g$ independent holomorphic
1-forms $\Omega _i$, $i=1,\cdots ,g$ (Abelian differentials of the
1-st kind) with periods
\be \label{abelian}
\omega _{ij} = \oint _{A_i} \Omega _j
\qquad \qquad
\omega_{Dij} = \oint _{B_i} \Omega _j\, .
\ee
The {\it period matrix} $\tau _{ij}$ is defined by
\be
\tau \equiv \omega _D \omega ^{-1}
\qquad \qquad
\tau _{ij} = \sum _k \omega _{Dik} \bigl ( \omega ^{-1} \bigr ) _{kj}\, .
\ee
The {\it Riemann Theorems} imply two key properties of the period matrix
\bea
&(1)& {\rm Im} \tau  >  0  \quad {\rm for \ non-degenerate} \ \Gamma \nonumber
\\
&(2)& \tau _{ij}  =  \tau _{ji}\, . 
\eea
Finally,  a change in homology basis  from $\{ A_i, \ B_i\}$ to $\{ A_i ', \
B_i'\}$ that preserves the intersection form $\#$ is an element of $Sp(2g, {\bf
Z})$, since the intersection matrix is equivalent to the symplectic matrix in
$2g$ dimensions,
\be
\left ( \matrix{B_i \cr A_i} \right )
\longrightarrow 
\left ( \matrix{B_i' \cr A_i '} \right ) = M
\left ( \matrix{B_i \cr A_i} \right )
\qquad \quad
M \in Sp(2g, {\bf Z})\, .
\ee
Thus, the monodromy group of the Abelian integrals  of
(\ref{abelian}) is a subgroup of $Sp(2g, {\bf Z})$.

\subsection{Identifying Seiberg-Witten and Riemann Surface Data}

A natural generalization of the Seiberg-Witten construction for $SU(2)$ gauge
group emerges in the following way. 

$\bullet$ The Seiberg-Witten curve is a family of Riemann surfaces $\Gamma
(u_1, \cdots , u_n)$, parametrized by $n$ complex parameters $u_1, \cdots ,
u_n$, which play the role of vacuum moduli and parametrize the flat directions
of $\N=2$ vacua. The family of curves may contain singular curves for certain
values of $u_i$, but these singularities should arise only when physical
massless particles appear.

$\bullet$ The curves $\Gamma (u_i)$  must be invariant under the Weyl group
Weyl($\G$), which is the residual gauge invariance after the gauge algebra
$\G$ has been broken down to its Cartan subalgebra, $U(1)^n$.

$\bullet$ The Seiberg-Witten differential $d\lambda$ is a meromorphic 1-form on
$\Gamma$ such that $\partial d\lambda / \partial u_i$ is a holomorphic 1-form
(or Abelian differential of the first kind) on $\Gamma$. The residues at the
poles of $d\lambda$ are then automatically independent of $u_i$, and are
linear combinations of the hypermultiplet mass parameters. Physically, the
fact that the residues of $d\lambda$ are independent of the $u_i$ means that
the hypermultiplet masses (entering in the $\N=1$ superpotential) are not
renormalized and thus are unchanged under any changes in $u_i$. Linearity of
the residues in terms of the hypermultiplet mass parameters is required by
the BPS mass formula, in which the hypermultiplet mass parameters enter
linearly.

$\bullet$ The periods of $\Gamma$ then enter in the following way. Since
$\partial d\lambda /\partial u_j$ are holomorphic 1-forms, a period matrix
$\tau_{ij}$ may be constructed from by integrating these 1-forms on a set of
homology cycles $A_i$, $B_i$, $i=1,\cdots ,n$ as follows. We begin by defining
the periods
\bea
\omega _{ij} & = & {\partial a_i \over \partial u_j} = 
{1 \over 2 \pi i} \oint _{A_i}  {\partial
d\lambda \over \partial u_j} \nonumber \\
\omega _{Dij} & = & {\partial a_{Di} \over \partial u_j} = 
{1 \over 2 \pi i} \oint _{B_i} 
{\partial d\lambda \over \partial u_j} 
\eea
where we have identified the periods $\omega$ and $\omega _D$ with the relevant
super-Yang-Mills quantities. The gauge coupling constant matrix $\tau$ may now
be naturally connected with the Riemann surface data by
\be
\tau = {\partial a_D \over \partial u} \bigg ( {\partial a \over \partial
u}\bigg ) ^{-1}\, .
\ee 
By making this identification, we guarantee that ${\rm Im} (\tau) >0$, $\tau
_{ij}=\tau_{ji}$ and holomorphicity of $\tau_{ij}$ as a function of $u_i$.
Recall that these were the key ingredients in the Seiberg-Witten construction.
Upon integrating in $u_k$, we obtain the fundamental relations
\be
a_i = {1 \over 2 \pi i} \oint _{A_i} d\lambda 
\qquad  \qquad
a_{Di} = {1 \over 2 \pi i} \oint _{B_i} d\lambda\, .
\ee 

$\bullet$ For the simplest cases, to be explained shortly, $\Gamma$ is
precisely of genus $g=n$, and $A_i$, $B_i$ $i=1,\cdots , n$ form a
{\it basis} of
$H_1(\Gamma, {\bf Z})$. But more generally, $n\leq g$, and
$A_i$, $B_i$ form a suitable subset of $H_1(\Gamma, {\bf Z})$.

\medskip

$\bullet$ A crucial requirement of the Seiberg-Witten and Riemann
surface correspondence is that the prepotential $\F$ defined by
the Riemann surface data, $a_{Dk}={\partial {\F}/ \partial a_k}$,
must have the logarithmic singularities ${\cal F}^{\rm pert}$
predicted by perturbative
field theory in the regime of large vacuum expectation values.
For $\G=SU(N)$ and no hypermultiplet, these singularities
have been described in (\ref{eff}). 
For general gauge algebra $\G$ with roots ${\cal R}(\G)$
and hypermultiplet of mass $m$
in a representation $R$ of $\G$ with weights ${\cal W}(R)$,
they are given by
\be\label{logsing}
\F^{\rm pert}(a)
={i\over 8\pi}
\bigg [ \sum_{\alpha\in{\cal R}(\G)}(\alpha\cdot a)^2\ln\,
{(\alpha\cdot a)^2\over\mu^2}
-
\sum_{\lambda\in{\cal W}(R)}(\lambda\cdot a+m)^2\ln\,
{(\lambda\cdot a+m)^2\over\mu^2}\bigg ]
\ee

\medskip

In view of the above correspondence, obtaining the Seiberg-Witten solution for
any given gauge algebra and hypermultiplet representation has been reduced to
obtaining the Seiberg-Witten curve $\Gamma(u_i)$, the Seiberg-Witten
differential $d\lambda$ and the cycles $A_i$, $B_i$ on $\Gamma(u_i)$. The
monodromy problem is in general very difficult to solve, and more indirect
arguments are needed to proceed. The key considerations that have been made
use of in the construction for $SU(N)$, $SO(N)$ and  $Sp(N)$ gauge algebras,
with either no hypermultiplets or hypermultiplets in the fundamental
representation of the gauge algebra are as follows.

\begin{itemize}

\item Matching the singularity structure of the curve $\Gamma (u_i)$ with
the appearance of massless particles in the $\N=2$ super-Yang-Mills theory;

\item $U(1)_R$ charge assignments;

\item Decoupling limits to smaller gauge algebras and smaller hypermultiplet
contents, including to the smallest of them all, namely the $SU(2)$ gauge
algebra which is known from the original Seiberg-Witten work;

\item Educated guesswork.    

\end{itemize}

\subsection{SU(N) Gauge Algebras, Fundamental Hypermultiplets}

We treat first the case of $\G = SU(N)$ gauge theory (for which
$n=N-1$), and $N_f$ hypermultiplets in the fundamental representation of
$SU(N)$. We shall limit ourselves to asymptotically free theories, for which
the number of fundamental hypermultiplets is constrained by $N_f < 2 N_c$. The
curve proposed in \cite{su(n)} is given by
\be\label{su(n)}
\Gamma (u_i) = \{ (k,y) \quad y^2 = A(k)^2 - \bar \Lambda ^2 B(k) \}
\qquad \quad 
\bar \Lambda \equiv  \Lambda ^{N-\12 N_f}
\ee
where the functions $A$ and $B$ are given by
\be
A(k) = \prod _{i=1} ^N (k-u_i)
\qquad \qquad
B(k) = \prod _{\alpha =1} ^{N_f}(k+m_\alpha)\, ,
\ee
and $m_\alpha$ are the hypermultiplet masses. The Seiberg-Witten
differential is defined to be
\be\label{su(n)diff}
d\lambda
=k\,d\ln\,(y+A(k))
\ee
We observe that 
even when $N_f=0$ and $N=2$, this curve and differential
are different from the 
curve found originally for $SU(2)$ in \S 3.8. Indeed, the
Seiberg-Witten curves associated with a given gauge theory
are not unique, although they may satisfy the same criteria
outlined above. We shall check below that these criteria
are satisfied by the curve (\ref{su(n)}) for the $SU(N)$ gauge theory
with $N_f$ hypermultiplets in the fundamental representation.  

\medskip

When $\Lambda =0$, the curve
degenerates to two copies of the complex plane given by $y = \pm A(k)$, which
have a double intersection precisely (and only) at the zeroes $u_i$ of $A(k)$.
As $\Lambda \not=0$, the double zeros open up into branch cuts. The end
points $x_k^{\pm}$ of these branch cuts are defined by
\be\label{branchcut}
A(x_k^{\pm})^2-B(x_k^{\pm})=0.
\ee
There
will be precisely $N$ branch cuts, so that the genus of the curve is 
$n=N-1$. Thus, the set
of cycles $A_i$ and $B_i$ may be chosen to be a basis of
 $H_1(\Gamma, {\bf Z})$.
The Seiberg-Witten differential may be written as 
\be
d\lambda = {k \ dk \over y} \biggl ( A'(k) - \12 A(k) {B'(k) \over B(k)} \biggr )
\, .
\ee
since this expression differs from the one in (\ref{su(n)diff})
by the term $\bar\Lambda^2\,kB'(k)dk/B(k)$,
which has zero periods.
It is easy to see that the residues of $d\lambda$ are independent of $u_i$,
since the only poles (at finite $k$) appear at $k=-m_\alpha$ through the term
inverse in $B(k)$. In fact, around $k = -m_\alpha$, we have $y \sim A(k)$, and
the behavior of $d\lambda$ is given by
\be
d\lambda \sim  \12 {m_\alpha \ dk \over k+m_\alpha}\, .
\ee 
Similarly, the pole at $k=\infty$ has a residue independent of $u_i$ and thus
$\partial d\lambda / \partial u_i$ is holomorphic.

\medskip

There are a number of simple checks on the validity of this curve. First, by
making a hypermultiplet mass, say $m_{N_f}$ large, while keeping $m_{N_f} \bar
\Lambda ^2$ fixed, we decouple precisely the hypermultiplet with mass
$m_{N_f}$. The limit of the corresponding curve agrees with the curve with one
less hypermultiplet. Second, by letting one of the vacuum expectation values
become large, one obtains a limit in which the gauge group $SU(N) \to
SU(N-1)$. Again the corresponding limits of the curves agree. Lastly, it was
argued based upon the Picard-Lefschetz construction, that the strong coupling
monodromies agree with the values expected from the dyon quantum numbers in
the theory.

\medskip

However, the true check which we need to carry out is that the
family of Riemann surfaces reproduce the correct logarithmic singularities
found in the perturbative regime for the prepotential $\cal F^{\rm pert}$
(c.f. (\ref{logsing})).
This is the regime where $\Lambda$
is small compared to the gauge scalar vacuum expectation values $a_i$ and the
hypermultiplet masses $m_\alpha$. 

\medskip

The first step is to derive the quantum order parameters $a_k={1\over 2\pi i}
\oint_{A_k}d\lambda$ in terms
of their classical limits $u_k$. Now the cycles $A_k$, $2\leq k\leq N$,
can be chosen to be contours surrounding the branch cuts between $x_k^-$ 
and $x_k^+$, which remain at a fixed distance from $u_k$
while the branch cuts themselves shrink to $u_k$ as $\bar\Lambda\to 0$.
This means that for $\Lambda$ small compared to $u_k$, we may
expand
\be\label{expansion}
d\lambda
=k({A'\over A}-{1\over 2}{B'\over B})
\sum _{m=0} ^\infty {\Gamma (m+\12) \over \Gamma (\12) m!} 
\ k  \biggl ({B\over A^2} \biggr ) ^mdk
\ee  
This reduces $d\lambda$ to a rational differential
on a sphere with punctures. The method of residues applies
readily to the evaluation of the periods around
$A_k$, and we find
\be\label{akuk}
a_i = u_i + \sum _{m=1} ^\infty {{\bar \Lambda }^{2m} \over 2^{2m} (m!)^2}
\biggl ( {\partial \over \partial u_i} \biggr )^{2m-1} S_i (u_i; u)^m\, .
\ee
The function $S_i(k;u)$ is defined by
\be \label{ess}
S_i(k;u) = \bar \Lambda ^{-2}(k-u_i)^2 {B(k) \over A(k)^2}
= {\prod _\alpha (k+m_\alpha) \over \prod _{l\not=i} (k-u_l)^2}
\, .
\ee

The evaluation of the dual variables $a_{Dk}={1\over 2\pi i}\oint_{B_k}
d\lambda$ is more difficult. This is because the cycles $B_k$
correspond to paths going from $x_1^-$ to $x_k^-$ on each sheet,
unlike the $A_k$ cycles which can be chosen at a fixed distance
from the cut $x_k^-,x_k^+$ cut as $\bar\Lambda\to 0$.
The expansion (\ref{expansion}) cannot be applied to
the evaluation of the $B_k$ periods as it stands.
However, we can restore its validity by introducing a parameter 
$\xi$ with $|\xi|$ small, and deform $d\lambda$ to
\bea \label{deflambda}
d\lambda (\xi) & = & 
k \biggl ( {A' \over A} - \12 {B' \over B} \biggr ) 
\biggl ( 1 - \xi^2{B \over A^2} \biggr ) ^{-\12} dk \nonumber \\
& = & 
\sum _{m=0} ^\infty {\Gamma (m+\12) \over \Gamma (\12) m!} 
\ k  \biggl ( {A' \over A} - \12 {B' \over B} \biggr ) \biggl (
{\xi^2 B\over A^2} \biggr ) ^mdk
\eea

The main advantage is that for fixed small $\xi$, the expansion
is reliable and the evaluation of the periods of $d\lambda (\xi)$
reduce to integrals of rational differentials, just as before
in the evaluation of the $A_k$-periods
(in practice, for each order
$\xi^{2m}$, it suffices to retain terms with poles of order up to $2m$).
The periods $a_{Dk}$ of $d\lambda$ are then obtained by
analytic continuation to $\xi=1$.

It is now easy to identify the logarithmic
terms in the resulting prepotential. Clearly, we have
\bea
d\lambda
=
{1\over 2}\sum_{\alpha =1}^{N_f} {m_\alpha \ dk\over k+m_\alpha }
+\sum_{l=1}^{N}a_l{dk\over k-u_l}+ \O(\Lambda)
\ea
since the residues of $d\lambda$ at $k+m_\alpha =0$ are known to be 
${1\over 2}m_\alpha$,
while the periods $a_i$ of $d\lambda$ around the $A_i$ cycles
tend to the residues of $d\lambda$ at $u_i$ as $\Lambda\to 0$.
The $B_i$ periods of $d\lambda$ are
\bea
\int_{x_1^-}^{x_i^-}
d\lambda
=
{1\over 2}\sum_{\alpha =1}^{N_f}m_\alpha \ln (x_i^-+m_\alpha )
+
\sum_{l=1}^{N}
a_l\ln (x_i^--u_l)+ \O(\Lambda)\, ,
\eea 
(where we have omitted, here as well as below, a similar expression with 
$x_k^-$
replaced by $x_1^-$, arising from the lower bound of
the integral).
This can be rewritten as
\be
\int_{x_1^-}^{x_i^-}
d\lambda
=
a_i\ln\bar\Lambda
+{1\over 2}\sum_{\alpha =1}^{N_f}(a_i+m_\alpha )\ln (x_i^-+m_\alpha)
-
\sum_{l=1}^{N}
(a_i-a_l)\ln (x_i^--u_l)+ \O(\Lambda)
\ee
since the branch points $x_i^-$ satisfy
\bea
\prod_{l=1}^{N}(x_i^--u_l)^2
=\bar\Lambda^2
\prod_{j=1}^{N_f}
(x_i^-+m_j)
\eea
Since $u_i$, $x_i^-$ are analytic functions of $a_j$ which
differ from $a_i$ by $\O(\bar\Lambda)$ only, it follows that
\bea\label{ad}
\int_{k_1^-}^{k_i^-}
d\lambda
=
a_i\ln\bar\Lambda
+{1\over 2}\sum_{\alpha =1}^{N_f}(a_i+m_\alpha )\ln (a_i +m_\alpha)
-
\sum_{l=1}^{N}
(a_i-a_l)\ln (a_i- a_l)+O(\Lambda)
\eea
where the terms $\O(\bar\Lambda)$ are analytic functions of $\bar\Lambda$.
Thus there are no higher powers of $\ln\bar\Lambda$ in the
prepotential, in agreement with the non-renormalization
theorem for $\N=2$ supersymmetric gauge theories.
In view of the constraint $\sum_{k=1}^Na_k=0$,
we may consider a prepotential ${\cal F}(a_1,\cdots,a_N)$
which is a function of all variables $a_k$, whose restriction
to the hyperplane $\sum_{k=1}^Na_k=0$ is the usual prepotential.
In terms of this prepotential ${\cal F}(a_1,\cdots,a_N)$,
we have $a_{Dk}={\partial\over\partial a_k}{\cal F}(a_1,\cdots,a_N)
-{\partial\over\partial a_1}{\cal F}(a_1,\cdots,a_N)$.
The expressions found in (\ref{ad}) identify the perturbative
part of the prepotential 
\be
{\cal F}^{\rm pert}
=
-{1\over 8\pi i}
\big(\sum_{l,m=1}^N(a_l-a_m)^2\ln {(a_l-a_m)^2\over\Lambda^2}
- 
\sum_{l=1}^N\sum_{j=1}^{N_f}(a_l+m_j)^2\ln{(a_l+m_j)^2\over\Lambda^2}
\big)
\ee
which is exactly the expression (\ref{logsing})
required for the $SU(N)$ gauge theory with $N_f$ hypermultiplets
of masses $m_j$, $1\leq j\leq N_f$, in the fundamental
representation.

\medskip

The same methods can be used to determine explicitly
the successive instanton corrections to the prepotential, corresponding
to the successive terms $\O(\bar\Lambda^d)$ in (\ref{deflambda}).
Clearly, the only remaining ingredient we need is a full
expansion of the branch points $x_k^-$ in terms of $u_k$ and $\Lambda$.
For this, we recast the defining equation (\ref{branchcut})
for the branch points as a fixed point equation
\be\label{fixedpoint}
x_k^{\pm}=u_k\pm\bar\Lambda S(x_k^{\pm})^{1\over 2}
\ee
The exact solution of this fixed point equation is given by
\be
x_k^{\pm}
=u_k+
\sum_{m=1}^{\infty}{(\pm)^m\bar\Lambda^m\over m!}
({\partial\over\partial u_k})^{m-1}S_k(u_k)^{m\over 2}
\ee

It is now clear that we can evaluate the periods of the Seiberg-Witten
differential to any order in $\bar\Lambda$. The answers will be
in terms of the {\it classical} moduli parameters $u_k$.
To get the correct prepotential, as we already saw in the
determination of ${\cal F}^{\rm pert}$, it is crucial
that all expressions be recast in terms of the {\it quantum}
moduli $a_k$. This can be done
routinely, since the equation (\ref{akuk})
allows to write $a_k$ and $u_k$ in terms of each other. 
Although
the calculations can become very cumbersome beyond one instanton order,
the methods are straightforward and have a certain flexibility.
Originally developed in the context of hyperelliptic curves
\cite{dkp1}\cite{dkp2}, they have been now extended successfully 
by I.P. Ennes, S.G. Naculich, H. Rhedin, and H.J. Schnitzer to 
the study of more complicated spectral curves than (\ref{su(n)}). These
include the ones corresponding to matter in the symmetric
and the antisymmetric representation of $SU(N)$
\cite{schnitzer}, as well as product groups \cite{feichtinger},
and other curves arising from M Theory.
Another method which has been used in the
determination of the prepotential from Seiberg-Witten curves
is the method of Picard-Fuchs equations. These methods
give exact differential equations satisfied by the periods of
the Seiberg-Witten differential, but they can get
complicated very quickly as the genus of the curve increases
\cite{picard}.

\medskip

However, the prepotential obeys a simple and remarkable
renormalization group type equation \cite{[25]} 
\be\label{rg}
{\partial \F \over \partial \Lambda} \bigg |_{a_i} = {2N- N_f \over 2\pi i}
\sum _{i=1} ^N u_i^2 \ ,
\ee
which can actually be used
to determine explicitly all instanton corrections,
once the dependence of $u_i$ is known in terms of the quantum 
moduli $a_i$ (Simpler cases of (\ref{rg}) had been found in
\cite{matone}. A broad framework for such equations had also
been provided in \cite{krichever1}, in the context of
the tau function for soliton equations).

\medskip

We shall limit ourselves here to exhibiting the expression 
for the prepotential
up to 2-instanton order, and to the cases where $N_f < 2N$\be
\F = \F ^{(0)} + \F ^{(1)} + \F ^{(2)} + {\cal O}(\bar \Lambda ^6)\, .
\ee
Here, $\F^{(0)}$ is the classical and perturbative contribution
\bea
2\pi i \F^{(0)} & = &
-{1 \over 4} \sum _{j\not=i} (a_i-a_j)^2 \ln {(a_i - a_j)^2 \over \Lambda ^2}
-(\ln 2 -2N+N_f) \sum _i a_i^2
\nonumber \\
&& +{1 \over 4} \sum _{i, \alpha} (a_i +m_\alpha)^2 \ln {(a_i + m_\alpha)^2
\over \Lambda ^2}\, ,
\eea
$\F^{(1)}$ is the one instanton contribution, given by
\be
2 \pi i \F^{(1)} 
=
{\bar \Lambda ^2 \over 4}  \sum _i S_i(a_i;a)\, ,
\ee
and $\F^{(2)}$ is the two instanton contribution, given by
\be
2 \pi i \F^{(2)}
=
{\bar \Lambda ^4 \over 16} \biggl [
\sum _{j \not=i} {S_i(a_i;a) S_j (a_j;a) \over (a_i-a_j)^2}
+{1 \over 4} \sum _i S_i(a_i;a) {\partial ^2 S_i (a_i;a) \over \partial a_i^2}
\biggr ]\, .
\ee 
Here, the function $S_i(k;a)$ is as defined in (\ref{ess}), but with $u_i$
replaced by $a_i$.   There is perfect agreement 
between the quantum field theory
calculations and these results derived directly from Seiberg-Witten theory
\cite{dkp1}.

\medskip

We note also that a
set of linear recursion relations was very recently
obtained for the prepotential, from which
instanton corrections to very high order may be read off almost 
without calculations \cite{chandh}.
In a related direction, the renormalization group equation
can be imbedded into a full hierarchy of equations satisfied
by the prepotential \cite{krichever1}.
These equations can also
be exploited to derive instanton corrections \cite{edelstein}.
For $SU(N)$ theories with hypermultiplets in the fundamental 
representation,
the prepotential has also been shown to satisfy WDVV equations
\cite{wdvv}. Clearly, all these properties are indicative
of a deeper underlying structure which is still yet to
be fully uncovered. 
Finally, we note that the renormalization group methods may also be used 
to perform systematic expansions at strong coupling, for small values of the
quantum moduli parameters \cite{dp4}. These results have found
exciting applications to twisted topological supersymmetric gauge
theories as well \cite{mm}. Various reviews may be found in \cite{[12]}.

\subsection{Classical Gauge Algebras, Fundamental Hypermultiplets}

The curves $\Gamma (u_i)$ for the other classical gauge groups $SO(N)$ and
$Sp(N)$ have also been constructed \cite{argyres}, and we have
\be
A(k) = k^a \prod _{i=1} ^n (k^2 - u_i^2)
\qquad \quad
B(k) = \Lambda ^c k^b \prod _{\alpha =1} ^{N_f} (k^2 - m_\alpha ^2)\, .
\ee
The values of the exponents $a$, $b$ and $c$ are given in table 6. 

\begin{table}[b]
\begin{center}
\begin{tabular}{|c||c|c|c|} \hline 
$\G$   & \ $a$ \ & \ $b$ \ & \ $c$ \  \\
\hline \hline
$SU(n+1)$  & $0$ & $0$ & \ $2n+2 -2N_f $ \ \\ \hline 
$SO(2n+1)$ & $0$ & $2$ & $4n-2 -2N_f $ \\ \hline
$Sp(2n)$   & $2$ & $0$ & $4n+4 -2N_f $ \\ \hline
$SO(2n)$   & $0$ & $4$ & $4n-4 -2N_f $ \\ \hline \hline
\end{tabular}
\end{center}
\caption{Exponents entering Seiberg-Witten curves for classical gauge algebras}
\label{table:6}
\end{table}

For simplicity, we have here restricted for $Sp(2n)$ to the case  where we
have at least two massless hypermultiplets. The general case was treated in
\cite{dkp2}.

\medskip

In the same paper \cite{dkp2}, it was shown that the prepotentials 
for the $SO(N)$ and
$Sp(N)$ cases could be simply expressed in terms of the prepotentials for the
$SU(N)$ cases. One finds for the case $\G = SO(2n+1)$,
\be
\matrix{
\F  _{SO(2n+1);N_f} (a_1, \cdots  , a_n; m_1, \cdots ,m_{N_f};\Lambda)
\cr
= 
\F _{SU(2n);2N_f+2} (a_1, \cdots  , a_n,-a_1, \cdots, -a_n; m_1,
\cdots ,m_{N_f},-m_1,\cdots,-m_{N_f}, 0,0;\Lambda)}
\ee
for $Sp(2n)$, with at least two massless hypermultiplets, 
\be
\matrix{
\F  _{Sp(2n);N_f} (a_1, \cdots , a_n; m_1, \cdots
,m_{N_f-2},0,0;\Lambda)
\cr
= 
\F _{SU(2n);2N_f-4} (a_1,  \cdots  , a_n,-a_1,\cdots, -a_n; m_1,
\cdots ,m_{N_f-2},-m_1,\cdots, -m_{N_f-2} ;\Lambda)}
\ee
and, finally, for $SO(2n)$, we have
\be
\matrix{
\F  _{SO(2n);N_f} (a_1, \cdots , a_n; m_1, \cdots ,m_{N_f};\Lambda)
\cr
= 
\F _{SU(2n);2N_f+4} (a_1, \cdots  , a_n,-a_1,\cdots, -a_n; m_1,
\cdots ,m_{N_f},-m_1,\cdots, -m_{N_f},0,0, 0,0;\Lambda)}
\ee
Again, these Seiberg-Witten expressions agree perfectly with the perturbative
results known directly from quantum field theory.

\vfill\eject

\setcounter{equation}{0}
\section{Mechanical Integrable Systems}

We consider Hamiltonian mechanical systems with a finite number of
degrees of freedom, namely positions $x_i$ and momenta $p_i$, $i=1,\cdots ,n$.
The Hamiltonian, Poisson bracket and Hamilton equations are as always
\bea \label{hamilton}
H(x_i,  p_i)  
& \qquad & 
\dot x_i = \{ x_i, H\} = {\partial H \over \partial p_i} 
\nonumber \\
\{ x_i, p_j  \} = \delta _{ij} 
& \qquad  &
\dot p_i =  \{ p_i , H \} = -{\partial H \over \partial x_i}\, .
\eea   
While for the purposes of classical mechanics, these quantities and equations
are customarily used with real valued $x_i$, $p_i$ and $H$, there is
absolutely no difficulty in letting $x_i$, $p_i$ and $H$ be complex. For our
purposes, derived from Seiberg-Witten theory, the natural setting is complex
analyticity and it will be crucial to allow for this complex generalization.

\medskip

A Hamiltonian mechanical system defined in (\ref{hamilton}) is {\it
(completely) integrable} if and only if there exist precisely $n$ functionally
independent {\it integrals of motion} $I_i (x,p)$, $i=1,\cdots ,n$, such that
\be
\dot I_i = \{ I_i, H\}=0 \qquad \qquad \{I_i, I_j\}=0\, .
\ee 
(More precisely, the system is integrable if the first relation holds,
while it is completely integrable when both equations hold.) For a
practical guide to integrable systems, see for example \cite{toda}.
As a result of
the existence of precisely as many integrals of motion as there are position
degrees of freedom $x_i$, it is possible to perform a canonical transformation
from the variables $x_i$, $p_i$ to {\it action-angle variables}, $I_i$,
$\psi_i$, $i=1,\cdots ,n$, for which time evolution is linear
\be
I_i(t) = I_i(0)
\qquad \qquad 
\psi _i(t) = \psi _i(0) + c_i I_i t
\ee
for some constants $c_i$. In other words, for a completely integrable system,
in action-angle variables, time evolution reduces to a linear flow on an
$n$-dimensional torus, parametrized by the angle variables $\psi _i$. 

\medskip

There is no general classification of Hamiltonians $H(x_i, p_i)$ which
correspond to completely integrable systems. There is also no systematic test
to determine whether a given Hamiltonian $H(x_i,p_i)$ is integrable or not.
Thus, deciding whether a given system is integrable is a difficult and
challenging problem in mechanics. 

\medskip

However, there is a general correspondence that has been 
used very fruitfully in
the study of integrable systems. This is the case whenever there exists a
{\it Lax pair}. A Lax pair is a pair of $N\times N$ matrix-valued functions
$L(x,p)$ and $M(x,p)$ which depend upon the dynamical variables $x$ and $p$,
such that the Lax equation $\dot L = [L,M]$ is equivalent to the Hamilton
equations of the Hamiltonian mechanical system associated with $H$,
\be
\dot L = [L,M] \quad \Longleftrightarrow \quad
\left ( \matrix{\dot x_i = \{ x_i ,H\} \cr \dot p_i = \{ p_i, H\} \cr}\right
)\, .
\ee
The dimension $N$ of the Lax matrices is not a priori known, and there is
again no systematic test or algorithm to establish the existence of a Lax pair,
even less to find it explicitly. 
A Lax pair is not unique, since one is always
free to perform a gauge transformation on $L$ and $M$ by
\be
L^S = S^{-1} L S
\qquad \qquad
M^S = S^{-1} M S - S^{-1} \dot S\, .
\ee

As soon as a Lax pair is known to exist for a given Hamiltonian system, 
integrals of motion immediately follow. Indeed, traces of powers of $L$ are
found to be time independent as well as gauge invariant,
\be
I_i \equiv \tr L^{n_i}
\qquad \Longrightarrow \qquad
\dot I_i = n_i \ \tr\bigl ( L^{n_i-1} [L,M] \bigr )=0\, .
\ee
Here, the traces of $L$ can be defined for all positive integer values $n_i$,
and all such $I_i$ are integrals of motion. However, since the matrix $L$ is of
dimension $N$, we know from Cayley's theorem that all $I_i$ with $n_i \geq N$
will be functionally dependent upon traces of lower powers. Thus, to get
``enough" integrals of motion for complete integrability, we need $n\leq N$.
For all known integrable systems, this condition appears to be fullfilled,
though we are not aware of any general theorem to that effect. Amongst all
traces of powers of $L$, we choose $n$ functionally independent 
$I_i$, $i=1,\cdots, n$ for
certain choises of integers $n_i$. In particular, 
the original Hamiltonian will
be functionally dependent upon these integrals of motion
\be
H= h(I_1, \cdots ,I_n)
\ee
since it is time-independent itself.

\subsection{Lax Pairs with Spectral Parameter - Spectral Curves}

A stronger form of integrability
which exists in many known integrable systems, 
is the fact that the Lax operators $L$ and $M$ may depend
upon an additional (complex) {\it spectral parameter}, 
usually denoted by $z$. The Lax operators become
then matrix-valued functions $L(z)$, $M(z)$ of $z$.
The Hamiltonian $H$ and the whole mechanical system does not 
depend upon $z$,
yet for all values of $z$, $L(z)$ and $M(z)$ are a Lax pair,   
\be
\dot L(z) = [L(z),M(z)] \quad \Longleftrightarrow \quad
\left ( \matrix{\dot x_i = \{ x_i ,H\} \cr \dot p_i = \{ p_i, H\} \cr}\right
)\, .
\ee
A Lax pair $L(z)$, $M(z)$ obeying these relations is called a {\it Lax pair
with spectral parameter}. Again, there is no systematic way to establish the
existence of a Lax pair with spectral parameter, and there is no algorithm for
finding it, even if the system is known to be integrable or even if a Lax pair
without spectral parameter is already known explicitly.
  
\medskip

To any Lax pair with spectral parameter, there is naturally associated a {\it
spectral curve}, defined by
\be \label{curve}
\Gamma = \big \{ (k,z) \in {\bf C} \times {\bf C};
\ \det \big ( kI - L(z) \big ) =0 \big \}\, .
\ee
and a natural one-form $d\lambda$, defined by
\be\label{kdz}
d\lambda=kdz
\ee
While $L(z)$ and $M(z)$ depend upon $2r$ phase space variables $x_i$ and $p_i$,
$i=1,\cdots ,n$, the spectral curve is time-independent and gauge invariant and
depends only upon the $n$ functionally independent integrals of motion $I_i$.
Furthermore, if $L(z)$ is a meromorphic function of $z$ (up to a gauge
transformation, which may or may not be meromorphic), then the form $d\lambda$
is a meromorphic form on the spectral curve
$\Gamma$. Depending on the periodicity properties of the
Lax pair $L(z)$, $M(z)$ with respect to $z$,
the domain of definition of $z$ may be more properly viewed
as the sphere of the torus, and a different variable than $z$
may be more suitable in the definition \ref{kdz}
for the Seiberg-Witten differential
(examples of this are discussed in subsequent sections).
The curve $\Gamma$ is correspondingly a branched covering of 
the sphere or of the torus.

\subsection{The Toda Systems}

The {\it non-periodic Toda system} is a non-relativistic system of $n+1$ points
on a linear chain, with exponential nearest neighbor interactions 
 (see for example 
\cite{toda}). The Hamiltonian is given by
\be
H= \12 \sum _{i=1} ^{n+1} p_i ^2 - M^2 \sum _{i=1} ^n e^{x_{i+1} - x_i}\, .
\ee
By translation invariance of the $x_i$ by a common constant, it is clear that
the center of mass $x_0 = \sum _i x_i$ degree of freedom may be decoupled from
the system, effectively leaving a system of $r$ coupled degrees of freedom. 
The simplest non-periodic system with $n=1$ reduces to the Liouville system.

\medskip

The {\it periodic Toda system} is a non-relativistic  system of $n+1$ points on
a circular chain, with exponential nearest neighbor interactions, given by
\be
H= \12 \sum _{i=1} ^{n+1} p_i ^2 - M^2 \sum _{i=1} ^{n+1} e^{x_{i+1} - x_i}\, ,
\ee
with the periodicity condition $x_{n+2}=x_1$. Again, the center of mass
coordinate decouples leaving a coupled system of $r$ degrees of freedom. The
simplest periodic Toda system is for $n=1$ and reduces to the Sine-Gordon
model.  

\medskip

Actually, both Toda systems admit a natural Lie algebraic interpretation. This
may be seen by viewing the linear chain of the non-periodic system as the
Dynkin diagram for the Lie algebra $A_n \sim SU(n+1)$, and the circular chain
of the periodic Toda system as the Dynkin diagram for the untwisted affine
Lie algebra $A_n ^{(1)} $. The set $\R_*$ of {\it simple roots} for each Lie
algebra is well-known,
\bea
A_n & \qquad & e_i - e_{i+1} \qquad i=1,\cdots ,n
\nonumber \\
A_n ^{(1)}  & \qquad & e_i - e_{i+1} \qquad
i=1,\cdots ,n+1, \qquad e_{r+2} =  e_1\, .
\eea
In terms of the simple roots, and the vector notation for position and momentum
variables, $x = (x_1, \cdots , x_{n+1})$ and $p = (p_1, \cdots ,p_{n+1})$, both
Toda Hamiltonians take on a unified form,
\be
H = \12 p^2 - M^2 \sum _{\alpha \in \R_*} e^{- \alpha \cdot x}\, .
\ee
Lie algebraically, the non-periodic Toda system is associated with the finite
dimensional Lie algebra $A_n$, while the periodic Toda system is associated
with the infinite dimensional untwisted affine Lie algebra $A_n ^{(1)}$.

\medskip

Now, for any finite-dimensional semi-simple Lie algebra $\G$ or one of the
associated untwisted $\G^{(1)}$ or twisted affine Lie algebras, we may
associate a Toda system based on the set of simple roots $\R_*$ of the Lie
algebra. For a Lie algebra $\G$ of rank $n$, we introduce the position
$x = (x_1, \cdots ,x_n)$ and momentum $p=(p_1, \cdots ,p_n)$
variables and define the Toda Hamiltonian associated with a Lie algebra $\G$
(which may be finite dimensional or affine) in terms of the set of simple
roots $\R_* (\G)$, as follows,
\be
H = \12 p^2 - M^2 \sum _{\alpha \in \R_*(\G)}
e^{-\alpha \cdot x}\, .
\ee
Remarkably, this Toda system is completely integrable for any Lie algebra
$\G$.

\medskip

In fact, for any finite dimensional Lie algebra $\G$, or its untwisted affine
extension $\G ^{(1)}$, the Toda system admits a Lax pair \cite{laxtoda}. To
exhibit this Lax pair, we need the following ingredients,

\begin{itemize}

\item $h = (h_1, \cdots ,h_n)$ is the array of Cartan generators of $\G$, with
$[h_i, h_j]=0$;

\item $E_\alpha$ is the generators of $\G$ associated with the root
$\alpha$;

\item $\R _* = \R_*(\G)$ is the set of simple roots of $\G$;

\item $\alpha _0$ is the affine root (to be included for $\G^{(1)}$);

\item a representation $\rho$ of $\G$, so that
all generators $h_i$, $E_{\alpha}$ can be viewed as $N$-dimensional
matrices, if $\rho$ is of dimension $N$.

\end{itemize}

\noindent
The Lax matrices then take on the form
\bea \label{todalax}
L & = &
p \cdot h + \sum _{\alpha \in \R_*} M e^{-\12 \alpha \cdot x} (E_\alpha  -
E_{-\alpha}) + \mu ^2 e^{\12 \alpha _0  \cdot x}  (z E_{-\alpha _0} - z^{-1}
E_{\alpha _0} )
\\
M & = &
-\12 \sum _{\alpha \in \R_*} M e^{-\12 \alpha \cdot x} (E_\alpha  +
E_{-\alpha}) + {\mu ^2 \over 2} e^{\12 \alpha _0  \cdot x}  (z E_{-\alpha _0} +
z^{-1} E_{\alpha _0})
\eea 

For a non-periodic Toda system, associated with a finite dimensional Lie
algebra $\G$, we set $\mu =0$ in (\ref{todalax}), and any dependence on the
spectral parameter $z$ disappears. For the periodic Toda system, associated
with the untwisted affine Lie algebra $\G ^{(1)}$, 
we have $\mu \not=0$, and the
Lax pair has a spectral parameter $z$.

\medskip

For non-simply laced Lie algebras $\G$ (i.e. for which not all roots have the
same lengths), one may also define a Toda system associated with {\it the
twisted affine Lie algebra} $(\G ^{(1)})^\dagger$. 
The roots of this algebra are
the duals (or co-roots) to the roots of $\G^{(1)}$, and the twisted affine Toda
system associated with these algebras is also integrable, and admit a Lax pair
with operators given by (\ref{todalax}) but with roots replaced with co-roots.

\medskip

For the periodic Toda systems, we have a Lax pair with spectral parameter and
thus we have a spectral curve, as defined in (\ref{curve}). For the case of
$A_n^{(1)}$ with $\rho$ the fundamental representation,
this curve may be worked out more explicitly. Indeed, the
determinant that enters (\ref{curve}) is 
\be \label{su}
\det \bigl ( kI - L(z) \bigr ) = A(k) - \12 (z + {\bar \Lambda ^2 \over z})\, .
\ee
The simplicity of the $z$-dependence is due to the fact that the generators
$E_{\alpha _0}$ and $E_{-\alpha _0}$ are matrices of rank 1, so that their
coefficients enter linearly in the expansion of the determinant. The
remaining polynomial $A(k)$ is independent of $z$ and is of degree $N=n+1$.
Since $\tr L=0$, and $k$ multiplies the identity matrix, 
the form of $A(k)$ is
\be \label{a}
A(k) = k^N + k^{N-2} J_2 + k^{N-3} J_3 
+ \cdots + J_N = \prod _{i=1} ^N (k-u_i)
\ee
where the parameters $J_i$ are functions of the integrals of motion $I_i = \tr
L^i$ only. The integrals of motion can 
in principle be expressed in terms of the
dynamical variables $x_i$ and $p_i$, 
but we shall not need these relations here.

\medskip

By a change of variables, $y= z-A(k)$, the equation for the curve
(\ref{curve}) may be evaluated explicitly using (\ref{a}), and we get
\be
\Gamma = \{ (k,y) \in {\bf C} \times {\bf C}, \qquad y^2 = A(k)^2 -\bar
\Lambda ^2\}\, .
\ee 
Remarkably, this curve is identical to the Seiberg-Witten curve for an $\N=2$
supersymmetric $SU(N)$ gauge theory without hypermultiplets. In fact, also the
Seiberg-Witten differential admits a very simple expression, given by
\be
d\lambda = k d\ln\,z= {A'(k)k dk \over \sqrt{A(k)^2 - \bar \Lambda ^2}}\, .
\ee  
The coincidence of the spectral curves of Toda systems and the Seiberg-Witten
curves of $\N=2$ supersymmetric Yang-Mills theory was first noted in
\cite{[2]}, in the case of $SU(2)$ without matter hypermultiplets.
It is natural to speculate that, for all gauge groups $\G$, a 
similar correspondence should hold between Toda systems
and Seiberg-Witten curves of $\N=2$ supersymmetric Yang-Mills
theories without matter hypermultiplets, for all gauge groups.
However, due to subtleties related to the renormalization group flow,
the correct correspondence is rather between Toda systems for the
dual group $(\G^{(1)})^{\vee}$ and the $\G$ Yang-Mills theory,
as shown in \cite{martinec}. We shall come back in \S 6
to this emergence of the dual group $(\G^{(1)})^{\vee}$, in the context
of Calogero-Moser systems.

\subsection{The Calogero-Moser Systems for SU(N)}

The Calogero-Moser systems are non-relativistic mechanical models of $n+1$
particles on a (complex line) with two body interactions, which are not
limited to nearest neighbor interactions. The original Calogero-Moser system
comes in three varieties
\cite{calogero}

\begin{itemize}

\item {\it The Rational Calogero-Moser System} 
\be
H= \12 \sum _{i=1} ^{n+1} p_i^2 - \12 m^2 \sum _{i\not= j} ^{n+1} 
{ 1\over (x_i
- x_j)^2}
\ee

\item {\it The Trigonometric Calogero-Moser System}
\be
H= \12 \sum _{i=1} ^{n+1} p_i^2 - \12 m^2 \sum _{i\not= j}^{n+1} 
{ 1\over \sin
^2(x_i - x_j)}
\ee

\item {\it The Elliptic Calogero-Moser System}
\be
H= \12 \sum _{i=1} ^{n+1} p_i^2 - \12 m^2 \sum _{i\not= j} ^{n+1} \wp (x_i -
x_j; \omega _1 , \omega_2)
\ee

\end{itemize}

As is clear from the above definitions, the interaction potential of the
rational and trigonometric systems is a rational and trigonometric (or simply
periodic) function respectively, while for the elliptic systems, the
potential is the Weierstrass elliptic (doubly periodic) function defined by
\be
\wp (x;\omega_1, \omega _2) \equiv
{1 \over x^2} 
+\sum _{(m,n) \not= (0,0)} \biggl (
{1\over (x+2m \omega_1 + 2n \omega _2)^2} -  {1\over (2m \omega_1 + 2n \omega
_2)^2} \biggr ) \, .
\ee  
Here, $2\omega _1$ and $2\omega _2$ are the periods, and $\wp$ may be thought
of as the electric field due to an electric dipole of unit strength on a torus
with sides $2 \omega_1$ and $2 \omega_2$. The three systems are related by 
the following limits : as $\omega _2 \to \infty$ with $\omega _1$ fixed, the
elliptic system tends to the trigonometric one, while as in addition $\omega _1
\to \infty$, the system tends to the rational case. In view of these limiting
relations, it suffices to carry out our analysis of integrability and
construction of Lax pairs (with spectral parameter) for the elliptic
Calogero-Moser systems. Integrability is preserved under these limits and will
smoothly carry over to the trigonometric and rational cases. 

\medskip

It turns out also that the elliptic Calogero-Moser system is the system of 
interest for Seiberg-Witten theory. An important clue is
the microscopic gauge coupling $\tau$ of (\ref{tau}). In theories where this
microscopic gauge coupling does not get renormalized, it can be expected to
play a central role in the prepotential and hence in the Seiberg-Witten curves.
Thus for theories such as super Yang-Mills with matter in the
adjoint representation (so that the beta function vanishes, in view
of (\ref{beta})), the correct corresponding integrable model
should involve a torus with moduli $\tau$. This requirement is satisfied
by the Calogero-Moser systems in their elliptic versions, but not
in their rational or trigonometric versions. The latter versions
have appeared however in many other physical contexts.
We refer to \cite{polychronakos} for some recent applications.

\medskip

The Calogero-Moser systems are naturally interpreted in terms of $A_n =
SU(n+1)$ Lie algebra data. Consider the set $\R$ of {\it all roots of} $A_n$;
in an orthonormal basis $e_i$, $i=1,\cdots ,n+1$, they are given by
\be
\R(A_n) = \{ e_i - e_j, \qquad i\not= j = 1,\cdots , n+1 \}\, .
\ee  
In terms of the set of all roots of $A_n$ and the position $x=(x_1, \cdots ,
x_{n+1})$ and momentum $p=(p_1,\cdots , p_{n+1})$ variables, we can recast the
Hamiltonian as follows
\be
H = \12 p^2 - \12 m^2 \sum _{\alpha \in \R} \wp (\alpha \cdot x; \omega_1,
\omega _2)
\ee
It is easy to generalize the above Hamiltonian to other Lie algebras, but we
shall postpone this discussion to \S 6, since a great number of subtleties
arise in the integrability properties of these systems.

\medskip

The elliptic Calogero-Moser system for $\G= A_n = SU(n+1)$ admits a Lax pair
{\it with spectral parameter} $z$, first derived in \cite{krichever2},
\bea
L_{ij}(z) & = & p_i \delta _{ij} - m (1-\delta _{ij}) \Phi (x_i-x_j;z) 
\\
M_{ij}(z) & = & d_i(x) \delta _{ij}
 + m (1-\delta _{ij}) \Phi ' (x_i-x_j;z)\, . 
\eea
Here $d_i(x)=m\sum_{k\not=i}\wp(x_i-x_k)$,
$\Phi '(x;z) = \partial _x \Phi (x;z)$, and $\Phi(x;z)$ is a Lam{\'e}
function, given by
\be
\Phi(x;z) = {\sigma(z-x) \over \sigma (x) \sigma (z)} e^{x\zeta(z)}
\ee
where the Weierstrass function $\zeta$ is related to $\wp$ by 
$\wp(z) = - \zeta
'(z)$ and $\zeta(z) = \sigma '(z) /\sigma (z)$. 
The Lam{\'e} function $\Phi(x;z)$
satisfies the Lam{\'e} equation,
\be
\biggl ( {d^2 \over dz^2} - \wp(x) \biggr ) \Phi (x;z)
 = 2\wp (z) \Phi (x;z)\, .
\ee 
This Lax pair and the corresponding spectral curve has sparked a rich
theory of elliptic solitons and spectral covers. We refer to
\cite{verdier} \cite{harnad} for some of these developments.
Complete integrability and the existence of a classical $R$-matrix
has been investigated in \cite{[18]} and \cite{[20]}.
For us, the above Lax pair will be of fundamental importance for a
different reason:
the associated spectral curve and differential
$d\lambda=kdz$ will turn out to give exactly the Seiberg-Witten curve
for the $SU(N)$ $\N=2$ super-Yang-Mills theory, with a hypermultiplet
in the adjoint representation.

\subsection{Relation between Calogero-Moser and Toda for SU(N)}

Remarkably, Toda systems may be obtained as a limit of elliptic Calogero-Moser
systems \cite{inozemtsev}. 
Qualitatively, the limit is expected to arise as
follows : in Calogero-Moser systems, all roots of the algebra are
being summed over, while for Toda systems, only the simple roots enter. Thus,
upon taking the limit, all interactions that are not nearest neighbor must
vanish, leaving only the nearest neighbor interaction of simple roots.

The limit is taken on the parameters $m$, $\omega _2$ and on the position
variables $x_i$ by letting $\omega _1 = -i \pi$ and 
\be \label{limit}
{\rm Re} (\omega_2) \to \infty
\qquad
\cases{m= M e^{\delta \omega _2} \cr x_j = X_j + 2 \omega_2\ \delta\ j \cr
M,\ X_j,\ 0 < \delta \leq {1 \over N} \ {\rm fixed} \cr}
\ee  
A very convenient way to investigate the limit is to expand the $\wp$-function
in terms of trigonometric functions
\be
\wp (x;-i\pi;\omega _2) = \12 \sum _{k=-\infty} ^\infty {1 \over \ch
(x+2k\omega _2) -1}\, .
\ee
(for a derivation of this formula, see Appendix C).
Since $\wp(x;\omega_1, \omega _2)$ is even in $x$, 
we restrict to $i<j$ without
loss of generality and we have
\be \label{sum}
m^2 \wp (x_i - x_j; -i\pi, 2 \omega _2)
=
\12 \sum _{k=-\infty} ^\infty {M^2 e^{2 \delta \omega _2} \over 
\ch \bigl [ X_i - X_j + 2 \omega _2 \bigl (\delta (i-j) +k \bigl ) \bigr ]
-1}\, .
\ee
When $\delta >1/N$, it is easy to see that no limit will exist, 
which is why we
have excluded this case from the outset in  (\ref{limit}). 

Because we have $0<\delta \leq 1/N$ all terms in the infinite sum (\ref{sum}),
except $n=0,1$, converge to 0. The limits of the remaining terms are
\be
k=0 \ \ {\rm term} \quad \longrightarrow \quad
\cases{M^2 \exp (X_j - X_i) & $j-i =1$ \cr
         0                  & $j-i\not= 1$ \cr}
\ee
and 
\be
k=1 \ \ {\rm term} \quad \longrightarrow \quad
\cases{M^2 \exp (X_N - X_1) & $\delta N=1 $ \cr
         0                    & $\delta N <1$ \cr}\, .
\ee
The contribution from the $k=0$ term emerges for any values of $\delta$, 
and it
should be manifest that this limit precisely corresponds to the Toda
exponential interaction between nearest neighbors, i.e. based on the simple
roots of $A_n = SU(n+1)$. The contribution from the $k=1$ term on the other
hand is absent for all values $\delta N<1$. In this case, the limiting Toda
system is based only on the simple roots of $A_n$ and corresponds to the
non-periodic case. When $\delta N=1$ on the other hand, the $k=1$ term survives
and yields the Toda exponential interaction for the affine root $- \alpha _0$.
In this case, the limiting Toda system is based on the simple roots of $A_n$ as
well as the affine root $-\alpha _0$, and corresponds to the periodic Toda
system, associated with the untwisted affine Lie algebra $A_n ^{(1)}$. In
summary
\be
\cases{\delta N <1 & $\longrightarrow \ {\rm non-periodic \ Toda \ for} \  A_n$
\cr
\delta N =1 & $\longrightarrow \ {\rm periodic \ Toda \ for} \ A_n ^{(1)}$ \cr}
\nonumber
\ee

\subsection{Relations with KdV and KP Systems}

The Toda and Calogero-Moser systems are also connected with higher dimensional
integrable systems. Some remarkable connections are: 

\begin{itemize}

\item {\it Toda Field Theory} : The basic fields are $\phi = (\phi _1,
\cdots , \phi _n)$ with $\phi _i(t,x)$  2-dimensional fields, $\partial _\mu =
(\partial _t, \partial _x)$, and Lagrangian
\be
\L = \12 \partial _\mu \phi \cdot \partial ^\mu \phi - M^2 \sum _{\alpha \in \R
_* } e^{\alpha \cdot \phi}\, .
\ee
These theories are integrable as field theories and the simplest cases are
Liouville theory, which enters into the study of non-critical string theory and
two-dimensional quantum gravity \cite{liouville},  as well as Sine-Gordon
theory which is equivalent to the massive Thirring model of fermions
\cite{sinegordon}.

\item {\it Korteweg-de Vries (KdV) equation} : The fundamental field is a
single scalar $u(t,x)$ in 2 dimensions and the field equation is
\be
\partial _t u = 6 u \partial _x u - \partial _x ^3 u
\ee
Of special interest are {\it rational solutions}, which are of the form
\be
u(t,x) = 2 \sum _j {1 \over \bigl ( x-x_j(t) \bigr ) ^2}\, .
\ee
By a theorem of Airault-McKean-Moser \cite{airault}, 
this Ansatz for $u$ solves the KdV
equation provided the positions $x_j(t)$ are subject to certain constraints as
well as to the following time-evolution equation \cite{krichever2}
\be
\partial _t x_j = \{ x_j , \tr L^3\}
\ee
where $L$ is the Lax operator for the rational Calogero-Moser system.

\item {\it Kadomzev-Petviashvili (KP) equation}  : The fundamental field is a
scalar $u(t,t',x)$ in three dimensions (one space and two times) with the field
equation
\be
3 \partial _{t'} ^2 u =
\partial _x \bigl ( \partial _t u - 6 u \partial _x u 
- \partial _x ^3 u \bigr )
\, .
\ee
The $t'$-independent solutions are governed by the KdV equation. It
has been shown by Krichever \cite{krichever2} that
the elliptic
solutions are of the form
\be
u(t,t',x) = 2 \sum _j \wp (x-x_j(t,t'))
\ee
provided the positions $x_j(t,t')$ obey the following time-evolution equations
\be
\partial _t x_j = \{ x_j , \tr L^3\} \qquad \qquad
\partial _{t'} x_j = \{x_j, \tr L^2 \}
\ee
where $L$ is the Lax operator for the elliptic Calogero-Moser system.  

\end{itemize}
So far, these properties have not yet entered directly
into the connection between integrable systems and Seiberg-Witten theory. 
However, it is now known that the symplectic forms arising in
Seiberg-Witten theory from the differential $d\lambda$
can be viewed as restrictions to the space of algebraic-geometric
solutions of symplectic forms for soliton equations
\cite{kp1}\cite{kp2}. It is likely that more connections
will emerge in the near future.
  
\subsection{Calogero-Moser Systems for General Lie Algebras}

As Olshanetsky and Perelomov \cite{olsha} realized very early on,
the Hamiltonian system (3.1) is only one
example of a whole series of Hamiltonian systems associated
with each simple Lie algebra.
More precisely, given any simple Lie algebra $\G$,
Olshanetsky and Perelomov \cite{olsha} introduced the system
with Hamiltonian
\be\label{untwisted}
H(x,p)
={1\over 2}\sum_{i=1}^np_i^2
-{1\over 2}
\sum_{\a\in{\cal R}(\G)}
m_{|\a|}^2\wp(\a\cdot x),
\ee
where $n$ is the rank of $\G$, ${\cal R}(\G)$
denotes the set of roots of $\G$, and the $m_{|\a|}$ are mass parameters.
To preserve the invariance of the Hamiltonian (3.5)
under the Weyl group, the parameters $m_{|\a|}$ depend only
on the orbit $|\a|$ of the root $\a$, and not on the root $\a$ itself.
(For $A_{N-1}= SU(N)$, 
it is common practice as we saw earlier to use $N$ pairs of
dynamical variables $(x_i,p_i)$, since the roots of $A_{N-1}$
lie conveniently on a hyperplane in ${\bf C}^N$.
The dynamics of the system are unaffected if we shift
all $x_i$ by a constant, and the number of degrees of freedom
is effectively $N-1=n$.) 
As in the original $SU(N)$ case, the elliptic systems (3.5)
admit rational and trigonometric limits.
Olshanetsky and Perelomov succeeded in constructing a
Lax pair for all these systems in the case of classical
Lie algebras, albeit without spectral parameter \cite{olsha}.

\bigskip

\noindent
{\bf  Twisted Calogero-Moser Systems defined by Lie Algebras}

\medskip

It turns out that the Hamiltonian systems (\ref{untwisted}) are not the
only natural extensions of the basic elliptic Calogero-Moser system. 
A subtlety arises for simple Lie algebras $\G$ which are not
simply-laced, i.e., algebras which admit roots of uneven
length. This is the case for the algebras $B_n$, $C_n$, $G_2$,
and $F_4$ in Cartan's classification.
For these algebras, the following {\it twisted} elliptic
Calogero-Moser systems were introduced by the authors in \cite{cm1,cm2}
\be\label{twistedh}
H_{\G}^{{\rm twisted}}
=
{1\over 2}\sum_{i=1}^np_i^2
-{1\over 2}
\sum_{\a\in{\cal R}(\G)}
m_{|\a|}^2
\wp_{\nu(\a)}(\a\cdot x).
\ee
Here the function $\nu(\a)$ depends only on the length of the root $\a$.
If $\G$ is simply-laced, we set $\nu(\a)=1$ identically. Otherwise,
for $\G$ non simply-laced, we set $\nu(\a)=1$ when $\a$ is a long root,
$\nu(\a)=2$ when $\a$ is a short root and $\G$ is one of the
algebras $B_n$, $C_n$, or $F_4$, and $\nu(\a)=3$ when $\a$ is a short root
and $\G=G_2$. The {\it twisted} Weierstrass function $\wp_{\nu}(z)$
is defined by
\be
\wp_{\nu}
(z)
=\sum_{\sigma=0}^{\nu-1}
\wp(z+2\omega_a{\sigma\over\nu}),
\ee
where $\omega_a$ is any of the half-periods $\omega_1$, $\omega_2$, or
$\omega_1+\omega_2$. Thus the twisted and untwisted Calogero-Moser systems
coincide for $\G$ simply laced. For $\G=B_n, \ C_n$, the twisted Calogero-Moser
models introduced here are equivalent to the systems introduced by Inozemtsev
\cite{inozemtsev}, while for $\G=F_4, \ G_2$, they are new integrable systems.
The original motivation for twisted Calogero-Moser systems was based on their
scaling limits (which will be discussed in the next section) \cite{cm1,cm2}.

\subsection{Scaling of Calogero-Moser to Toda for General Lie Algebras}

The key feature of the scaling limit of the $SU(N)$
Calogero-Moser system to the $SU(N)$ Toda system 
was the collapse
of the sum over the entire root lattice of $A_{N-1}$
in the Calogero-Moser Hamiltonian to the
sum over only simple roots in the Toda Hamiltonian for the
Kac-Moody algebra $A_{N-1}^{(1)}$.
Our task is to extend this mechanism to general Lie algebras.
It turns out that there are two possible extensions, depending
on whether we use the Coxeter number or the dual Coxeter number
\cite{cm2}.

\bigskip

\noindent
{\bf  Scaling Limits based on the Coxeter Number}

\medskip

For this, we consider the following generalization  of the preceding scaling
limit
\bea\label{coxeterscaling}
m &= & Mq^{-{1\over 2}\d},\\
x &= & X-2\omega_2\d\rho^{\vee},
\eea
Here $x=(x_i)$, $X=(X_i)$ and $\rho^{\vee}$ are $n$-dimensional vectors.
The vector $x$ is the dynamical variable of the Calogero-Moser system.
The parameters $\d$ and $\rho^{\vee}$
depend on the algebra $\G$ and are yet to be chosen.
As for $M$ and $X$, they have the same interpretation as
earlier, namely as respectively the mass parameter
and the dynamical variables of the limiting system.
Setting $\omega_1=-i\pi$, the contribution of each root $\a$ to the
Calogero-Moser potential can be expressed as
\be \label{lim}
m^2\wp(\a\cdot x)
=
{1\over 2}M^2
\sum_{k=-\infty}^{\infty}
{e^{2\d\omega_2}\over
{\rm ch}(\a\cdot x-2k\omega_2)-1}
\ee
It suffices to consider positive roots $\a$.
We shall also assume that $0\leq \d\,\a\cdot\rho^{\vee} \leq 1$. The
contributions of the $k=0$ and $k=-1$ summands in (\ref{lim}) are proportional
to $\exp \{ 2\omega_2(\d-\d\,\a\cdot\rho^{\vee}) \}$
and $\exp \{ 2\omega_2(\d-1+\d\,\a\cdot\rho^{\vee})\}$ respectively.
Thus the existence of a finite scaling limit requires that
\be\label{cox}
\d\,\leq\d\,\a\cdot\rho^{\vee}\leq 1-\d.
\ee
Let $\a_i$, $1\leq i\leq n$ be a basis of simple roots
for $\G$. If we want all simple roots $\a_i$
to survive in the limit, we must require that
\be
\a_i\cdot\rho^{\vee}=1,\ \
1\leq i\leq n.
\ee
This condition characterizes the vector
 $\rho^{\vee}$ as the {\it level vector}.
Next, the second condition in (\ref{cox})
can be rewritten as $\d\{1+\max_{\a}\,(\a\cdot\rho^{\vee})\} \leq 1$. But
\be \label{level}
h_{\G}=1+ \max_{\a}\,(\a\cdot\rho^{\vee})
\ee
is precisely the Coxeter number of $\G$, and we must have 
$\d\leq {1\over h_{\G}}$. Thus when $\d<{1\over h_{\G}}$,
the contributions of all the roots except
for the simple roots of $\G$ tend to $0$.
On the other hand, when $\d={1\over h_{\G}}$,
the highest root $\a_0$ realizing the maximum over
$\a$ in (\ref{level}) survives.
Since $-\a_0$ is the additional simple root for the affine Lie algebra
$\G^{(1)}$, we arrive in this way at the following theorem,
which was proved in \cite{cm2}

\bigskip 

\noindent
{\bf Theorem 1}.
{\it Under the limit (\ref{coxeterscaling}), with $\d={1\over h_{\G}}$,
and $\rho^{\vee}$ given by the level vector,
the Hamiltonian of the elliptic Calogero-Moser system
for the simple Lie algebra $\G$
tends to the Hamiltonian of the Toda system
for the affine Lie algebra $\G^{(1)}$.}

\bigskip

\noindent
{\bf  Scaling Limit based on the Dual Coxeter Number}

\medskip

If the Seiberg-Witten spectral curve of the $\N=2$
supersymmetric gauge theory with a hypermultiplet in
the adjoint representation is to be realized as
the spectral curve for a Calogero-Moser system,
the parameter $m$ in the Calogero-Moser system
should correspond to the mass of the hypermultiplet.
In the gauge theory, the dependence of the coupling
constant on the mass $m$ is given by
\be
\tau={i\over 2\pi}h_{\G}^{\vee}{\rm ln}\,{m^2\over M^2}
\qquad
\Longleftrightarrow
\qquad
m=Mq^{-{1\over 2h_{\G}^{\vee}}}
\ee
where $h_{\G}^{\vee}$ is the quadratic Casimir of the
Lie algebra $\G$. This shows that the correct physical
limit, expressing the decoupling of the hypermultiplet
as it becomes infinitely massive,
is given by (\ref{coxeterscaling}), but with $\d={1\over h_{\G}^{\vee}}$.
To establish a closer parallel with our preceding discussion,
we recall that the quadratic Casimir $h_{\G}^{\vee}$
coincides with the {\it dual Coxeter number} of $\G$,
defined by
\be
h_{\G}^{\vee}=1+\max_{\a}\,(\a^{\vee}\cdot\rho),
\ee
where $\a^{\vee}=2\a / \a^2$ is the coroot associated
to $\a$, and 
\be 
\rho={1\over 2}\sum_{\a>0}\a
\ee
is the well-known Weyl vector.

For simply laced Lie algebras $\G$ (ADE algebras),
we have $h_{\G}=h_{\G}^{\vee}$, and the preceding scaling limits
apply. However, for non simply-laced algebras
($B_n$, $C_n$, $G_2$, $F_4$),  we have $h_{\G}>h_{\G}^{\vee}$,
and our earlier considerations show that the untwisted
elliptic Calogero-Moser Hamiltonians do not tend to
a finite limit under (\ref{coxeterscaling}), $q\to 0$, $M$ is kept fixed.
This is why the twisted Hamiltonian systems (\ref{twistedh})
have to be introduced. The twisting produces precisely
to an improvement in the asymptotic behavior
of the potential which allows a finite, non-trivial limit.
More precisely, we can write
\be
m^2\wp_{\nu}(x)
=
{\nu^2\over 2}
\sum_{n=-\infty}^{\infty}
{m^2\over {\rm ch}\,\nu(x-2n\omega_2)-1}.
\ee
Setting $x=X-2\omega_2\d^{\vee}\rho$, we
obtain the following asymptotics
\be
m^2\wp_{\nu}(x)
=\nu^2M^2
\cases{e^{-2\omega_2(\d^{\vee}\a^{\vee}\cdot\rho-\d^{\vee})-\a^{\vee}\cdot 
X}
+e^{-2\omega_2(1-\d^{\vee}\a^{\vee}\cdot\rho-\d^{\vee})+\a^{\vee}\cdot X},
&if $\a$ is long;\cr
e^{-2\omega_2(\d^{\vee}\a^{\vee}\cdot\rho-\d^{\vee})-\a^{\vee}\cdot X},
&if $\a$ is short.\cr}
\ee
This leads to the following theorem \cite{cm2}

\bigskip

\noindent
{\bf Theorem 2}.
{\it 
Under the limit $x=X+2\omega_2 {1\over h_{\G}^{\vee}} \rho$,
$m=Mq^{-1 / (2h_{\G}^{\vee})}$,
with $\rho$ the Weyl vector and $q\to 0$,
the Hamiltonian of the twisted elliptic Calogero-Moser system
for the simple Lie algebra $\G$
tends to the Hamiltonian of the Toda system
for the affine Lie algebra $(\G^{(1)})^{\vee}$.}

\bigskip

So far we have discussed only the scaling limits of the Hamiltonians.
However, similar arguments show that the Lax pairs constructed below
also have finite, non-trivial scaling limits whenever
this is the case for the Hamiltonians.
The spectral parameter $z$ should scale as $e^z=Zq^{1\over 2}$,
with $Z$ fixed. The parameter $Z$ can be identified
with the loop group parameter
for the resulting affine Toda system.

\vfill\eject

\setcounter{equation}{0}
\section{Calogero-Moser Lax pairs for general Lie algebras}

While the Lax pair for the elliptic $SU(N)$ Calogero-Moser system
was constructed \cite{krichever2} 
shortly after the proposal of the Calogero-Moser
systems was made \cite{olsha}, the Lax pairs $L(z)$, $M(z)$
for elliptic Calogero-Moser systems associated with 
general Lie algebras were constructed only recently \cite{cm1}.

The key complication encountered when passing from $SU(N)$ 
to other Lie algebras is the following. For $SU(N)$, the Lax
operators $L$ and $M$ are $N\times N$ matrices, and thus 
belong to the (complexified) Lie algebra of $SU(N)$. In
other words, the existence and calculation of a Lax pair is 
essentially a problem in the Lie algebra of $SU(N)$. 
For general Lie algebra $\G$ however, it will turn out that
$L$ and $M$ cannot belong to $\G$ for generic values of the
spectral parameter. Thus, the problem of the existence
and calculation of the Lax pairs becomes a problem in a
more general algebra in which $\G$ will have to be embedded.
 
Before presenting the Lax pairs with spectral
parameter for general simple Lie algebras,
we describe them first in a concrete and relatively simpler case,
namely when $\G$ is one of the classical algebras $B_n$, $C_n$,
or $D_n$, and the elliptic Calogero-Moser system is untwisted.
In these cases, the operators $L(z)$ and $M(z)$ have a suggestive
expression in terms of matrices.

\subsection{Lax Pairs with Spectral Parameter for Classical
Lie Algebras}

 Following Olshanetsky
and Perelomov \cite{olsha}, it is convenient to consider
the Lax pair for the root system ${\cal R}(BC_n)
\equiv {\cal R}(B_n)\cup {\cal R}(C_n)$, although
this set is not strictly speaking the root system of an algebra.
The expression (\ref{untwisted}) defines then the corresponding
(untwisted) elliptic Calogero-Moser system, with
mass parameters $m_1$, $m_2$, and $m_4$. The cases $B_n$, $C_n$,
and $D_n$ are recovered with the following choices of masses
\bea\label{bcn}
B_n\quad & &m_4=0 \nonumber\\
C_n\quad & &m_1=0 \nonumber\\
D_n\quad & &m_1=m_4=0
\eea
We claim that the elliptic Calogero-Moser system associated to
${\cal R}(BC_n)$ is integrable and admits a Lax pair with spectral
parameter if 
\be\label{integrability}
m_1(m_1^2-2m_2^2+\sqrt 2\, m_2m_4)=0
\ee
In this case, the Lax pair $L(z)$, $M(z)$ is of the form
\be\label{laxbc}
L(z)=P+X,\quad\quad M(z)=D+Y
\ee
Here the matrices $P,X,D,Y$ are all $(2n+1)\times (2n+1)$
dimensional, with $P,D$ diagonal matrices
\bea
P&=&{\rm diag}(p_1,\cdots,p_n;-p_1,\cdots,-p_n;0)\nonumber\\
D&=&{\rm diag}(d_1,\cdots,d_n;+d_1,\cdots,+d_n;0)\nonumber\\
\eea
and $X, Y$ of the form
\be
X=\pmatrix{A & B_1 & C_1\cr
B_2 & A^T & C_2\cr
C_2^T& C_1^T & 0\cr},
\quad\quad
Y=\pmatrix{A' & B_1' & C_1'\cr
B_2' & A'^T & C_2'\cr
C_2'^T& C_1'^T & 0\cr}
\ee
The entries of the matrix $D$ are given by
\be\label{d(x)}
d_i=-2m_2\sum_k\wp(x_k)
+
{m_1^2\over m_2}\wp(x_i)
+
\sqrt 2\,m_4\wp(2x_i)
+
m_2\sum_{k\not=i}[\wp(x_i-x_k)+\wp(x_i+x_k)]
\ee
those of the matrix $X$ by
\bea
A_{ij}&=&m_2(1-\delta_{ij})\Phi(x_i-x_j,z)\nonumber\\
B_{1ij}&=&m_2(1-\delta_{ij})\Phi(x_i+x_j,z)
+\sqrt 2\,m_4\delta_{ij}\Phi(2x_i,z)\nonumber\\
B_{2ij}&=&m_2(1-\delta_{ij})\Phi(-x_i-x_j,z)
+\sqrt 2\,m_4\delta_{ij}\Phi(-2x_i,z)\nonumber\\
C_{1i}&=&m_1\Phi(x_i,z)\nonumber\\
C_{2i}&=&m_1\Phi(-x_i,z)
\eea
and those of the matrix $Y$ are given by similar formulas,
with $A,B,C$, and $\Phi(x,z)$ replaced respectively
by $A',B',C'$, and $\Phi'(x,z)$.
We note that for $z=\omega_a$, where $\omega_a$ is one of the
half periods, then the preceding Lax pair reduces to the one
found by Olshanetsky and Perelomov \cite{olsha}.

\medskip

The matrices $L(z)$, $M(z)$ can now be verified directly
to be a Lax pair by calculation. However, as preparation for the
general Ansatz for the Lax pair for general simple Lie algebras,
we note that they arise by the following representation
theoretic construction. The verification that they form a Lax pair
for the elliptic $BC_n$ Calogero-Moser system can 
then be carried out in the general framework provided in
the next section \S 6.2.

\medskip

We embed $\G=B_n$ into $GL(N,{\bf C})$, with $N=2n+1$,
by embedding the fundamental representation of $B_n$ into
the fundamental representation of $GL(N,{\bf C})$.
The weights of $B_n$ obtained by the decomposition of the adjoint
representation of $GL(N,{\bf C})$ automatically contain
all the roots of the $BC_n$ system. Let $\lambda_I$, $1\leq I\leq N$,
be the weights of the fundamental representation of $B_n$.
In terms of an orthonormal basis of vectors $e_i$, $1\leq i\leq n$,
we have
\be
\lambda_i=e_i;
\quad
\lambda_{i+n}=-e_i\quad 1\leq i\leq n;
\quad
\lambda_{2n+1}=0.
\ee
Let $u_I$, $1\leq I\leq N$, be the weights of the fundamental
representation of $GL(N,{\bf C})$. They are vectors in ${\bf C}^N$.
They can be verified to decompose orthogonally as
\be
\sqrt 2 \,u_I=\lambda_I+v_I,
\ee
for vectors $v_I$ which are orthogonal to $\lambda_J$,
and satisfy $v_i=v_{i+n}$, $1\leq i\leq n$. We can deduce
the decomposition of the roots of $GL(N,{\bf C})$ into weights
of $B_n$. There are three orbits, consisting of the weights
of $B_n$ with length${}^2=2$, which may be viewed as the roots
of $D_n$
\bea
\sqrt 2\,(u_i-u_j)&=&e_i-e_j+v_i-v_j\nonumber\\
\sqrt 2\,(u_{n+j}-u_{n+i})&=&e_i-e_j-v_i+v_j\nonumber\\
\sqrt 2\,(u_i-u_{n+j})&=&e_i+e_j+v_i-v_j\nonumber\\
\sqrt 2\,(u_{n+i}-u_j)&=&-e_i-e_j+v_i-v_j,\quad i\not=j,
\eea
the weights of length${}^2=4$ (additional roots of $C_n$)
\bea
\sqrt 2\,(u_i-u_{n+i})&=&2e_i\nonumber\\
\sqrt 2\,(u_{n+i}-u_i)&=&-2e_i
\eea
and the weights of length${}^2=1$ (additional roots of $B_n$)
\bea
\sqrt2\, (u_i-u_N)&=&e_i+v_i-v_{2n+1}\nonumber\\
\sqrt2\, (u_N-u_{n+i})&=&e_i-v_i+v_{2n+1}\nonumber\\
\sqrt2\, (u_{n+i}-u_N)&=&-e_i+v_i-v_{2n+1}\nonumber\\
\sqrt2\, (u_N-u_i)&=&-e_i-v_i+v_{2n+1}
\eea
Then the matrices $X,Y$ are of the form
\bea\label{XY}
X&=&\sum_{\alpha\in{\cal R}(BC_n)}\Phi(\alpha\cdot x)
\big(\sum_{\lambda_I-\lambda_J=\alpha}
C_{IJ}E_{IJ}\big)\nonumber\\
Y&=&\sum_{\alpha\in{\cal R}(BC_n)}\Phi'(\alpha\cdot x)
\big(\sum_{\lambda_I-\lambda_J=\alpha}
C_{IJ}E_{IJ}\big)
\eea
where the matrices $E_{IJ}$ are given by 
$(E_{IJ})_{LM}=\delta_{IL}\delta_{JM}$,
and $C_{IJ}$ are constants proportional to the masses $m_1$, $m_2$,
and $m_4$.

\medskip

We can now generalize this set up to the case of an arbitrary
simple Lie algebra $\G$.

\subsection{The General Ansatz}

Let the rank of $\G$ be $n$, and $d$ be its dimension.
Let $\Lambda$ be a representation of $\G$ of dimension $N$,
of weights $\l_I$, $1\leq I\leq N$. 
We fix a Cartan subalgebra
${\cal H}$ for $GL(N,{\bf C})$ which contains
the Cartan subalgebra ${\cal H}_{\G}$ of $\G$.
A basis for ${\cal H}$ consists then of
a basis $h_i$, $1\leq i\leq n$, of Cartan generators
for $\G$, together with a complementary set $\tilde h_i$,
$n+1\leq i\leq N$, of generators of $GL(N,{\bf C})$ satisfying
$[h_i,h_j]=[h_i,\tilde h_j]=[\tilde h_i,\tilde h_j]=0$.
By adding to the $\tilde h_i$ suitable linear
combinations of the $h_i$, we may assume that the $h_i$ and $\tilde h_i$
are mutually orthogonal with respect to the Cartan-Killing
form
$tr[Ad_{h_i},Ad_{\tilde h_j}]=0$.

\medskip

Let $u_I\in {\bf C}^N$
be the weights of the fundamental
representation of $GL(N,{\bf C})$. Project
orthogonally the $u_I$'s onto the $\l_I$'s as
\be
su_I=\l_I+v_I,
\ \
\l_I\perp v_J.
\ee
The coefficient $s$ is determined by the normalizations of the various
weights. Let $e_i$, $1\leq i\leq n$, be an orthonormal basis
for the weight space of $\G$, and set $\l_I
=\sum_{i=1}^n\l_{Ii}e_i$.
Then $s\,u_I\cdot e_k=\l_I\cdot e_k=\l_{Ik}$. In particular,
the vector $s\,e_k$, viewed as a vector in the weight space
of $GL(N,{\bf C})$ can be expressed as
$s\,e_k=\sum_{I=1}^N(s\,e_k\cdot u_I)\,u_I
=\sum_{I=1}^N\l_{Ik}u_I$.
Squaring and summing over $k$ gives
\be\label{dynkin}
\,s^2
={1\over n}\sum_{I=1}^N\l_I^2=I_2(\Lambda)
\ee
where $I_2(\Lambda)$ is by definition the second Dynkin index
of the representation $\Lambda$ of $\G$.
We note that
\be
\a_{IJ}=\l_I-\l_J
\ee
is a weight of $\Lambda \otimes \Lambda ^*$ associated to the
root $u_I-u_J$ of $GL(N,{\bf C})$.
More precisely, let $E_{IJ}=u_Iu_J^T$ be the 
usual generators for $GL(N,{\bf C})$. Then
the basic commutation relations in terms of $E_{IJ}$,
$h_i$, $\tilde h_i$ are

\be\label{commutation}
[E_{IJ},E_{KL}]=\delta_{JK}E_{IL}-\delta_{IL}E_{KJ}
\ee
\be
[h,E_{IJ}]=(\l_I-\l_J)E_{IJ},\quad
[\tilde h,E_{IJ}]=
(v_I-v_J)E_{IJ}
\ee
The generators $E_{II}$ form an alternative basis for ${\cal H}$.
Comparing the commutators of $E_{II}$
and $(h,\tilde h)$ with $E_{IJ}$, $I\not=J$,
we obtain easily the relation between $E_{II}$
and $(h,\tilde h)$
\be\label{eii}
E_{II}
={1\over s^2}(\l_I\cdot h+v_I\cdot\tilde h)
\ee 
The centralizer of ${\cal H}_{\G}$ in $GL(N,{\bf C})$
may be larger that the Cartan subalgebra ${\cal H}$
of $GL(N,{\bf C})$. We denote it by ${\cal H}\oplus GL_0$.
For all simple Lie algebras in their lowest dimensional
faithful representation, we have $GL_0=0$, except
in the cases of $F_4$ and $E_8$, where the dimension of
$GL_0$ is $2$ and $56$ respectively. 

\bigskip

The Lax pairs
for both untwisted and twisted Calogero-Moser systems
will be of the form
\be\label{lax}
L=P+X,
\ \
M=D+Y,
\ee
where the matrices $P,X,D$, and $Y$ are given by
\be
X=\sum_{I\not=J}C_{IJ}\P_{IJ}(\a_{IJ},z)E_{IJ},
\ \ \
Y=\sum_{I\not=j}C_{IJ}\P'_{IJ}(\a_{IJ},z)E_{IJ}
\ee
and by
\be
P=p\cdot h,
\ \ \ \
D=d\cdot (h\oplus\tilde h)+\Delta.
\ee
Here $\Delta$ is an element of $GL_0$. 
The functions $\P_{IJ}(x,z)$ and the coefficients
$C_{IJ}$ are yet to be determined.
We begin by stating the necessary and sufficient
conditions for the pair $L(z)$, $M(z)$ of (4.1)
to be a Lax pair for the
(twisted or untwisted) Calogero-Moser
systems. For this, it is convenient to
introduce the following notation
\bea\label{wp}
\P_{IJ} &= & \P_{IJ}(\a_{IJ}\cdot x),
\quad 
\P_{IJ}'=\P_{IJ}'(\a_{IJ}\cdot x)\\
\wp_{IJ}'
&= & \P_{IJ}(\a_{IJ}\cdot x,z)\P_{JI}'(-\a_{IJ}\cdot x,z)
-\P_{IJ}'(\a_{IJ}\cdot x,z)
\P_{JI}'(-\a_{IJ}\cdot x,z).
\eea
Then the Lax equation $\dot L(z) =[L(z),M(z)]$ implies the
Calogero-Moser system if and only
if the following three identities are satisfied
\be \label{ansatz1}
\sum_{I\not=J}C_{IJ}C_{JI}\wp_{IJ}'\a_{IJ}
=
s^2\sum_{\a\in {\cal R}(\G)}
m_{|\a|}^2\wp_{\nu(\a)}(\a\cdot x)
\ee
\be \label{ansatz2}
\sum_{I\not=J}C_{IJ}C_{JI}
\wp_{IJ}'(v_I-v_J)
=0
\ee
\bea \label{ansatz3}
\sum_{K\not= I,J}
C_{IK}C_{KJ}(\P_{IK}\P_{KJ}'-\P_{IK}'\P_{KJ})
&= &
sC_{IJ}\P_{IJ}d\cdot (u_I-u_J)
+
\sum_{K\not= I,J}
\Delta_{IJ}C_{KJ}\P_{KJ} \nonumber \\
& & \qquad\qquad\qquad
- \sum_{K\not= I,J} C_{IK}\P_{IK}\Delta_{KJ}
\eea
In fact, applying the commutation relations (\ref{commutation})
shows readily that the conditions $\dot X=[P,Y]$ and $\dot x=p$ 
are equivalent. The remaining terms in the equation
$\dot L=[L,M]$ decompose into two equations,
$\dot P=[X,Y]_{\cal H}$ and $[X,Y]_{GL(N,{\bf C})\ominus
{\cal H}}+[X,D]=0$. Now the commutator $[X,Y]$
is given by
\be
[X,Y]
=\sum_{J\notin\{I,L\}}C_{IJ}C_{JL}
(\Phi_{IJ}\Phi'_{JL}-\Phi_{IJ}'\Phi_{JL})E_{IL}
\nonumber
\ee
The component $[X,Y]_{\cal H}$ is obtained by retaining only
the terms $E_{II}$. Expressing $E_{II}$ in terms of $(h,\tilde h)$
as in (\ref{eii}), we find that the equation $\dot P=[X,Y]_{\cal H}$
is equivalent to
\be
\dot p\cdot h={1\over s^2}\sum_{I\not=J}
C_{IJ}C_{JI}\wp_{IJ}'(\l_I\cdot h+v_I\cdot \tilde h).
\ee 
In view of (\ref{wp}), $\wp_{IJ}'=-\wp_{JI}'$,
and we may antisymmetrize in $I$ and $J$.
Since $h,\tilde h$ are linearly independent, the resulting equation
can only hold if the second identity (\ref{ansatz2}) is satisfied. 
Assuming this,
the equation reduces
then to the Calogero-Moser equation of motion
\be
\dot p
={1\over 2}\sum_{\a\in {\cal R}(\G)}m_{|\a|}^2\wp_{\nu(\a)}'(\a\cdot x)
\ee
if the first identity (\ref{ansatz1}) is imposed.
Finally, the terms in the third identity (\ref{ansatz3}) are recognized
as just the coefficients of $E_{IJ}$, $I\not=J$, in
$[X,Y]_{GL(N,{\bf C})\ominus{\cal H}}$, $[d\cdot(h\oplus\tilde h,X]$,
and $[\Delta,X]$ respectively. Thus the identity (\ref{ansatz3}) is
equivalent to the vanishing of $[X,Y]_{GL(N,{\bf C})\ominus{\cal H}}
+[X,D]$, completing the argument.

\bigskip
 
The following theorem was established in \cite{cm1}:

\bigskip

\noindent
{\bf Theorem 3}. {\it A representation $\Lambda$, functions
$\Phi_{IJ}$, and coefficients $C_{IJ}$ with a spectral parameter $z$
satisfying (\ref{ansatz1}), (\ref{ansatz2}), (\ref{ansatz3}) 
can be found for all twisted  and untwisted elliptic
Calogero-Moser systems associated with a simple Lie algebra
$\G$, except possibly in the case of twisted $G_2$.
In the case of $E_8$, we have to assume the existence of 
a $\pm1$-valued cocycle.}   

\bigskip

\subsection{Lax Pairs for Untwisted Calogero-Moser Systems}

\medskip

Returning to the case of the untwisted elliptic Calogero-Moser
system for the $BC_n$ system, we can now at the
same stroke explain how the matrices (\ref{XY}) are found,
and verify that they lead to a Lax pair.
Given the Ansatz (\ref{laxbc}), the embedding of the fundamental
representation of $B_n$
into the fundamental representation of $GL(N,{\bf C})$ ($N=2n+1$)
described in \S 6.1, and the choice
$\Phi_{IJ}(x,z)=\Phi(x,z)$ for all $I,J$,
the main issue reduces to the existence
and determination of the constants $C_{IJ}$.

\medskip

For each $i\not=j$, there are two distinct roots of
$GL(N,{\bf C})$ which project to $e_i-e_j$,
namely $\sqrt2(u_i-u_j)$ and $\sqrt2(u_{n+j}-u_{n+i})$.
The conditions (\ref{ansatz1}) and (\ref{ansatz2})
become
\bea
2m_2^2&=& C_{ij}^2+C_{n+j,n+i}^2\nonumber\\
0&=&C_{ij}^2(v_i-v_j)+C_{n+j,n+i}^2(-v_i+v_j)
\eea
Similar equations are obtained for the roots $e_i+e_j$.
Using the linear independence of the vectors $v_i$, we find
the conditions
\bea
m_2^2&=&C_{ij}^2=C_{n+i,n+j}^2=C_{n+i,j}^2\nonumber\\
2m_4^2&=&C_{i,n+i}^2\nonumber\\
m_1^2&=&C_{iN}^2=C_{n+i,N}^2
\eea
These equations can be solved by taking
\bea
m_2&=&C_{ij}=C_{n+i,n+j}=C_{n+i,j}\nonumber\\
2m_4&=&C_{i,n+i}\nonumber\\
m_1&=&C_{iN}=C_{n+i,N}
\eea
With this choice of coefficients $C_{IJ}$, we turn to the
third condition (\ref{ansatz3}). Letting
$i$ and $j$ take values between $1$ and $n$,
there are 6 cases to be considered:
(1) $I=i,\quad J=j$; (2) $I=i,\quad J=N$;
(3) $I=i,\quad J=n+j$ with $i\not=j$; (4)  $I=i,
\quad J=n+i$;
(5) $I=N,\quad J=n+j$;
and (6) $I=n+i,\quad J=n+j$.
We set $d\dot e_i=0$. Then case (4) is satisfied,
cases (3) and (6) give the same equations as case (1),
and case (5) gives the same equation as case (2).
The cases (1) and (2) themselves lead to the following
equations
\bea
m_2d\cdot(v_i-v_j)&=&
\sum_{k\not=i,j}m_2^2[\wp(x_i-x_j)
-
\wp(x_k-x_j)
+
\wp(x_i+x_k)
-
\wp(x_k+x_j)]\nonumber\\
& &+m_1^2[\wp(x_i)-\wp(x_j)]
+
\sqrt2 m_2m_4[\wp(2x_i)-\wp(2x_j)],\nonumber\\
m_1d\cdot(v_i-v_N)&=&
\sum_{k\not=i}m_1m_2[\wp(x_i-x_j)
+
\wp(x_i+x_k)
-
2\wp(x_k)]\nonumber\\
& &+\sqrt2 m_1m_4[\wp(2x_i)-\wp(x_i)]
\eea
We may assume that $m_2\not=0$, since otherwise the system
decomposes trivially into a set of non-interacting
one-dimensional systems.
Then the most general solution of the first equation
in the preceding condition is
\be
d\cdot v_i
=d_0
+{m_1^2\over m_2}\wp(x_i)
+
\sqrt2 m_4\wp(2x_i)
+
\sum_{k\not=i}m_2[\wp(x_i-x_k)
+
\wp(x_i+x_k)]
\ee
where $d_0$ is an arbitrary function of $x$
which is independent of $i$.
Substituting this into the second equation yields
\be
m_1d\cdot v_N
=
m_1d_0
+
m_1(-2m_2+\sqrt2 m_4+{m_1^2\over m_2})
\wp(x_i)
+
2m_1m_2\sum_k\wp(x_k)
\ee
Since the left hand side is independent of $i$,
our construction works only when $m_1(m_1^2-2m_2^2+\sqrt2 m_2m_4)=0$,
which is the condition we stated earlier in \S 6.1.
When it is satisfied, a vector $d$ can clearly be found.
Choosing $d_0$ so that $d\cdot v_N=0$, we obtain the
Lax pair announced previously for the $BC_n$ system.

\medskip

We summarize now some important
features of the Lax pairs we obtain in Theorem 3.

\begin{itemize}

\item In the case of the {\it untwisted} Calogero-Moser systems,
we can choose $\P_{IJ}(x,z)=\P(x,z)$,
$\wp_{IJ}(x)=\wp(x)$ for all $\G$.

\item $\Delta=0$ for all $\G$, except for $E_8$.

\item For $A_n$, the Lax pair (\ref{lax}) corresponds
to the choice of the fundamental representation for $\Lambda$.
A different Lax pair can be found by taking
$\Lambda$ to be the antisymmetric representation.

\item As we saw before, for the $BC_n$ system, the Lax pair is
obtained by imbedding $B_n$ in $GL(N,{\bf C})$
with $N=2n+1$. When $z=\omega_a$ (half-period),
the Lax pair obtained this way reduces to
the Lax pair obtained by Olshanetsky and Perelomov \cite{olsha}.

\item For the $B_n$ and $D_n$ systems,
additional Lax pairs with spectral parameter
can be found by taking $\Lambda$ to be the spinor representation.

\item For $G_2$, a first Lax pair with spectral parameter can be
obtained by the above construction with $\Lambda$ chosen to be the ${\bf 7}$ of
$G_2$. A second Lax pair with spectral parameter can be obtained by restricting
the {\bf 8} of $B_3$ to the ${\bf 7}\oplus{\bf 1}$ of $G_2$.

\item For $F_4$, a Lax pair can be obtained by
taking $\Lambda$ to be the ${\bf 26}\oplus{\bf 1}$
of $F_4$, viewed as the restriction of
the {\bf 27} of $E_6$ to its $F_4$ subalgebra.

\item For $E_6$, $\Lambda$ is the {\bf 27} representation.

\item For $E_7$, $\Lambda$ is the {\bf 56} representation.

\item For $E_8$, a Lax pair with spectral parameter
can be constructed with $\Lambda$ given by the {\bf 248} representation,
if coefficients $c_{IJ}=\pm 1$ exist with the following cocycle conditions
\bea
c(\lambda,\lambda-\d)c(\lambda-\d,\mu) & =&
c(\lambda,\mu+\d)c(\mu+\d,\mu) \nonumber \\
& & {\rm \ when\ \d\cdot\lambda=-\d\cdot\mu=1,
\ \lambda\cdot\mu=0}\\
c(\l,\mu)c(\l-\d,\mu) & =&c(\l,\l-\d) \nonumber \\
&& {\rm \ when\ \d\cdot\l=\l\cdot\mu=1,
\ \d\cdot\mu=0}\\
c(\l,\mu)c(\l,\l-\mu) & =&
-c(\l-\mu,-\mu) \nonumber \\
&& {\rm \ when\ \l\cdot\mu=1}.
\eea
The matrix $\Delta$ in the Lax pair is then the $8\times 8$ matrix
given by
\bea
\Delta_{ab} &=&
\sum_{\d\cdot\b_a=1\atop \d\cdot\b_b=1}
{m_2\over 2}
\big(c(\b_a,\d)c(\d,\b_b)
+
c(\b_a,\b_a-\d)c(\b_a-\d,\b_b)\big)
\wp(\d\cdot x) \nonumber \\
&&
-\sum_{\d\cdot\b_a=1\atop \d\cdot\b_b=-1}
{m_2\over 2}
\big(c(\b_a,\d)c(\d,\b_b)
+
c(\b_a,\b_a-\d)c(\b_a-\d,\b_b)\big)
\wp(\d\cdot x) \nonumber \\
\Delta_{aa} &=&
\sum_{\b_a\cdot\d=1}
m_2\wp(\d\cdot x)
+2m_2\wp(\b_a\cdot x),
\eea
where $\b_a$, $1\leq a\leq 8$, is a maximal set of 8 mutually
orthogonal roots. The vector $d$ and coefficients $C_{IJ}$ are given by
\bea
C_{\lambda\mu}& = &
\left\{ \matrix{
m_2\, c(\lambda,\mu) & \lambda\cdot\mu=1 & c(\lambda,\mu)=\pm 1 \cr
0 & {\rm otherwise} & \cr}\right . \\
C_{\lambda,c} & =&
\left\{ \matrix{\sum_{a=1}^8{1\over 2}(\lambda\cdot\beta_a)c(\lambda,\beta_a(\lambda\cdot
\beta_a))C_{\beta_a,c} & \lambda\not=\pm \beta_b \cr
\pm C_{\beta_b,c} & \lambda=\pm \beta_b\cr}
\right . \\
\sqrt 60\, d\cdot u_{\lambda}
& = & \sum_{\delta\cdot\lambda=1}
m_2\wp(\delta\cdot x)+2m_2\wp(\lambda\cdot x).
\eea

\end{itemize}

Explicit expressions for the constants $C_{IJ}$ and the functions
$d(x)$, and thus for the Lax pair are particularly simple when the
representation $\Lambda$ consists of only a single Weyl orbit of weights. 
This
is the case when $\Lambda$ is either

\begin{itemize}

\item the defining representation of $A_n$, $C_n$ or $D_n$;

\item any rank $p$ totally anti-symmetric representation of $A_n$;

\item an irreducible fundamental spinor representation of $B_n$ or $D_n$;

\item the ${\bf 27}$ of $E_6$; the ${\bf 56}$ of $E_7$.

\end{itemize}

Then, the weights $\lambda$ and $\mu$ of $\Lambda$ provide unique labels instead
of $I$ and $J$, and the values of $C_{IJ}=C_{\lambda \mu}$ are given by a
simple formula
\be\label{cij}
C_{\lambda \mu} = \left \{ \matrix{
\sqrt{{\alpha ^2 \over 2}} m_{|\alpha|} &
{\rm when} \ \alpha =\lambda - \mu \ {\rm is \ a \ root} \cr
&\cr
0 & {\rm otherwise} \cr} \right .
\ee
The expression for the vector $d$ may be summarized by
\be
s d \cdot u_\lambda = \sum _{\lambda \cdot \delta =1;\ \delta ^2=2}
m_{|\delta |} \wp (\delta \cdot x)
\ee
(For $C_n$, the last equation has an additional term, as given in (\ref{d(x)}).)
In each case, the number of independent couplings $m_{|\alpha|}$ equals the
number of different root lengths.

In the other remaining cases, the vector $d$ and coefficients $C_{IJ}$
are given as follows. For $G_2$ in the {\bf 7} representation,
we have
\bea
C_{\lambda,\mu} & = & \left\{\matrix
{m_2 & \lambda\cdot\mu=\pm{1\over 3} \cr
0 & {\rm otherwise} \cr}\right . \\
C_{\lambda,7} & = & \sqrt2\, m_2 \\
\sqrt 2 \, d\cdot u_{\lambda}
& = & \sum_{\delta^2=1,\lambda\cdot\delta=1}
m_2\wp(\delta\cdot x)+m_2\wp(\lambda\cdot x)\\
\sqrt 2\, d\cdot u_7
& = & {1\over 2}\sum_{\kappa^2=2/3}m_2\wp(\kappa\cdot x).
\eea
For $F_4$ in the ${\bf 26\oplus 1}$, we have
\bea
C_{\lambda\mu} & = & \left\{ \matrix{
m_2 & \lambda\cdot\mu=0,{1\over 2} \cr
0 & {\rm otherwise}\cr}\right . \\
C_{\lambda,a} & = & m_2(1-\delta_{[\lambda],a})\\
C_{a,b} & = & 0 \\
\sqrt 6\, d\cdot u_{\lambda}&=&
2m_2\wp(\lambda\cdot x)
+
m_2\sum_{\delta^2=1,\lambda\cdot\delta}\wp(\delta\cdot x)
-
{1\over 2}m_2\sum_{\kappa\in[\lambda]}\wp(\kappa\cdot x)\\
\sqrt 6\, d\cdot u_a & = &
-m_2\sum_{[\kappa]=a}\wp(\kappa\cdot x)
+
{1\over 2}m_2\sum_{\kappa}\wp(\kappa\cdot x).
\eea
Here we have grouped the 24 short roots of $F_4$ into three classes 
$8^v=\{\pm e_i;\ 1\leq i\leq 4\}$, $8^s
=\{{1\over 2}\sum_{i=1}^4\epsilon_ie_i;\ \prod_{i=1}^4\epsilon_i=1\}$,
$8^c=\{{1\over 2}\sum_{i=1}^4\epsilon_ie_i;
\ \prod_{i=1}^4\epsilon_i=-1\}$,
and the index $a$ runs through the index set $\{v,s,c\}$.

\medskip

We shall illustrate the methods in a particularly simple case,
namely $E_6$. Recall that the root system of $E_6$ is
\bea
 \pm e_i\pm e_j, \quad & 1\leq i<j\leq 5 \nonumber\\
 \pm {1\over 2}(\sqrt 3 \, e_6+\sum_{i=1}^5
\epsilon_ie_i), \quad & \prod_{i=1}^5\epsilon_i=1.
\eea
The simple roots are chosen as in table 8 in Appendix B.
The weights of the ${\bf 27}$ representation of $E_6$
are given by
\bea
{2\over \sqrt 3}e_6 &\nonumber\\
{1\over 2\sqrt 3}e_6-{1\over 2}\sum_{i=1}^5\epsilon_ie_i,
\quad &\prod_{i=1}^5\epsilon_i=1\nonumber\\
-{1\over 2\sqrt 3}e_6\pm e_i, \quad &1\leq i\leq 5.
\eea
The Weyl orbit structure of ${\bf 27}\otimes{\bf 27^*}$ is

\begin{table} [h]
\begin{center} 
\begin{tabular}{|c|c|c|c| }
\hline
Weyl Orbit & Multiplicity & $\#$ Weights & Length${}^2$ \\
\hline
\hline
[000000] & 27 & 1 & 0\\
\hline
[000001] & 6 & 72 & 2\\
\hline
[100010] & 1 & 270 & 4\\
\hline
\end{tabular}
\end{center}
\caption{Weyl orbit structure of ${\bf 27}\otimes {\bf 27^*}$ of $E_6$.}
\label{table:7}
\end{table}

We shall require the following simple property of the roots $\a$
and weights $\l$ of $E_6$ in the ${\bf 27}$ representation: 
\be\label{weight}
\l\pm\a={\rm weight}
\quad
\Longleftrightarrow
\quad
\l\cdot\a=\mp1
\ee 
When $\l\pm\a$ is not a weight, $\l\cdot\a=0$. To see this, we note
that $|\l\cdot\a|\leq {2 \over \sqrt 3} \times \sqrt 2<2$,
since all weights $\l$ of ${\bf 27}$ satisfy $\l^2={4\over 3}$.
As $\l\cdot\a$ is an integer, it can only assume the values
$0,1$ or $-1$. If $\l-\a=\mu$ is a weight, then
$\l\cdot\a-\mu\cdot\a=2$. Since $\mu\cdot\a\in \{0,\pm1\}$,
we must have $\l\cdot \a=1$. The argument for $\l+\a$ is similar.
The reverse implication in (\ref{weight}) is a basic fact of the theory
of Lie algebras.

\medskip

We return now to the construction of the Lax pair. Since $E_6$ is
simply laced, $\wp_{IJ}(x)=\wp(x)$ for all $I,J$. With $C_{IJ}$
given by (\ref{cij}),
the first condition (\ref{ansatz1}) for a Lax pair reduces to
\be
\sum_{\a\in{\cal R}(\G)}\wp'(\a\cdot x)
\sum_{\l_I-\l_J=\a}1
=
s^2
\sum_{\a\in{\cal R}(\G)}\wp'(\a\cdot x)
\ee
However, it follows from (\ref{dynkin}) that 
$s^2={1\over 6}27\times {4\over 3}=6$.
On the other hand, for all roots $\a$, the multiplicity  $\sum_{\l_I-\l_J=\a}1$
of the Weyl orbit is also $6$. 
Thus the condition (\ref{ansatz1}) is satisfied.

\medskip

To obtain the second condition (\ref{ansatz2}), it suffices to verify that
\be
\sum_{\l_I-\l_J=\a}(v_I-v_J)=0
\ee
for each root $\a$. Pairing the left hand side with
an arbitrary vector $u_K$ gives
\be
\sum_{\l_I-\l_J=\a}(v_I-v_J)\cdot u_K
=
\sum_{\l_I-\l_J=\a}(s(\delta_{IK}-\delta_{JK})
-{1\over s}\a\cdot \l_K)
\ee
Using again the fact that $s^2$ and the multiplicity of the Weyl orbit 
are both $6$, this can be rewritten as
\be\label{e6}
\sum_{\l_I-\l_J=\a}(v_I-v_J)\cdot u_K
=
\sqrt 6 \bigg\{\big(\sum_{\l_I-\l_J=\a}(\delta_{IK}-\l_{JK})\big)
-\a\cdot\l_K\bigg\}
\ee
When $\a\cdot\l_K=0$, $\l\pm\a$ is not a weight, and
both expressions on the right hand side of (\ref{e6}) vanish.
When $\a\cdot\l_K=\pm 1$, there exists a unique
weight so that $\l_K\mp\a$ is a weight.
Thus both expressions on the right hand side of (\ref{e6})
are $\pm1$, and the right hand side still vanishes.
This establishes (\ref{ansatz2}).

\medskip

Recall that in the case of $E_6$, we choose 
$\Delta=0$ and $\Phi_{IJ}(x,z)=\P(x,z)$ for all $I,J$,
$I\not=J$.
In view of the key identity (\ref{functional}) for $\P(x,z)$, the
third identity (\ref{ansatz3}) for the Lax pair can be rewritten as
\be\label{ansatz31}
\sqrt 6 C_{IJ}\,d\cdot (u_I-u_J)
=m_{\sqrt 2}^2[\sum
[\wp(\a_{IK}\cdot x)
-
\sum\wp(\a_{KJ}\cdot x)],
\ee
where both sums on the right hand side are restricted
to those $K\not=I,J$ for which $\l_I-\l_K$ and $\l_K-\l_J$
are both roots. This implies that $(\l_I-\l_K)^2
=(\l_J-\l_K)^2=2$, and hence
$\l_I\cdot\l_K=\l_J\cdot\l_K={1\over 3}$.

\medskip

We consider separately the two 
possibilities where $(\l_I-\l_J)^2=4$ and $(\l_I-\l_J)^2=2$.

\medskip
In the first case, $\l_I-\l_J$ is not a root, and $C_{IJ}=0$.
In particular, $(\l_I-\l_J)^2=4$, and hence $\l_I\cdot\l_J=-{2\over 3}$.
By setting $\beta=\l_I-\l_K$,
the first sum on the right of (\ref{ansatz31}) can be viewed as the sum
over all roots $\beta$ with $\beta\cdot\l_I=1$,
$\beta\cdot\l_J=-1$. Indeed, in this case, $\l_I-\l_J$
is not a root, and the restriction $K\not=I,J$
is automatically satisfied. Furthermore,
$\beta\cdot\l_I=\l_I\cdot\l_I-\l_K\cdot\l_I
=1$, $\beta\cdot\l_J=\l_I\cdot\l_J-\l_K\cdot\l_J=-1$.
Conversely, a root $\beta$ with these properties
can be written as $\beta=\l_I-\l_K=\l_K-\l_J$.
Similarly, by introducing $\beta=\l_K-\l_J$,
the second sum on the right of (\ref{ansatz31}) can be
viewed as the sum over all roots $\beta$
with $\beta\cdot\l_I=-1$, $\beta\cdot\l_J=1$.
The right hand side of (\ref{ansatz31}) becomes
\be
\sum_{\beta\cdot\l_I=-\beta\cdot\l_J=1}\wp(\beta\cdot x)
-
\sum_{\beta\cdot\l_I=-\beta\cdot\l_J=-1}\wp(\beta\cdot x),
\ee
which vanishes since $\wp(x)$ is even. Thus the third condition
for the Lax pair is satisfied when $\l_I-\l_J$ is not a root.

\medskip

Assume now that $\l_I-\l_J$ is a root. Then $(\l_I-\l_J)^2=2$
and $\l_I\cdot\l_J={1\over 3}$. 
Introducing $\beta=\l_I-\l_K$ and arguing as before,
we can view the first sum on the right hand side of (\ref{ansatz31})
as the sum over all roots $\beta$ with
$\beta\cdot\l_I=-1$, $\beta\cdot\l_J=0$,
and $\beta\not=\l_I-\l_J$.
Expressing the other sum in the same way, we obtain
\be
\sqrt 6 m_{\sqrt 2}d\cdot (u_I-u_J)
=
m_{\sqrt 2}^2
\big(\sum_{\beta\cdot\l_I=-1,\beta\cdot\l_J=0}\wp(\beta\cdot x)
-
\sum_{\beta\cdot\l_J=-1,\beta\cdot\l_I=0}\wp(\beta\cdot x)
\big)
\ee
where we have added and subtracted the terms corresponding
to $\beta=\l_I-\l_J$. Now recall that the only possible values 
for $\beta\cdot\l$ are
$0$ or $\pm1$. Thus the above right hand side can be rewritten as
\be
m_{\sqrt 2}^2
\big(\sum_{\beta\cdot\l_I=-1}\wp(\beta\cdot x)
-
\sum_{\beta\cdot\l_J=-1}\wp(\beta\cdot x)
\ee
up to the following combination of terms
\be\label{error}
-m_{\sqrt 2}
\big(\sum_{\beta\cdot\l_I=-1;\beta\cdot\l_J=\pm1}
\wp(\beta\cdot x)
-
\sum_{\beta\cdot\l_I=-1;\beta\cdot\l_J=\pm1}
\wp(\beta\cdot x)\big).
\ee
But in this last combination, the terms with $\beta\cdot\l_I
=\beta\cdot\l_J=-1$ cancel out. The remaining sums
are over $\beta\cdot\l_I=-\beta\cdot\l_J=1$
and $\beta\cdot\l_I=-\beta\cdot\l_J=1$ respectively.
Since the function $\wp(x)$ is even, the flip $\beta\to-\beta$
shows that the expression (\ref{error}) cancels.
Thus the third condition for the Lax pair is
satisfied by setting 
$d\cdot u_I=m_{\sqrt 2}\sum_{\beta\cdot\l_I=-1}
\wp(\beta\cdot x)$, which is the prescription
we gave earlier.
  
\subsection{Lax Pairs for Twisted Calogero-Moser Systems}

Recall that the twisted and untwisted Calogero-Moser systems
differ only for non-simply laced Lie algebras, namely
$B_n$, $C_n$, $G_2$ and $F_4$.
These are the only algebras we discuss in this paragraph.
The construction (\ref{lax}) gives then Lax pairs for all of them, 
with the possible exception of twisted $G_2$.
Unlike the case of untwisted Lie algebras however,
the functions $\P_{IJ}$ have to be chosen
with care, and differ for each algebra. More specifically,

\begin{itemize}

\item For $B_n$, the Lax pair is of dimension $N=2n$,
admits two independent couplings $m_1$ and $m_2$,
and
\be
\P_{IJ}(x,z)
=
\cases{
\P(x,z), &if $I-J\not= 0,\pm n$\cr
\P_2({1\over 2}x,z), &if $I-J=\pm n$\cr}.
\ee
Here a new function $\P_2(x,z)$ is defined by
\be
\P_2({1\over 2}x,z)
={\P({1\over 2}x,z)\P({1\over 2}x+\omega_1,z)
\over
\P(\omega_1,z)}
\ee
The vector $d$ and coefficients $C_{IJ}$ are given by
\bea
C_{IJ} & = & \left\{\matrix{
m_2 & I-J\not=0,\pm n \cr
m_1 & I-J=\pm n \cr}\right .\\
d\cdot v_i& = &
m_2\sum_{J-i\not=0,n}\wp((e_i-\lambda_J)\cdot x)
+{1\over 2}m_1\wp_2(e_i\cdot x).
\eea

\item For $C_n$, the Lax pair is of dimension $N=2n+2$,
admits one independent coupling $m_2$,
and
\be
\P_{IJ}(x,z)
=
\P_2(x+\omega_{IJ},z),
\ee
where $\omega_{IJ}$ are given by
\be
\omega_{IJ}
=
\cases{0, &if $I\not=J=1,2,\cdots,2n+1$;\cr
\omega_2, &if $1\leq I\leq 2n,\ J=2n+2$;\cr
-\omega_2, &if $1\leq J\leq 2n,\ I=2n+2$.\cr}
\ee
The vector $d$ and coefficients $C_{IJ}$ are given by
\bea
C_{IJ} & = & \left\{\matrix{
m_2 & I,J=1,\cdots,2n;\ I-J\not=\pm n \cr
{1\over\sqrt 2}m_4=\sqrt 2 m_2 & I=1,\cdots,2n;\ J=2n+1,2n+2;
\ I \leftrightarrow J \cr
2m_2 & I=2n+1,\ J=2n+2;
\ I \leftrightarrow J \cr}\right .\\
\sqrt 2\, d\cdot u_I & = &
\sum_{J-I\not=0,\pm n}m_2\wp_2((\lambda_I-\lambda_J)\cdot x)
+8m_2\wp(2\lambda_I\cdot x);
\ I=1,\cdots,2n\\
\sqrt 2 \, d\cdot u_{2n+1}& =&
\sum_{J=1}^{2n}\wp_2(\lambda_J\cdot x)+2m_2\wp_2(\omega_2)\\
\sqrt 2\, d\cdot u_{2n+2} &=&
\sum_{J=1}^{2n}\wp_2(\lambda_J\cdot x+\omega_2)
+2m_2\wp_2(\omega_2).
\eea

\item For $F_4$, the Lax pair is of dimension $N=24$,
two independent couplings $m_1$ and $m_2$,
\be
\P_{\l\mu}(x,z)
=
\cases{\P(x,z), &if $\l\cdot\mu=0$;\cr
\P_1(x,z), &if $\l\cdot\mu={1\over 2}$;\cr
\P_2({1\over 2}x,z), &if $\l\cdot\mu=-1$.\cr}
\ee
where the function $\P_1(x,z)$ is defined by
\be
\P_1(x,z)
=
\P(x,z)
-
e^{\pi i\zeta(z)+\eta_1z}
\P(x+\omega_1,z)
\ee
Here it is more convenient to label
the entries of the Lax pair directly by the weights
$\lambda=\lambda_I$ and $\mu=\lambda_J$ instead of $I$ and $J$.
The vector $d$ and coefficients $C_{IJ}$ are given by
\bea
C_{\lambda\mu}& =&\left\{\matrix{
m_2 & \lambda\cdot\mu=0 \cr
{1\over \sqrt 2}m_1 & \lambda\cdot\mu={1\over 2}\cr
0 & \lambda\cdot\mu=-{1\over 2}\cr
\sqrt 2 m_1 & \lambda\cdot\mu=-1\cr}\right .\\
\sqrt 6\,d\cdot v_{\lambda}
& = & \sum_{\delta\,{\rm long}\atop\delta\cdot\lambda=1}
m_2\wp(\delta\cdot x)
-
\sum_{\kappa\in[\lambda]}{1\over 2\sqrt 2}
m_1\wp_2(\kappa\cdot x)
+
{1\over\sqrt 2}m_1\wp_2(\lambda\cdot x).
\eea
Here $[\lambda]$ is the classification of the 24 short roots of $F_4$
into $8^v$, $8^s$, and $8^c$ introduced in the previous section.

\item For $G_2$, candidate Lax pairs
can be defined in the {\bf 6} and {\bf 8}
representations of $G_2$, but it is still unknown whether
elliptic functions $\P_{IJ}(x,z)$
exist which satisfy the required identities.

\end{itemize}

\medskip

We illustrate the construction of Lax pairs for twisted Calogero-Moser
systems by reproducing here the construction of \cite{cm1} 
in the simplest case
of $B_n$.
Since we are dealing here with the twisted Calogero-Moser system for a
non-simply laced algebra, we let $\G=B_n^{\vee}$, and choose
$\Lambda$ to be the fundamental representation of $B_n^{\vee}$,
which is of dimension $N=2n$. The weights $\lambda$ of $\Lambda$
are
\be
\lambda_i=-
\lambda_{i+n}
=e_i,
\quad
1\leq i\leq n.
\ee
Via the Ansatz described in (\ref{ansatz1}-\ref{ansatz3}), the 
representation $\Lambda$ is
imbedded in $GL(N,{\bf C})$ with $s^2=2$, and
\be
\sqrt 2\,u_i=e_i+v_i,
\quad
\sqrt 2\, u_{n+i}=-e_i+v_i,
\quad 1\leq i\leq n.
\ee
The roots of $GL(N,{\bf C})$ decompose into short roots of $B_n^{\vee}$
\bea
\sqrt 2\, (u_i-u_j)&=& e_i-e_j+v_i-v_j,
\quad i\not=j\\
\sqrt 2\, (u_{\nu+i}-u_{\nu+j})&=& e_i-e_j+v_i-v_j,
\quad i\not=j\\
\sqrt 2\, (u_i-u_{\nu+j})&=& e_i+e_j+v_i-v_j,
\quad i\not=j\\
\sqrt 2\, (u_{\nu+i}-u_j)&=& -e_i-e_j+v_i-v_j,
\quad i\not=j
\eea
and long roots of $B_n^{\vee}$
\bea
\sqrt 2\,(u_i-u_{\nu+i}&=& 2e_i\\
\sqrt 2\,(u_{\nu+i}-u_i)&=& -2e_i.
\eea
Recall that the constants $C_{IJ}$ are given in this case by
\be
C_{IJ}=\cases{m_2, &if $I-J\not=0,\pm n$\cr
m_1, &if $I-J=\pm n$\cr}
\ee
Then the condition (\ref{ansatz2}) in the requirements for Lax pairs
is satisfied, since each short root of $B_n^{\vee}$
has two roots of $GL(N,{\bf C})$ as preimages,
and they come with opposite values of $v_i-v_j$.
Each long root of $B_n^{\vee}$ has no $v$-dependence at all,
and thus does not enter (\ref{ansatz2}). The condition (\ref{ansatz1})
in the requirements for Lax pairs is an easy consequence
of the functional equations (\ref{functional1}) and (\ref{functional}) for the
$\P(x,z)$ and $\Lambda(x,z)$ functions.
Thus our main task is to find coefficients $d_I$
so that the last condition (\ref{ansatz3}) for the Lax pair
be satisfied.

\medskip

We set $\Delta=0$. Using the antisymmetry of the right hand
side of (\ref{ansatz3}) under $x\to -x$ and $I\to J$ on the left hand side,
we see that $d(-x)=d(x)$. Using this symmetry, we may restrict
attention to the cases $I<J$. Two cases arise,
$J=n+I$ and $J-I\not=0,n$, which we discuss separately.
Consider first the case $J=n+I$, in which (\ref{ansatz3}) reduces to
\bea
m_I\Lambda(2x_i)\, d\cdot 2e_i
&=&\sum_{K\not=i,n+i}
m_2^2\{
\P(\alpha_{iK}\cdot x)
\P'(\alpha_{K(n+i)}\cdot x)
-
\P'(\alpha_{iK}\cdot x)
\P(\alpha_{K(n+i)}\cdot x)\}\nonumber\\
&=& m_2^2\P(2x_i)
\sum_{k\not=i}^n\{\wp(\alpha_{ik}\cdot x)-
\wp(\alpha_{k(n+i)}\cdot x)\\
&\quad& +\wp(\alpha_{i(n+k)}\cdot x)
-\wp(\alpha_{(n+k)(n+i)}\cdot x)\}
\eea
The right hand side is easily seen to vanish, which simply requires
that $d\cdot e_i=0$. Next, we consider the case $J-I\not=0,n$,
for which (\ref{ansatz3}) becomes
\bea
&{}&m_2\P(\alpha_{IJ}\cdot x)\,s\, d\cdot (u_I-u_J)\nonumber\\
&=&\sum_{I-K\not=0,\pm n\atop K-J\not=0,\pm n}
m_2^2\{\P(\alpha_{IK}\cdot x)
\P'(\alpha_{KJ}\cdot x)
-
\P'(\alpha_{IK}\cdot x)
\P(\alpha_{KJ}\cdot x)\}\nonumber\\
&\quad&+\sum_{I-K=\pm n\atop K-J\not=0,\pm n}
m_1m_2\{\Lambda(\alpha_{IK}\cdot x)
\P'(\alpha_{KJ}\cdot x)
-
\Lambda'(\alpha_{IK}\cdot x)
\P(\alpha_{KJ}\cdot x)\}\nonumber\\
&\quad&+\sum_{I-K\not0,\pm n\atop K-J=\pm n}
m_1m_2\{\P(\alpha_{IK}\cdot x)
\Lambda'(\alpha_{KJ}\cdot x)
-
\P'(\alpha_{IK}\cdot x)
\Lambda(\alpha_{KJ}\cdot x)\}\nonumber\\
&\quad&+\sum_{I-K=\pm n\atop K-J=\pm n}
m_1^2\{\Lambda(\alpha_{IK}\cdot x)
\Lambda'(\alpha_{KJ}\cdot x)
-
\Lambda'(\alpha_{IK}\cdot x)
\Lambda(\alpha_{KJ}\cdot x)\}
\eea
The last sum in the preceding equation vanishes identically, because 
the conditions
$I-K=\pm n$ and $K-J=\pm n$ imply that $I-J=0,\pm 2n$,
which is impossible since $I\not=j$.
By noticing that if $I-K=\pm n$, we have that $\lambda_I=-\lambda_K$
for all $I$ and $K$, we can easily make the second and third sum collapse
to single terms. Thus we obtain
\bea \label{next}
&{}& m_2\P(\alpha_{IJ}\cdot x)\,s\, d\cdot(u_I-u_J)
\nonumber \\
&&  \quad =
\sum_{I-K\not=0,\pm n\atop K-J\not=0,\pm n}
m_2^2\{\P(\alpha_{IK}\cdot x)
\P'(\alpha_{KJ}\cdot x)
-
\P'(\alpha_{IK}\cdot x)
\P(\alpha_{KJ}\cdot x)\}
\nonumber\\
&& \qquad 
+m_1m_2\{\Lambda(2\lambda_I\cdot x)
\P'(-(\lambda_I+\lambda_J)\cdot x)
-
\Lambda'(2\lambda_I\cdot x)
\P(-(\lambda_I+\lambda_K)\cdot x)\}
\nonumber\\
&&\qquad 
+m_1m_2\{\P((\lambda_I+\lambda_J)\cdot x)
\Lambda'(-2\lambda_J\cdot x)
-
\P'((\lambda_I+\lambda_J)\cdot x)
\Lambda(-2\lambda_J\cdot x)\} \qquad 
\eea
We now make use of the relations (\ref{functional1}) and (\ref{functional2}) 
for the functions
$\Phi$ and $\Lambda$ to simplify the right hand side of (\ref{next}).
Omitting an overall factor of $m_2\P(\alpha_{IJ}\cdot x)$,
(\ref{next}) is reduced to
\be
s\, d\cdot(u_I-u_J)
=
\sum_{I-K\not=0,\pm n\atop K-J\not=0,\pm n}
\{
\wp(\alpha_{IK}\cdot x)
-\wp(\alpha_{KJ}\cdot x)\}
+
{1\over 2}
m_1\{\wp_2(\lambda_I\cdot x)-\wp_2(\lambda_J\cdot x)\}
\ee
This is solved by setting
\be
d\cdot v_i
=
\sum_{J-i\not=0,n}
m_2\wp((e_i-\lambda_j)\cdot x)
+
{1\over 2}
m_1\wp_2(e_I\cdot x).
\ee

\subsection{Scaling Limits of Lax Pairs}

All the Lax pairs we constructed have finite scaling limits.
This follows from the asymptotic behavior
of the function $\P(u,z)$ 
\be
\P(u,z)\rightarrow
\cases{+e^{-{1\over 2}u}(1-Z^{-1}e^{u-\omega_2}),
& ${\rm Re}(u)\to +\infty$\cr
-e^{{1\over 2}u}(1-Ze^{-u-\omega_2}),
& ${\rm Re}(u)\to-\infty$\cr}
\ee
in the range $|{\rm Re}\, u|<2\omega_2$.
This asymptotic behavior results in finite limits for
the expression $C_{IJ}\P_{IJ}(\alpha\cdot x)$ under the scalings 
defined in \S 5.7 for the Hamiltonians. The behavior of the functions
$C_{IJ}\P'_{IJ}(\alpha\cdot x)$ is similar, and thus both operators
$L(z)$ and $M(z)$ have well defined and finite limits.

\medskip

More precisely, consider first the scaling limits of the
untwisted Calogero-Moser systems defined with $\delta=1 /  h_{\G}$,
where $h_{\G}$ is the Coxeter number.
This means that we let $x=X+2\omega_2\delta\rho^{\vee}$,
$e^z=Zq^{-{1\over 2}}$ as in
Theorem 1, and set
\be
C_{IJ}=\cases{M_{|\alpha|}e^{\delta\omega_2}c_{IJ},
& when $\alpha_{IJ}=\alpha\in{\cal R}(\G)$\cr
0, & when $\alpha_{IJ}\notin {\cal R}(\G)$\cr}
\ee
Then the matrices $L(z)$, $M(z)$ converge as $\omega_2\to +\infty$
to matrices $L(Z)$, $M(Z)$ of the form (\ref{todalax}), where the root system
is the one associated with the affine algebra $\G^{(1)}$,
and the generators $E_{\alpha}$ are given by
\be
E_{\alpha}
=\sum_{\alpha_{IJ}=\alpha}c_{IJ}E_{IJ}.
\ee
The matrices $L(Z)$, $M(Z)$ form a Lax pair for the Toda system
associated with the affine algebra $\G^{(1)}$.

\medskip

Similarly, consider now the scaling limits of the twisted Calogero-Moser
systems defined with $\delta^{\vee}=1 / h_{\G}^{\vee}$, where
$h_{\G}^{\vee}$ is the dual Coxeter number.
This means that we let
$x=X+2\omega_2\delta^{\vee}\rho^{\vee}$,
$e^z=Zq^{-{1\over 2}}$ as in
Theorem, and set
\be
C_{IJ}=\cases{M_{|\alpha|}e^{\delta^{\vee}\omega_2}c_{IJ},
& when $\alpha_{IJ}=\alpha\in{\cal R}(\G)$\cr
0, & when $\alpha_{IJ}\notin {\cal R}(\G)$\cr}
\ee
Then the matrices $L(z)$, $M(z)$ of the Lax pair for the twisted 
Calogero-Moser systems $B_n$, $C_n$, and $F_4$ all have finite limits.
The entries of their limits are given as follows. For $B_n$
and $F_4$, the entries of the limit of $L(z)$ and $M(z)$
are given respectively by
\be
C_{IJ}\P_{IJ}(\alpha\cdot x,z)
\to
\cases
{\pm\kappa_{\G}M_{|\alpha|}c_{IJ}e^{\mp{1\over 2}\alpha^{\vee}\cdot X},
&if $l^{\vee}(\alpha^{\vee})=\pm1$;\cr
\mp\kappa_{\G}M_{|\alpha|}c_{IJ}e^{\pm{1\over 2}\alpha_0^{\vee}\cdot X},
&if $l^{\vee}(\alpha^{\vee})=\pm l_0^{\vee}$;\cr
0 &otherwise\cr}
\ee
and by
\be
C_{IJ}\P_{IJ}'(\alpha\cdot x,z)
\to
\cases
{-{1\over 2}\kappa_{\G}M_{|\alpha|}c_{IJ}e^{\mp{1\over 2}\alpha^{\vee}\cdot X},
&if $l^{\vee}(\alpha^{\vee})=\pm1$;\cr
-{1\over 2}\kappa_{\G}M_{|\alpha|}c_{IJ}e^{\pm{1\over 2}\alpha_0^{\vee}\cdot X},
&if $l^{\vee}(\alpha^{\vee})=\pm l_0^{\vee}$;\cr
0 &otherwise.\cr}
\ee
Here $\kappa_{\G}$ is a constant depending on the Lie algebra,
with $\kappa_{B_n}=1$ and $\kappa_{F_4}=2$. As for the case of $C_n$,
the entries of the limits of the matrices $L(z)$ and $M(z)$
are given respectively by
\be
C_{IJ}\P_{IJ}(\alpha\cdot x,z)
\to
\cases
{\pm2
M_{|\alpha|}c_{IJ}e^{\mp{1\over 2}\alpha^{\vee}\cdot X},
&if $l^{\vee}(\alpha^{\vee})=\pm1$;\cr
\mp2
M_{|\alpha|}c_{IJ}e^{\pm{1\over 2}\alpha_0^{\vee}\cdot X}
Z^{-{1\over 2}\mp{1\over 2}},
&if $l^{\vee}(\alpha^{\vee})=\pm l_0,
I<J$;\cr
\mp2
M_{|\alpha|}c_{IJ}e^{\pm{1\over 2}\alpha_0^{\vee}\cdot X}
Z^{{1\over 2}\mp{1\over 2}},
&if $l^{\vee}(\alpha^{\vee})=\pm l_0,
J<I$;\cr
0 &otherwise,\cr}
\ee
and by
\be
C_{IJ}\P_{IJ}'(\alpha\cdot x,z)
\to
\cases
{-2
M_{|\alpha|}c_{IJ}e^{\mp{1\over 2}\alpha^{\vee}\cdot X},
&if $l^{\vee}(\alpha^{\vee})=\pm1$;\cr
-2
M_{|\alpha|}c_{IJ}e^{\pm{1\over 2}\alpha_0^{\vee}\cdot X}
Z^{-{1\over 2}\mp{1\over 2}},
&if $l^{\vee}(\alpha^{\vee})=\pm l_0,
I<J$;\cr
-2
M_{|\alpha|}c_{IJ}e^{\pm{1\over 2}\alpha_0^{\vee}\cdot X}
Z^{{1\over 2}\mp{1\over 2}},
&if $l^{\vee}(\alpha^{\vee})=\pm l_0,
J<I$;\cr
0 &otherwise.\cr}
\ee
In each case, the resulting matrices $L(Z)$, $M(Z)$ form a Lax pair
for the Toda system associated with $(\G^{(1)})^{\vee}$.
In other words, starting with the Lax pair for the twisted
Calogero-Moser system for $B_n$, $C_n$, and $F_4$,
we obtain respectively a Lax pair for the Toda system
associated with $(B^{(1)})^{\vee}=A_{2n-1}^{(2)}$,
$(C_n^{(1)})^{\vee}=D_{n+1}^{(2)}$,
and $(F_4^{(1)})^{\vee}=E_6^{(2)}$.

\bigskip

We note that recently Lax pairs of root 
 type have been considered \cite{sasaki1}
which correspond, in the above Ansatz (5.3-5),
 to $\Lambda$ equal to the adjoint
representation of $\G$ and the coefficients $C_{IJ}$ vanishing for $I$ or $J$
associated with zero weights. This choice yields another Lax pair for the case
of $E_8$. However, as the authors have themselves pointed out \cite{sasaki2},
the Lax pairs they obtained do not tend to finite limits
under the scalings defined in Theorems 1 and 2. Thus the corresponding spectral
curves do not appear to be suitable as Seiberg-Witten curves
for supersymmetric Yang-Mills theories.

\vfill\eject

\setcounter{equation}{0}
\section{Super-Yang-Mills and Calogero-Moser Systems}

In this section, we begin by discussing the general correspondence 
between integrable systems and $\N=2$ super Yang-Mills theory
and then treat the correspondence in detail for $SU(N)$ gauge group
and a hypermultiplet in the adjoint representation of the gauge group.
Why the effective prepotential of $\N=2$ supersymmetric gauge
theories can be realized by a fibration of spectral curves
remains one of the most important unanswered
questions in Seiberg-Witten theory.
However, assuming this fact, it is easy to see, as Donagi and
Witten pointed out, the emergence of integrable models.

\subsection{Correspondence Seiberg-Witten and Integrable Systems}

The key ingredient is the symplectic form $\omega$
on the fibration of Jacobians (or Prym varieties) 
associated to the spectral curves
$\Gamma$
\be\label{omega}
\omega=\delta \bigg (\sum_{i=1}^n d \lambda(z_i) \bigg )
\ee
Here we have identified the Jacobian (or Prym) of $\Gamma$
with the symmetric product of $r$ copies of $\Gamma$,
and the differential $\delta$ in (\ref{omega}) is taken with respect to
both the $z_i$ variables and the vacuum parameters
$u_i$ for the base of the fibration. Evidently,
$\omega$ is then of the form $\omega=\omega_{ji}du_j\wedge dz_i$,
with no components of the form $du_i\wedge du_j$ or
$dz_i\wedge dz_j$. It follows that the Poisson brackets
$\{u_k,u_l\}$ all vanish. Thus the vacuum moduli parameters
of the gauge theory can be viewed as a maximal pairwise commuting
set of Hamiltonians with respect to the symplectic form $\omega$.
This correspondence between $\N=2$ supersymmetric gauge theories
and integrable models is quite attractive in its generality.
However, it should be stressed that it is probably incomplete.
Indeed, in all known cases, one particular Hamiltonian amongst the 
infinite set of Hamiltonians seems to
play a special role, namely as a beta function for the gauge theory.
Also, there is at this moment no systematic rule for
how to identify the integrable model
corresponding to a given gauge theory.

\medskip

Schematically, the aspects of the correspondence between
the data that arise in both settings can be summarized
as follows in table \ref{table:8}. 

\begin{table}[h]
\begin{center}
\begin{tabular}{|c |c|} \hline
$\N=2$ super-Yang-Mills Data      & Integrable Systems Data  \\ \hline \hline
& Integrable Hamiltonian system associated \\
$(\G,\ \R, \ \ \tau, \ m,\ \Gamma, \ \lambda)$ & with Lie algebra $\G$ \\ 
& $x_i, \ p_i \ \in {\bf C}, \ i=1,\cdots , n={\rm rank}(\G)$ \\ \hline
S.W. curve $\Gamma$ & spectral curve $\det \big ( kI - L(z) \big ) =0$ \\ 
\hline
S.W. differential $d\lambda$ & $d\lambda=k\,dz$ \\ \hline
$\omega=\delta (\sum_{i=1}^n d\lambda(z_i))$ & symplectic form \\ \hline
vacuum moduli $a_i$ & integrals of motion $I_i$ \\ \hline
beta function of RG equation & Hamiltonian  \\ \hline
\end{tabular}
\end{center}
\caption{Data map between Seiberg-Witten theory and Integrable systems}
\label{table:8}
\end{table}

\medskip

In the table, $\tau$ stands for the complex Yang-Mills coupling, $\R$ and $m$
stand for the hypermultiplet representation and mass respectively. The exact
integrable system that arises in the correspondence will depend upon the gauge
algebra $\G$ and the
representation $\R$ of the hypermultiplet.
Its systematic identification is one of the major problems
of the study of supersymmetric Yang-Mills theories. 

\medskip

One of the first cases obtained in all generality is when 
the gauge algebra is simple and no hypermultiplets are
present \cite{martinec}.
We have established previously that the Seiberg-Witten curve for $\N=2$
super-Yang-Mills for $SU(N)$ gauge group and no hypermultiplets, arrived at by
arguments of singularity analysis, $R$-symmetry properties and limiting
behavior, is precisely of the same form as the spectral curve for the period
Toda  system associated with the Lie algebra $A_{N-1} = SU(N)$. 
The correct generalization of this correspondence involves the periodic Toda 
system associated with the twisted affine Lie algebra $(\G ^{(1)})^\vee$. 
The appearance of the dual affine algebra $(\G^{(1)})^\vee$ 
is due to the grading of the term $z+{\mu\over z}$
in the spectral curve $\det (kI-L(z))=0$. When $L(z), M(z)$ is the
Lax pair for the $\G^{(1)}$ Toda system, this term will appear
with the grading $h_{\G}$, if $k$ and the Casimirs $u_l$ of $\G$
are given gradings $1$ and $l$ respectively.
Identifying $u_l$ with the vacuum moduli $u_l$ of the gauge theory,
we find that the instanton generated term $z+{\mu\over z}$
should have instead grading $h_{\G}^{\vee}$,
where $h_{\G}^{\vee}$ is the dual Coxeter number.
This is why the correct integrable system
for pure super-Yang-Mills theory with gauge algebra $\G$
should be the periodic Toda system associated with
$(\G ^{(1)})^\vee$. 
For the
classical Lie algebras $\G$, the spectral curves derived from the 
$(\G ^{(1)})^\vee$ Toda system
were found to be in precise agreement with the preceding 
constructions in terms
of the singularity structure of the Seiberg-Witten curves, $R$-symmetry
properties and the matching of various decoupling limits. 

\medskip

Thus the case of pure Yang-Mills theories, with simple gauge algebra
$\G$ and no additional hypermultiplets, has been settled.
The main problem we wish to address is the identification
of the integrable models corresponding to $\N=2$ supersymmetric
Yang-Mills either with product gauge algebras,
or with additional matter hypermultiplets. Among these,
one theory is particularly attractive and has a certain universal
aspect, in the sense that many other theories follow from it
in suitable decoupling limits. This is the Yang-Mills theory
with simple gauge algebra $\G$, and one matter hypermultiplet
in the adjoint representation of $\G$. 

\medskip

The first case to be treated successfully was $\G=SU(N)$,
where the integrable model was identified by Donagi and Witten
\cite{donagi} as the $SU(N)$ Hitchin system \cite{hitchin}.
Several authors \cite{nekrasov} subsequently recognized
the spectral curves of the $SU(N)$ Hitchin system as
identical to the spectral curves of the $SU(N)$ elliptic
Calogero-Moser system. This correspondence between
$SU(N)$ Seiberg-Witten theory and $SU(N)$ Calogero-Moser
system was however still rather rudimentary, since
the vacuum parameters of the Yang-Mills theory were still
obscure, and the prepotential arising from the Calogero-Moser
system were not yet known to satisfy the monodromy
properties required by field theory. Our purpose in this section
is to address these issues in the case of $\G=SU(N)$, and to show 
that for general $\G$, the integrable model corresponding
to the Yang-Mills theory with a hypermultiplet in the adjoint
representation is the twisted Calogero-Moser system associated to the
Lie algebra $\G$.

\subsection{Calogero-Moser and Seiberg-Witten Theory for $SU(N)$}

The full correspondence between Seiberg-Witten 
theory for $\N=2$ $SU(N)$ super-Yang-Mills theory with one hypermultiplet 
in the adjoint representation of the gauge algebra, and the elliptic 
$SU(N)$ Calogero-Moser 
systems was obtained in \cite{dp1}. We describe it here in some detail.

\medskip

All that we shall need here of the elliptic Calogero-Moser system is its Lax
operator $L(z)$, whose $N\times N$ matrix elements are given by
\be\label{L(z)}
L_{ij}(z) = p_i\delta _{ij} - m(1-\delta _{ij}) \Phi (x_i-x_j,z)
\ee
Notice that the Hamiltonian is simply given in terms of $L$ by $H(x,p) =
\half \tr L(z)^2 + C\wp (z)$ with $C=-\half m^2 N(N-1)$.

\medskip

The correspondence between the data of the elliptic Calogero-Moser system and 
those of the Seiberg-Witten theory is as follows. 

\begin{itemize}

\item The parameter $m$ in (\ref{L(z)}) is the hypermultiplet mass;

\item The gauge coupling $g$ and the $\theta$-angle are related to the modulus
of the torus (or elliptic curve) $\Sigma ={\bf C} /(2\omega _1 {\bf Z} + 2
\omega _2 {\bf  Z})$ by 
\be
\tau = {\omega _2 \over \omega _1} = {\theta \over 2 \pi} + {4 \pi i\over g^2}
\, ;
\ee

\item The Seiberg-Witten curve $\Gamma$ is the spectral curve of the 
elliptic Calogero-Moser model, defined by
\be
\Gamma = \{ (\tilde k,z) \in {\bf C} \times \Sigma, \ \det\bigl (\tilde
kI-L(z)\bigr )=0\}
\ee
and the Seiberg-Witten 1-form is $d\lambda = k \ dz$. $\Gamma$ is invariant 
under the Weyl group of $SU(N)$. 
It can be viewed as $N$ copies of the torus $\Sigma$,
glued along suitable cuts on each copy.
We use here the notation $\tilde k$ in the equation for the
spectral curve, in order to reserve the notation $k$ for a more
convenient variable to be introduced later.

\item Using the Lax equation $\dot L = [L,M]$, it is clear that the 
spectral curve is independent of time, and can be dependent only upon the 
constants of motion of the Calogero-Moser system, of which there are only $N$.
These integrals of motion may be viewed as parametrized by the quantum moduli
of the Seiberg-Witten system.

\item Finally, $d\lambda =\tilde kdz$ is meromorphic, with a simple pole on 
each of the $N$ sheets above the point $z=0$ on the base torus. The residue at 
each of these poles is proportional to $m$, as required by the general set-up 
of Seiberg-Witten theory, explained in \S 2.

\end{itemize}

\subsection{Four Fundamental Theorems}

While the above mappings of the Seiberg-Witten data onto the Calogero-Moser 
data are certainly natural, there is no direct proof for them, and it is important 
to check that the results inferred from it agree with known facts from quantum 
field theory. To establish this, as well as a series of further predictions
from the correspondence, we give four theorems (the
proofs may be found in \cite{dp1} for the first three theorems,
and in \cite{dp2} for the last one).

\bigskip

\noindent
{\bf Theorem 4.} {\it The spectral curve equation $\det (\tilde kI-L(z))=0$ 
can be expressed as
\be\label{H(k)} 
det(\tilde k-L(z))
={\vartheta _1 \biggl (
{1 \over 2\omega _1}(z-m{\partial \over \partial k}) \big | \tau \biggr )
\over
\vartheta_1({z\over 2\omega_1}|\tau)} H(k)
\bigg | _{k=\tilde k+m\partial_z\ln\vartheta_1({z\over 2\omega_1}|\tau)}=0
\ee
where $H(k)$ is a monic polynomial in $k$ of degree $N$, whose zeros (or 
equivalently whose coefficients) correspond to the moduli of the gauge theory. 
If $H(k)=\prod_{i=1}^N(k-k_i)$, then
\be
\lim_{q\to 0}{1\over 2\pi i}\oint_{A_i}\tilde kdz=k_i-{1\over 2}m.
\ee}

\bigskip

Here, $\vartheta _1$ is the Jacobi $\vartheta$-function, which admits 
the following simple 
series expansion in powers of the instanton factor $q=e^{2\pi i \tau}$
\be
\vartheta_1(u|\tau)
=
\sum_{r\in{1\over 2}+{\bf Z}}
q^{{1\over 2}r^2}e^{2\pi i r(u+{1\over 2})}
\ee
Thus, in terms of the new variable $k$ defined by
\be
k=\tilde k+m\partial_z\ln\vartheta_1({z\over 2\omega_1}|\tau)+{1\over 2}m
\ee 
the equation of the spectral curve equation  can be written 
perturbatively in $q$ as
\be\label{series}
\sum _{n \in {\bf Z}} (-)^n q ^{\half n(n-1)} e^{nz} H(k-n \cdot m) =0
\ee
where we have set $\omega _1 = -i\pi$ without loss of generality. 
The series 
expansion (\ref{series}) is superconvergent and sparse in the 
sense that it receives
contributions only at integers that grow like $n^2$.
But more important, it provides a geometric interpretation of the
order parameters of the super Yang-Mills theory: at $q=0$,
the base torus $\Sigma$ degenerates into a sphere with two
punctures at $w=0$ and $w=\infty$, if we introduce the new
variable
\be
w=e^z
\ee
Each of the copies of the
torus $\Sigma$ making up the spectral curve $\Gamma$ upstairs
degenerates correspondingly into a sphere with
two punctures. Now the equation (\ref{series}) of the spectral curve $\Gamma$
at $q=0$ becomes
\be \label{H(k)1}
H(k)-wH(k-m)=0
\ee
Thus the punctures lying above $w=0$ are given by the $N$ zeroes
$k_i$ of the polynomial $H(k)$, while those
lying above $w=\infty$ are given by $k_i-m$.
Now the $A$-cycles on $\Gamma$ can be chosen to be
circles of a fixed radius around each of the points $k_i$.
At the degeneration point $q=0$, the Seiberg-Witten differential
$d\lambda=\tilde kdz$ can be written as
\be
d\lambda=\tilde k\,d\ln\,w=kd\ln {H(k)\over H(k-m)}
-md\ln\,\vartheta_1({z\over 2\pi i}|\tau)
-{1\over 2}m\, dz
\ee
The second terms on the right does not contribute to $A$-periods,
while the other two terms readily give
\be
{1\over 2\pi i}\oint_{A_i}d\lambda_{\bigg \vert_{q=0}}=k_i-{1\over 2}m.
\ee
as asserted in Theorem 4. This relation provides the key starting point of the
correspondence between Calogero-Moser systems and supersymmetric
gauge theories, by identifying the classical vacua parameters
in terms of the moduli of the Calogero-Moser spectral curves.

\medskip

The parametrization of spectral curves by polynomials $H(k)$ has been
extended from $SU(N)$ Calogero-Moser systems to $SU(N)$ spin Calogero-Moser
systems in \cite{dp3}, using methods of quantum field theory on
Riemann surfaces \cite{determinant}. For $SU(N)$ Calogero-Moser systems,
a different derivation from \cite{dp1} has been recently obtained in
\cite{vaninsky}. 

\medskip

The expansion (\ref{series}) actually allows us to determine the exact value
of the quantum vacua parameters $a_i$ as power series in $q$.
Geometrically, for $q\not=0$, each copy of the torus making up $\Gamma$
has a branch cut of length $O(|q|^{1/2})$
which shrinks to the puncture $k_i$ as $q\to 0$. 
But the contours $A_i$ can still be chosen at a fixed, nonvanishing
distance from $k_i$. Thus the $A_i$-periods
of $d\lambda$ can be evaluated by residue formulas, just as
in the case of the fundamental representation. More specifically,  
applying the methods in \cite{dkp1} for the perturbative solution
of fixed point equations, we find
\be
w=
{H(k)\over H(k-m)}
\biggl [ 1+\sum_{n=1}^{\infty}{q^n\over n!}{\partial^n\over\partial y^n}
F^n(y)_{\vert_{y=1}} \biggr ]
\ee
where the function $F(y)$ is defined by
\bea\label{F}
F(y)
&=&
\sum_{n=1}^{\infty}
q^{{1\over 2}n(n+1)}
(-)^n
\biggl [ y^{-n}\eta_n ^+ (k)-y^{n+1}\eta_n ^- (k- m) \biggr ]
\nonumber \\
\eta_n ^\pm (k)
&=&
{H(k \pm  mn)H(k \mp m)\over H(k)^{n+1}}
\eea
We would like to stress that the expansion (\ref{F}) is a reliable expansion
only when the denominators of the expressions $\eta_n ^\pm (k)$
and $\eta_n ^\pm (k-m)$ are bounded away from $0$.
As explained earlier, this can be achieved in the evaluation
of the $A$-periods, by fixing
the $A$-cycles away from the points $k_i$. 
We can deduce then the full expansion in $q$
of the Seiberg-Witten differential $d\lambda=\tilde kd\ln w$. Each term
in the expansion is rational in $k$ with poles at $k_i$
and $k_i-\beta m$, and the corresponding residues are easily
evaluated.
For example, we obtain to order $q^2$
\bea\label{ak}
a_i&=&k_i+q\bar S_i(k_i)+{1\over 4}q^2(\bar S_i)^{'''}(k_i) + \O(q^3) 
\nonumber \\
\bar S_i(k)&\equiv&
{H(k+m)H(k-m)\over \prod_{j\not=i}(k-k_j)^2}
\eea 
For the super Yang-Mills theory with matter in the adjoint representation,
the microscopic gauge coupling $q=e^{2\pi i\tau}$ 
is analogous to the renormalization
scale $\Lambda$ encountered earlier for the fundamental representation,
and the full expansion in $q$ of the vacua parameters $a_i$
is required to evaluate the contributions to the prepotential
${\cal F}$ of instanton processes to an arbitrary order.

\bigskip

\noindent
{\bf Theorem 5.} {\it The prepotential of the Seiberg-Witten theory obeys a 
renormalization group-type equation that simply relates $\F$ to the 
Calogero-Moser Hamiltonian, expressed in terms of the quantum order parameters 
$a_j$
\be\label{rgcm}
a_j = {1\over 2\pi i}\oint _{A_j} d \lambda
\qquad \qquad
{\partial \F \over \partial \tau} \bigg | _{a_j } 
= H(x,p) = \12 \sum _{i=1} ^N k_i ^2
\ee
Furthermore, in an expansion in powers of the instanton factor $q=e^{2\pi i 
\tau}$, the quantum order parameters $a_j$ may be computed by residue methods
in terms of the zeros of $H(k)$.} (The polynomial $H(k)$ is not to be confused
with the Calogero-Moser Hamiltonian $H(x,p)$; the notation is conventional.)

\bigskip

The microscopic gauge coupling $\tau$
corresponds geometrically to the moduli of the base torus $\Sigma$,
so that the renormalization group equation (\ref{rgcm}) is mathematically
a statement about deformations of complex structures.
The main underlying equation is the following
\be\label{rgcm1}
\delta a_{Di}
=
\sum_{j=1}^N\oint_{A_j}\tilde k \Omega _i
\ee
where $\Omega_i$, $1\leq i\leq N$, is the
basis of holomorphic Abelian integrals dual to the
$A_i$ basis of cycles. It can be derived by showing that the
deformation $\delta d\lambda$ of the Seiberg-Witten differential
satisfies a $\bar\partial$-equation, and solving this equation
in terms of the prime form \cite{dp1}. Now the Abelian differentials
$\Omega_i$ can also be written as
\be
\Omega_i={1\over 2\pi i}{\partial\over\partial a_i}d\lambda_i
\ee
The equation (\ref{rgcm1}) implies
\be
{\partial\over\partial\tau}
\biggl ( {\partial{\cal F}\over\partial a_i} \biggr )
=
{1\over 4\pi i}{\partial\over\partial a_i}
\biggl ( \sum_{j=1}^N\oint_{A_j}\tilde k^2 dz \biggr )
\ee
from which the renormalization group equation (\ref{rgcm}) follows.

\medskip

The general form of the prepotential ${\cal F}$ for the $SU(N)$ 
super Yang-Mills
theory with matter in the adjoint representation is
\be
{\cal F}
={\cal F}^{\rm (pert)}+\sum_{n=1}^{\infty}q^n{\cal F}^{(n)},
\ee
where ${\cal F}^{\rm (pert)}$ is the perturbative part
which can be determined by standard methods of quantum field theory
\be \label{pertpart}
\F ^{({\rm pert})} = {\tau \over 2} \sum _i a_i ^2
- {1 \over 8 \pi i} \sum _{i,j} \biggl [ (a_i -a_j)^2 \ln (a_i - a_j)^2
- (a_i -a_j -m)^2 \ln (a_i -a_j -m)^2 \biggr ],
\ee
and ${\cal F}^{(n)}$ are the terms due to instanton processes.
The renormalization group equation (\ref{rgcm}) shows that all the coefficients
${\cal F}^{(n)}$ can just be read off from an expansion
of the Hamiltonian $H(x,p)$ in terms of the vacua parameters $a_i$.
Such an expansion is easily obtained by writing $H(x,p)$ in terms of
the Calogero-Moser integrals of motion $k_i$, and inverting the
relations (\ref{ak}) giving $a_i$ in terms of $k_i$.  
For example, to two-instanton order, the results can be described
as in Part (b)
of the following theorem \cite{dp1}:
\bigskip

\noindent
{\bf Theorem 6.} {\it {\rm (a)} The perturbative part
${\cal F}^{\rm (pert)}$ of the prepotential
given by the Calogero-Moser spectral curves
is indeed of the form (\ref{pertpart});

{\rm (b)} To two-instanton order, the instanton corrections 
can be expressed in terms of a single function
\be
S_i(a) = {\prod _{j=1} ^N \big [ (a_i-a_j)^2-m^2 \bigr ] \over 
\prod _{j\not=i} (a-a_j)^2}.
\ee
as follows}
\bea
\F ^{(1)} & = & {1 \over 2 \pi i} \sum _i S_i(a_i) \nonumber \\
\F ^{(2)} & = & {1 \over 8 \pi i} \biggl [
  \sum _i S_i(a_i) \partial _i ^2 S_i(a_i)
  + 4 \sum _{i\not= j} {S_i(a_i)S_j(a_j) \over (a_i -a_j)^2} - { 
S_i(a_i)S_j(a_j) \over (a_i -a_j-m)^2}  \biggr ] 
\eea

Part (a) of Theorem 6 cannot be obtained from the renormalization
group equation. We discuss briefly its proof,
which is important, because it
is the defining criterion for the Calogero-Moser spectral
curves to provide an exact solution of the four-dimensional
supersymmetric gauge theory. The main problem in establishing
(a) is to determine the leading terms in the periods $a_{Di}$
of the Seiberg-Witten differential $d\lambda$. Recall that as $q$
moves away from $0$, each puncture $k_i$ and $k_i+ m$
on each sheet above the base torus $\Sigma$ opens into a cut.
Let $k_i^{\pm}$ be the end points of the cut near $k_i$.
They can be identified as the solutions of the equation
\be
{dk\over dz}=0
\ee
This equation can be solved perturbatively in $q$. To $O(q)$ order,
we find
\be\label{turning}
k_i^{\pm}=k_i\pm q^{1/2}k_i^{(1)},
\quad
k_i^{(1)}=2{H^{1/2}(k_i- m)H^{1/2}(k_i+ m)
\over \prod_{j\not=i}(k_i-k_j)}
\ee
Similarly, the end points of the cut near $k_i+\beta m$ can be 
identified as $k_i^{\pm}+ m$.

The $B$-cycles for the spectral curve $\Gamma$ can be chosen as
paths from $k_i^+$ to $k_i^++ m$. Thus the dual periods
$a_{Di}$ are given by
\be\label{aD}
a_{Di}={1\over 2\pi i}\int_{k_i^+}^{k_i^+ +m}d\lambda
={1\over 2\pi i}\int_{k_i^+}^{k_i^+ + m}
k\,d\ln\, w
\ee
The most difficult step in the evaluation of (\ref{aD}) is
how to express the differential $d\, \ln w$ in terms of $dk$.
For the periods $a_i$, this step was easy, because the $A$-cycles
could be kept at fixed distance from $k_i$. Here, the $B$-cycles
end at $k_i^+$, which is at distance $q^{1/2}\to 0$ from $k_i$,
in view of (\ref{turning}). Thus the expansion (\ref{F}) for $w$ does 
not apply in the
present case. However, it turns out that an approximation for $w$
as a function of $k$ can be found to order $O(q)$. It is given by
\be
w={H(k)\over H(k- m)}\times
{1+\sqrt{1-4q\eta_1 ^+ (k)}
\over 
1+\sqrt{1-4q\eta_1 ^-(k-m)}}
\ee
where the functions $\eta_1 ^+(k)$ and $\eta_1 ^-(k-m)$
are defined in (\ref{F}).
Expanding in $q$, the differential $d\lambda$ can now be expressed
in terms of rational functions of $k$. The same analytic continuation
methods used in the case of the fundamental representation
apply, giving the leading terms of $a_{Di}$. The perturbative part
${\cal F}^{({\rm pert})}$ can be read off from $a_{Di}={\partial {\cal F}
\over\partial a_i}$, establishing (a).

\medskip

The perturbative corrections to the prepotential are of course 
predicted by asymptotic freedom. The formulas
in (b) for the instanton corrections 
$\F^{(1)}$ and $\F^{(2)}$ are new, as they
have not yet been computed by direct field theory methods.
Perturbative expansions of the prepotential in powers
of $m$ have also been obtained in \cite{minahan}.

\medskip

The moduli $k_i$, $1\leq i\leq N$, of the gauge theory are 
evidently integrals of motion
of the system. To identify these integrals of motion, denote by $S$ be
any subset of $\{1,\cdots,N\}$, and let $S^*=\{1,\cdots,N\}\setminus S$,
$\wp(S)=\wp(x_i-x_j)$ when $S=\{i,j\}$. Let also $p_S$ denote
the subset of momenta $p_i$ with $i\in S$. We have \cite{dp2}

\medskip

\noindent
{\bf Theorem 7}. {\it For any $K$, $0\leq K\leq N$,
let $\sigma_K(k_1,\cdots,k_N)=
\sigma_K(k)$ be the $K$-th symmetric polynomial of $(k_1,\cdots,k_N)$,
defined by $H(u)=\sum_{K=0}^N(-)^K\sigma_K(k)u^{N-K}$. Then}
\be
\sigma_K(k)
=
\sigma_K(p)
+
\sum_{l=1}^{[K/2]}m^{2l}
\sum_{|S_i\cap S_j|=2\delta_{ij}\atop 1\leq i,j\leq l}
\sigma_{K-2l}(p_{(\cup_{i=1}^lS_i)^*})
\prod_{i=1}^l[\wp(S_i)+{\eta_1\over\omega_1}]
\ee

\medskip
The proof of Theorem 7 requires some new elliptic function
identities, linking combinations of the function $\Phi(x,z)$
with determinants of the $\bar\delta$-operator on the torus
\cite{determinant}. These identities indicate a close
relationship between elliptic Calogero-Moser systems and
free fermions.

\subsection{Partial Decoupling of Hypermultiplet, Product Groups}

The spectral curves of certain gauge theories can be easily derived
from the Calogero-Moser curves by a partial decoupling of
the hypermultiplet. Indeed, 

\begin{itemize}

\item the masses of the gauge multiplet
and hypermultiplet are $|a_i-a_j|$ and $|a_i-a_j+m|$. In suitable limits,
some of these masses become $\infty$, and states with infinite mass decouple.
The remaining gauge group is a subgroup of $SU(N)$. 

\item When the effective coupling of a gauge subgroup
is $0$, the dynamics freeze and the gauge states become non-interacting.

\end{itemize}

Non-trivial decoupling limits arise when $\tau\to\infty$ and $m\to \infty$.
When all $a_i$ are finite, we obtain the pure Yang-Mills theory.
When some hypermultiplets masses remain finite, the $U(1)$
factors freeze, the gauge group $SU(N)$ is broken down to
$SU(N_1)\times\cdots\times SU(N_p)$,
and the remaining hypermultiplets are in e.g. fundamental or
bifundamental representations. For example,
let $N=2N_1$ be even, and set
\be
k_i=v_1+x_i,\ \
k_{N_1+j}=v_2+y_j,
\ \
1\leq i,j\leq N_1,
\ee
with $\sum_{i=1}^{N_1}x_i=\sum_{j=1}^{N_1}y_j=0$. (The term $v=v_1-v_2$
is associated to the $U(1)$ factor of the gauge group).
In the limit $m\to \infty$, $q\to 0$, with $x_i$, $y_j$, $\mu=v-m$
and $\Lambda=mq^{1\over N}$ kept fixed,
the theory reduces to a $SU(N_1)\times SU(N_1)$ gauge theory,
with a hypermultiplet in the bifundamental
$(N_1,\bar N_1)\oplus (\bar N_1,N_1)$, and
spectral curve
\be
A(x)-t(-)^{N_1}B(x)
-2^{N_1}\Lambda^{N_1}({1\over t}-t^2)=0,
\ee
where $A(x)=\prod_{i=1}^{N_1}(x-x_i)$, $B(x)=\prod_{j=1}^N(x+\mu-y_j)$,
$t=e^z$. This agrees with the curve found by Witten \cite{witten}
using M Theory, and by Katz, Mayr, and Vafa \cite{katz} using
geometric engineering.

\medskip
The prepotential of the $SU(N_1)\times SU(N_1)$ theory can be
also read off the Calogero-Moser prepotential. It is convenient to
introduce $x_i^{(I)}$, $I=1,2$, by $x_i^{(1)}=x_i$, $x_i^{(2)}=y_i$,
$1\leq i\leq N_1$. Set
\be
A_i^I=\prod_{j\not=i\atop j\in I}(x-x_j^{(I)}),
\ \
B^I(x)=\prod_{j\in J\atop |I-J|=1}(\mu\pm(x-x_j^{(J)})),
\ \
S_i^I(x)={B^I(x)\over A_i^I(x)^2},
\ee
where the $\pm$ sign in $B^I(x)$ is the same
as the sign of $J-I$. Then the the first two
orders of instanton corrections to the prepotential
for the $SU(N_1)\times SU(N_1)$ theory are given by
\bea
\F_{SU(N_1)\times SU(N_1)}^{(1)}
&=&
{(-2\Lambda)^{N_1}\over 2\pi i}\sum_{I=1,2}\sum_{i\in I}S_i^I(x_i^{(I)})\\
\F_{SU(N_1)\times SU(N_1)}^{(2)}
&=&
{(-2\Lambda)^{2N_1}\over 8\pi i}\sum_{I=1,2}\sum_{i\in I}S_i^I(x_i^{(I)})
{\partial^2S_i^I(x_i^{(I)})\over\partial x_i^{(I)2}}
+
\sum_{i\not=j\atop i,j\in I}
{S_i^I(x_i^{(I)})S_j^I(x_j^{(I)})\over (x_i^{(I)}-x_j^{(I)})^2}.
\nonumber
\eea

These formulas can serve as useful checks on candidates
for Seiberg-Witten curves for theories with product
gauge groups obtained by other methods
\cite{witten, feichtinger}.

\vfill\eject

\setcounter{equation}{0}
\section{Calogero-Moser and Seiberg-Witten for General G}

We consider now the $\N=2$ supersymmetric gauge theory for
a general simple gauge algebra $\G$ and 
a hypermultiplet of mass $m$ in the adjoint representation.

\subsection{The General Case}

Then, we have the following results, established in  \cite{[9]}

\medskip

$\bullet$ the Seiberg-Witten curve of the theory
is given by the spectral curve $\Gamma=\{(k,z)\in{\bf C}\times\Sigma;
\det(kI-L(z))=0\}$ of the {\it twisted} elliptic
Calogero-Moser system associated to the Lie algebra $\G$.
The Seiberg-Witten differential $d\lambda$ is given by $d\lambda=kdz$.

\medskip

$\bullet$ The function $R(k,z)=\det(kI-L(z))$ is polynomial
in $k$ and meromorphic in $z$. The spectral curve $\Gamma$
is invariant under the Weyl group of $\G$. It depends
on $n$ complex moduli, which can be thought
of as independent integrals of motion of the Calogero-Moser system.

\medskip

$\bullet$
The differential $d \lambda=kdz$ is meromorphic on $\Gamma$,
with simple poles. The position and residues of the poles are
independent of the moduli. The residues are linear
in the hypermultiplet mass $m$. (Unlike the case of
$SU(N)$, their exact values
are difficult to determine for general $\G$).

\medskip

$\bullet$ In 
the $m \to 0$ limit, the Calogero-Moser system reduces to
a free system, the spectral curve $\Gamma$ is just the producr
of several unglued copies of the base torus $\Sigma$,
indexed by the constant eigenvalues of $L(z)=p\cdot h$.
Let $k_i$, $1\leq i\leq n$, be
$n$ independent eigenvalues, and $A_i,B_i$ be the $A$
and $B$ cycles lifted to the corresponding sheets.
For each $i$, we readily obtain 
\bea
a_i & = & {1\over 2\pi i}\oint_{A_i}d\lambda
={k_i\over 2\pi i}\oint_Adz={2\omega_1\over 2\pi i}k_i
\nonumber \\
a_{Di} & = & {1\over 2\pi i}\oint_{B_i}d\lambda
={k_i\over 2\pi i}\oint_Bdz={2\omega_1\over 2\pi i}\tau k_i
\eea
Thus the prepotential $\F$ is given by
$\F={\tau\over 2}\sum_{i=1}^na_i^2$. This is the classical prepotential
and hence the correct
answer, since in the $m\to 0$ limit, the theory
acquires an $\N=4$ supersymmetry, and receives no
quantum corrections.

\medskip

$\bullet$ The $m\to\infty$ limit is the crucial consistency check,
which motivated the introduction of the {\it twisted} Calogero-Moser
systems in the first place \cite{cm1,cm2}. In view of Theorem 2
and subsequent comments, in the limit $m\to\infty$, $q\to 0$,
with 
\bea 
x & = & X+2\omega_2{1\over h_{\G}^{\vee}}\rho \nonumber \\
m & = & Mq^{-{1\over 2h_{\G}^{\vee}}}
\eea
with $X$ and $M$ kept fixed,
the Hamiltonian and spectral curve for the twisted
elliptic Calogero-Moser system with Lie algebra $\G$
reduce to the Hamiltonian and spectral curve for the
Toda system for the affine Lie algebra $(\G^{(1)})^{\vee}$.
This is the correct answer. Indeed,
in this limit, the gauge theory with adjoint hypermultiplet
reduces to the pure Yang-Mills theory,
and the Seiberg-Witten spectral curves for
pure Yang-Mills with gauge algebra $\G$ have been shown by
Martinec and Warner \cite{martinec} to be
the spectral curves of the Toda system
for $(\G^{(1)})^{\vee}$.

\medskip

$\bullet$ As in the known correspondences between Seiberg-Witten theory
and integrable models \cite{dp1}, we expect the following equation
\be
{\partial \F\over\partial\tau}=H_{\G}^{twisted}(x,p),
\ee
to hold. Note that the left hand side can be
interpreted in the gauge theory as a renormalization group equation.

\medskip

$\bullet$ For simple laced $\G$, the curves $R(k,z)=0$ are modular
invariant. Physically, the gauge theories for these Lie algebras
are self-dual. For non simply-laced $\G$,
the modular group is broken to the congruence subgroup
$\Gamma_0(2)$ for $\G=B_n,C_n$, $F_4$, and to $\Gamma_0(3)$
for $G_2$. The Hamiltonians of the twisted Calogero-Moser systems
for non-simply laced $\G$ are also transformed under Landen
transformations into the Hamiltonians of the twisted Calogero-Moser system
for the dual algebra $\G^{\vee}$. It would be interesting to determine whether
such transformations exist for the
spectral curves or the corresponding gauge theories themselves.

\bigskip

Spectral curves for certain gauge theories with classical gauge algebras and
matter in the adjoint representation have also
been proposed in \cite{witten} and \cite{katz}, based on branes and
M-theory.
Some generalizations of the construction by Witten \cite{witten} for
$SU(N)$ were given in \cite{m}, \cite{uranga} and \cite{yokono}. Relations between
the dynamics of gauge theories from branes and integrable systems were
proposed and analyzed in \cite{cherkis}.
A possible role in Seiberg-Witten theory
for Ruijsenaars-Schneider and related relativistic
systems has been investigated in \cite{ohta}.

\subsection{Spectral Curves for Low Rank}

In the case of $\G=D_n$, the trigonometric limit ($q\to 0$)
of the spectral curve $R(k,z)=0$ can be written down
explicitly and takes a particularly simple form (see
(\ref{trigo}) below). Now the equation $R(k,z)=0$ for the spectral curve
is polynomial in $p_j$, $\wp$, $g_2$, $g_3$, and $\Delta$,
where $g_2$, $g_3$ are the usual Eisenstein series of weights
4 and 6, and $\Delta=g_2^3-27g_3^2$ is the discriminant.
In the limit $q\to 0$, only $\Delta$ vanishes.
Since the weight of $\Delta$ is $12$
and the degree of the spectral curves is $2n$,
the spectral curves are determined by their trigonometric limit
when $n\leq 5$.

\medskip

The derivation of the trigonometric limit is as follows.
Let $Z$ be the spectral parameter defined by
\be\label{zZ}
{1\over Z}=
{1\over 2}{\rm coth}\,{z\over 2}
-{1\over z}
\ee
Then in the trigonometric limit $\tau\to i\infty$, the elliptic
functions $\wp(z)$ and $\Phi(x,z)$ reduce to
\bea
\wp(z)&\to &{1\over Z^2}-{1\over 6}\\
\Phi(x,z)&\to &{1\over 2}{\rm coth}\,{x\over 2}
-{1\over Z}
\eea
Now the spectral curve $\Gamma$ depends on the dynamical
variables $p_i,x_i$, but only through $n$ combinations
$u_i=u_i(m)$. At $m=0$, $u_i(m)$ are just the $n$ independent
Casimirs of $\G$, which are polynomials in $p$
of degrees $\gamma_i+1$. As $m$ is deformed away from $0$,
the $u_i(m)$ are still recognizable by their leading $p$
behavior. Thus we may carry out our calculations with
any choice of the variables $x_i$. A particularly convenient choice
is
\be
x=\xi\rho^{\vee},
\quad
\a\cdot x=\xi\, l(\a),
\quad \xi\to\infty.
\ee
Then
\be
\P(\a\cdot x,z)\to -{1\over Z}+\cases{+{1\over 2}, &if $\a>0$\cr
-{1\over 2}, &if $\a<0$.\cr}
\ee
Let $P={\rm diag}(p_1,\cdots,p_n)$, $\mu=\mu^++\mu^-$, with
$\mu^{\pm}$ given by
\be
\mu_{ij}^+=\cases{1, &if $i<j$\cr
0, &if $i\geq j$\cr}
\quad\quad
\mu_{ij}^-=\cases{1, &if $i>j$\cr
0, &if $i\leq j$\cr}
\ee
Then the function $R(k,z)$ can be expressed as
\be
R(k,z)
=
{\rm det}
\pmatrix{kI-P+{m\over Z}\mu-{m\over 2}(\mu^+-\mu^-)
& ({m\over Z}-{m\over 2})\mu\cr
({m\over Z}+{m\over 2})\mu
&
kI+P+{m\over Z}\mu+{m\over 2}(\mu^+-\mu^-)\cr}
\ee
By taking column and row linear combinations, we find
\be
R(k,z)
=
{\rm det}[(kI+P-m\mu^-)(kI-P-m\mu^+)
+k(m+2{m\over Z})\mu]
\ee
Although each of the two factors in the above determinant is triangular,
the determinant is still difficult to evaluate due to the
presence of the third term $k(m+{m\over Z})\mu$.
It is here that we must introduce the rank $1$
matrix $\mu+I$, and make a shift similar to the key one in 
the study of $SU(N)$ Calogero-Moser systems: defining
a new variable $A$ by
\be\label{kA}
0=A^2+mA+2k{m\over Z}-k^2
\ee
we can write
\be
R(k,z)={\rm det}[(AI+P-m\mu^-)(AI-P-m\mu^-)
+(mA+2k{m\over Z})(\mu+I)]
\ee
Since $\mu+I=uu^T$ with $u^T=(1,\cdots,1)$, and 
${\rm det}(M+uu^T)={\rm det}\, M (1+u^TM^{-1}u)$
for any invertible matrix $M$, () can be expressed as
\be
R(k,z)
=
\prod_{j=1}^n(A^2-p_j^2)
+(mA+2{m\over Z})\sum_{j=1}^n\prod_{i=1}^{j-1}((A+m)^2-p^2)
\prod_{i=j+1}^n(A^2-p_i^2)
\ee
In analogy with the $SU(N)$ case, we introduce a polynomial $H(A)$ by
\be
H(A)=\prod_{j=1}^n(A^2-p_j^2)=\sum_{j=0}^n(-1)^{n-j}A^{2j}u_{2n-2j}.
\ee
The final expression for $R(k,z)$ is in terms of $H(A)$
\be\label{trigo}
R(k,z)
=
{m^2+mA-2k{m\over z}\over m^2+mA}H(A)
+
{mA+2k{m\over Z}\over m^2+2mA}H(A+m).
\ee
The similarity with the equation (\ref{H(k)}) for the $SU(N)$ Calogero-Moser
spectral curves is now manifest.
Just as in the case of $SU(N)$, the parameters $p_j$ can be
identified with the classical vacuum parameters of the gauge theory.

\medskip

We derive now explicit formulas for $R(k,z)$ for $n\leq 5$.
Let
\be
R(k,z)=\sum_{j=0}^n(-1)^{n-j}P_{2j}u_{2n-2j}
\ee
Then (\ref{trigo}) implies the following recursive relation
\be\label{trigoterms}
0=P_{2(j+1)}
-(2k^2+m^2
-4k{m\over Z})P_{2j}
+k^2(k-2{m\over Z})^2P_{2(j-1)}
\ee
with $P_0=1$, $P_2=k^2$. This works out to
\bea
P_4&=&k^4-4k^2{m^2\over Z^2}+m^2k^2\\
P_6&=&k^6-12k^4{m^2\over Z}^2+16k^3{m^3\over Z^3}
-4k^3{m^3\over Z}-4k^2{m^4\over Z^2}
+3k^4m^2+k^2m^4\\
P_8&=&k^8-24k^6{m^2\over Z^2}+6k^6m^2+64k^5{m^3\over Z^3}-16k^5{m^3\over Z}
-48k^4{m^4\over Z^4}\\
&\quad&-8k^4{m^4\over Z^2}+5k^4m^4
+32k^3{m^5\over Z^3}
-8k^3{m^5\over Z}
-4k^2{m^6\over Z^2}+k^2m^6\\
P_{10}&=&k^{10}-40k^8{m^2\over Z^2}+160 k^7{m^3\over Z^3}
-240 k^6{m^4\over Z^4}+128 k^5{m^5\over Z^5}\\
&\quad&+m^2[10k^8-40 k^7{m\over Z}+160k^5{m^3\over Z^3}
-160k^4{m^4\over Z^4}]\\
&\quad&+m^4[15k^6-48k^5{m\over Z}+12k^4{m^2\over Z^2}
+48k^3{m^3\over Z^3}]\\
&\quad&+m^6[7k^4-12k^3{m\over Z}-4k^2{m^2\over Z^2}]+m^8k^2.
\eea
Let now $0=R(k,z)=\sum_{j=0}^nQ_{2j}(k)u_{2n-2j}$ be the equation
of the spectral curve for general $q$. Using the limit of $\wp$
and its derivatives, we may identify the functional dependence
on $\wp(x)$ which gave rise to each of the terms in (\ref{trigoterms}).
The result is
\bea
Q_0&=&1\\
Q_2&=&k^2\\
Q_4&=&k^4-4k^2m^2\wp\\
Q_6&=&k^6-12k^4m^2\wp-8k^3m^3\wp'\\
Q_8&=&k^8-24k^6m^2\wp-32k^5m^3\wp'-48k^4m^4\wp^2+64g_2k^2m^6\wp\\
Q_{10}&=&k^{10}-40k^8m^2\wp-80k^7m^3\wp'-240k^6m^4\wp^2
-64k^5m^5\wp\wp'\\
&\quad&+704g_2k^4m^6\wp+512g_2k^3m^7\wp'-768k^2m^8g_3\wp.
\eea
This completes our derivation of the spectral curves for $D_n$
for $n\leq 5$.

\subsection{Perturbative Prepotential for SO(2n)}

With the parametrization (\ref{trigo}) in terms of the polynomial
$H(A)$, it is now easy to evaluate the logarithmic terms of
the prepotential ${\cal F}$. We have already shifted the variable
$k$ to the variable $A$ defined by (\ref{kA}). A complete shift
of the variables $(k,z)$ to new variables $(A,u)$ is obtained by
setting
\be\label{zu}
e^u={k^2-(A+m)^2\over k^2-A^2}
\ee
The equation () of the curve becomes
\be
e^u={H(A+m)\over H(A)}.
\ee
It remains to determine the Seiberg-Witten differential $d\lambda
=kdz$ in terms of $A$ and $u$. Evidently, $k$ can be solved in terms
of $A$ and $u$ using (\ref{zu}). To write $z$ in terms of $A$ and $u$,
we note that $z$ can be expressed in terms of $Z$ by (\ref{zZ}), which can
be solved in terms of $A$ and $k$ using (\ref{kA}). The final outcome is
\bea
e^z&=&{{2\over Z}+1\over{2\over Z}-1}=
{(k-A)(k+A+m)\over (k+A)(k-A-m)}\\
d\lambda
&=&-Adu-md\ln (k^2-(A+m)^2)
\eea
This is now essentially the same set-up as the spectral curves
and Seiberg-Witten differential for the $SU(N)$ Calogero-Moser system. 
The same methods used earlier
give immediately the logarithmic terms in the prepotential
\bea
{\cal F}^{\rm 1-loop}
=
-{1\over 8\pi i}
\sum_{\alpha\in{\cal R}(D_r)}
(\alpha\cdot a)^2\ln (\alpha\cdot a)^2
-(\alpha\cdot a+m)^2\ln (\alpha\cdot a+m)^2
\eea 
These are the logarithmic singularities expected
from field theory considerations.

\section*{Acknowledgments}

These notes are based on lectures delivered by E.D. at the 1999
Banff Summer School, and by D.H.P. at the meeting
{\it Mathematics from Physics} in May 1999
at the University of Illinois, Urbana-Champaign. 
The authors would like to thank respectively Jacques Hurtubise, Yvan 
Saint-Aubin, Luc Vinet for their invitation to lecture at Banff,
and Steve Bradlow, John D'Angelo, Robert Leigh,
and Mike Stone for their invitation to
lecture at Urbana. They would like to express their
appreciation for the very warm hospitality extended to
them at each place. E.D. also acknowledges the 
Aspen Center for Physics and the
Laboratoires de Physique Th\' eorique at Ecole Polytechnique and at Ecole
Normale Sup\' erieure, where part of these notes were drafted. 
The research of
E.D. is supported in part by NSF Grants No. PHY-95-31023 and
PHY-98-19686, as well as by the Centre National de Recherche
Scientifique (CNRS). The
work of D.H.P. is supported in part by the NSF grant DMS-98-00783.

\vfill\eject

\setcounter{equation}{0}
\section{Appendix A : Notations and Conventions}

Vector indices run over the following values $ \mu ,~\nu ,~\cdots= 0,1,2,3$ and
$i, ~ j, ~\cdots = 1,2,3$. The flat Minkowski space-time metric is given by
$-\eta _{00} =\eta _{11} =\eta _{22} =\eta _{33} =1$,
which is invariant under translations and under the Lorentz group $SO(1,3)$.
Three- and four dimensional totally anti-symmetric Levi-Civita symbols are
defined by
\be
\epsilon _{ijk} = \epsilon ^{ijk}, 
\qquad 
\epsilon ^{123}=1;
\qquad \qquad
\epsilon _{\mu \nu \rho \sigma} =- \epsilon ^{\mu \nu \rho \sigma},
\qquad \epsilon ^{0123}=1\, .
\ee
The Poincar{\'e} dual of an anti-symmetric rank 2 tensor is defined by
\be
\tilde F_{\mu \nu} = \12 \epsilon_{\mu \nu \rho \sigma} F^{\rho \sigma}
\ee
and the duality operation squares to minus the identity : $\tilde{\tilde
F}=-F$. 

\subsection{Spinors}

\medskip

\noindent
The {\it Pauli matrices} are defined by
\be
\sigma ^0 = \pmatrix{-1 &  0 \cr 0 & -1 \cr}
\qquad
\sigma ^1 = \pmatrix{ 0 &  1 \cr 1 & 0  \cr}\qquad
\sigma ^2 = \pmatrix{ 0 & -i \cr i & 0  \cr}\qquad
\sigma ^3 = \pmatrix{ 1 & 0 \cr  0 & -1 \cr}
\ee
and their ``conjugates" are given by $ \bar \sigma ^0 = \sigma ^0$ and 
$\bar \sigma ^i =- \sigma ^i$.
The {\it Clifford-Dirac} $\gamma$ matrices obey the Clifford algebra relations
\be
\{ \gamma ^\mu , \gamma ^\nu\} = \gamma ^\mu \gamma ^\nu 
+ \gamma ^\nu  \gamma
^\mu  = - 2 \eta ^{\mu \nu} I
\ee
with $\gamma ^0$ anti-Hermitian and $\gamma ^i$ Hermitian. The matrix $\gamma
_5$ obeys
\be
\{\gamma _5, \gamma ^\mu\}=0 \qquad \qquad \gamma _5 = \gamma ^5 = (\gamma _5
)^\dagger = i \gamma ^0 \gamma ^1 \gamma ^2
\gamma ^3 \qquad \qquad (\gamma _5)^2 = I\, .
\ee
Representation matrices for the (reducible) spinor representation of the
Lorentz group $SO(1,3)$ are defined by
\be
\Sigma ^{\mu \nu}= {i \over 2} [\gamma ^\mu , \gamma ^\nu ]
\qquad {\rm with} \qquad
[\gamma _5 , \Sigma ^{\mu \nu}]=0 \, .
\ee
The {\it charge conjugation} matrix $C$ is defined by
\be
C\gamma _\mu C^{-1} =  - (\gamma _\mu )^T 
\ee
which implies the useful identities
\be
C\Sigma _{\mu \nu} C^{-1} =  - (\Sigma _{\mu \nu} )^T\, , 
\qquad
C\gamma _\mu \gamma _5 C^{-1} =   (\gamma _\mu \gamma _5 )^T\, , 
\qquad
C\gamma _5 C^{-1} =   (\gamma _5 )^T 
\ee

\subsection{Dirac matrices in a Weyl basis}

In a Weyl basis $\gamma ^5$ is chosen to be diagonal; this basis is convenient
for chiral fermions.
\be
\gamma ^\mu = \pmatrix{ 0  & \sigma ^\mu \cr \bar \sigma^\mu  & 0 \cr }, 
\qquad \qquad 
\gamma _5 = \pmatrix{ 1  & 0 \cr 0  & -1 \cr } \qquad \qquad 
C =i \gamma ^2 \gamma ^0
\ee

\subsection{Dirac matrices in a Majorana basis}

In a Majorana basis all $\gamma ^\mu$ matrices are chosen to be real. This basis
is particularly convenient when dealing with Majorana fermions, or when
converting from the two-component notation. 
\bea
\gamma ^0 = \pmatrix{ 0  & i\sigma ^2 \cr i\sigma ^2   & 0 \cr }
\quad \
\gamma ^1 = \pmatrix{ -\sigma ^3  &  0 \cr  0 & -\sigma ^3 \cr } 
& \quad &
\gamma ^2 = \pmatrix{ 0  & i\sigma ^2 \cr -i\sigma^2  & 0 \cr }
\quad \
\gamma ^3 = \pmatrix{ \sigma ^1  & 0 \cr 0  & \sigma ^1 \cr } 
\nonumber \\
\gamma ^5 = \pmatrix{ \sigma ^2  & 0 \cr 0  & - \sigma ^2 \cr }
& \qquad &
C=\pmatrix{ 0 & -i\sigma ^2 \cr i\sigma ^2  & 0 \cr  } 
\eea

\subsection{Two-Component Spinors}

The correspondence between two-component and four-component spinors will be
made in the Weyl basis, given previously. Left-handed spinors transform
under the $(1/2,0)$ representation of the Lorentz group and are denoted by
$\chi _\alpha$ or their transpose by $\chi ^\alpha$. Right-handed spinors
transform under the $(0, 1/2)$ representation of the Lorentz group and are
denoted by $\bar \psi ^{\dot \alpha}$ or their transpose by $\bar \psi _{\dot
\alpha}$. One has the following relations 
\bea
\chi ^\alpha & = & \epsilon ^{\alpha \beta} \chi _\beta \qquad \qquad 
\chi _\alpha = \epsilon _{\alpha \beta} \chi ^\beta 
\nonumber \\
\bar \psi ^{\dot \alpha} & = & \epsilon ^{\dot \alpha \dot \beta} \bar \psi
_{\dot \beta} 
\qquad \qquad 
\bar \psi _{\dot \alpha} = \epsilon _{\dot \alpha \dot \beta} \bar \psi ^{\dot
\beta} \eea
with the following conventions for $\epsilon$ :
\bea
\epsilon ^{12} & = & \epsilon _{21}= -\epsilon ^{21}=-\epsilon _{12}=1,
\qquad
\epsilon ^{11}=\epsilon ^{22}=0; \qquad \qquad \epsilon _{\alpha \beta}
\epsilon ^{\beta \gamma} = \delta _\alpha {}^{\gamma} 
\nonumber \\
\epsilon ^{\dot 1 \dot 2} & = & \epsilon _{\dot 2 \dot 1}= - \epsilon
^{\dot 2 \dot 1}=-\epsilon _{\dot 1 \dot 2}=1, \qquad
\epsilon ^{\dot 1 \dot 1}=\epsilon ^{\dot 2 \dot 2}=0, 
\qquad \qquad 
\epsilon _{\dot \alpha \dot\beta}
\epsilon ^{\dot \beta \dot \gamma} = \delta _{\dot \alpha} {}^{\dot \gamma}
\eea
{\it Contraction conventions} are 
\bea
\psi \chi = \psi ^\alpha \chi _\alpha, 
\qquad & \qquad & \qquad 
\bar \psi \bar \chi = \bar \psi _{\dot \alpha } \bar \chi ^{\dot \alpha}
\nonumber \\
\chi \sigma ^m \bar \psi = \chi ^\alpha \sigma ^m _{\alpha \dot \alpha} \bar
\psi ^{\dot \alpha}, \qquad & \qquad & \qquad
\bar \chi \bar \sigma ^m \psi = \bar \chi _{\dot \alpha} \bar \sigma ^{m \dot
\alpha \alpha} \psi _{\alpha} 
\eea
{\it Transposition Identities}
\bea
\psi \chi = \chi \psi 
\qquad & \qquad & \qquad 
(\chi \psi )^\dagger = \bar \chi \bar \psi = \bar \psi \bar \chi 
\nonumber \\
\chi \sigma ^m \bar \psi = - \bar \psi \bar \sigma ^m \chi, 
\qquad & \qquad  & \qquad
(\chi \sigma ^m \bar \psi )^\dagger = \psi \sigma ^m \bar \chi 
\nonumber \\
\chi \sigma ^m \bar \sigma ^n \psi = \psi \sigma ^n \bar \sigma ^m \chi
\qquad & \qquad  & \qquad 
(\chi \sigma ^m \bar \sigma ^n \psi ) ^\dagger = \bar \psi \bar
\sigma ^n \sigma ^m \bar \chi 
\eea
{\it Fierz Identity}
\be
(\psi \varphi ) \bar \chi _{\dot \beta} = -{1 \over 2} (\varphi \sigma ^m \bar
\chi) (\psi \sigma _m) _{\dot \beta} 
\ee
{\it Relating Four-component Spinors and Two-component Spinors} in a Weyl basis
\bea
{\rm Left ~Weyl} \quad : \quad \Psi _L  =  \pmatrix{ \chi
_\alpha \cr 0 \cr}
& \qquad & 
 {\rm Dirac} \quad : \quad \Psi _D  =  \pmatrix{\chi
_{ \alpha} \cr \bar \psi ^{\dot \alpha} \cr } 
\nonumber \\
{\rm Right ~Weyl} \quad : \quad \Psi _R  =  \pmatrix{0 \cr \bar \psi ^{\dot
\alpha}\cr }
& \qquad & 
{\rm Majorana} \quad : \quad \Psi _M  =  \pmatrix{ \psi _{\alpha} \cr \bar \psi
^{\dot \alpha} \cr }
\eea
{\it Transposition identities in four-component form}
\bea
\bar \psi _1 \Gamma \psi _2 = \bar \psi _2 \tilde \Gamma \psi _1 \qquad
{\rm with} \qquad 
\Gamma & = & +\tilde \Gamma \qquad \Gamma = 1,~\gamma _5,
\gamma ^m \gamma _5 
\nonumber \\
\Gamma & = & -\tilde \Gamma \qquad \Gamma = \gamma ^m, ~\gamma _{[\mu } \gamma
_{\nu ]} 
\eea

\vfill\eject

\setcounter{equation}{0}
\section{Appendix B : Lie Algebra Theory}

In Figure 1, we give the Dynkin diagrams for the finite dimensional simple Lie
algebras; for the untwisted affine Lie algebras (left
column) and for the twisted affine Lie algebras (right column). The simple
roots are labeled following Dynkin notation, and are given in an orthonormal
basis in 
Table 9, 
where we also list the dimension, the Coxeter and dual
Coxeter numbers (to be defined below). We list the set of all roots in Table 
\ref{table:10}, and of the highest roots in Table \ref{table:11}. 
Below we provide additional notations 
and definitions \cite{olive}.

Let $\G$ be one of the finite dimensional simple Lie algebras of rank $n$,
let $\alpha _i$, and $\alpha _i ^\vee \equiv 2\alpha _i /\alpha _i ^2$,
$i=1,\cdots, n$ be its simple roots and coroots respectively. 
The coroot $\alpha ^\vee$ of any root is defined by $\alpha ^\vee = 2 \alpha 
/\alpha ^2$. Any (co-)root admits a unique decomposition into a sum of simple 
(co-)roots, with integer coefficients $l_i$ and $l_i ^\vee$.
\be
\alpha = \sum _{i=1} ^n l_i \alpha _i
\qquad \qquad
\alpha ^\vee = \sum _{i=1} ^n l_i ^\vee \alpha _i ^\vee.
\ee
The coefficients $l_i$ and $l_i ^\vee$ are either all positive or all 
negative 
according to whether $\alpha$ (or $\alpha ^\vee$) is positive or negative 
respectively. They are related by
\be
l_i ^\vee = { \alpha _i ^2 \over \alpha ^2} l_i,
\qquad i=1,\cdots,n.
\ee
The {\it highest root} $\alpha _0$ and co-root $\alpha _0 ^\vee$ play special 
roles. The extension of the simple root system of an algebra $\G$ by $\alpha 
_0$ generates the {\it untwisted affine Lie algebra} $\G ^{(1)}$, while the 
extension of the simple coroot system of $\G$ by $\alpha _0 ^\vee$ generates 
the {\it dual affine Lie algebra} $(\G ^{(1)})^\vee$. When $\G$ is non-simply 
laced, $(\G ^{(1)})^\vee$ coincides with one of the {\it twisted affine Lie 
algebras}. The Dynkin diagrams of these various Lie algebras are given in 
Table 
1. The decompositions of $\alpha _0$ and $\alpha _0 ^\vee$ onto roots or 
coroots 
\be
\alpha _0 
= \sum _{i=1} ^n a_i \alpha _i 
\qquad \qquad 
\alpha _0 ^\vee
= \sum _{i=1} ^n a _i ^\vee  \alpha _i ^\vee.
\ee
define the {\it marks} $a_i$ and the {\it comarks} $a_i ^\vee$, which are 
given  in Table 4.
The {\it Coxeter number} $h_\G$ and the {\it dual Coxeter number} $h_\G 
^\vee$ are defined by
\be
h_\G = 1 + \sum _{i=1} ^n a_i
\qquad \qquad
h _\G ^\vee = 1 + \sum _{i=1} ^n a_i ^\vee,
\ee
and their values are given in Table 9.
For simply laced Lie algebras, for which all roots have the same length 
(normalized to $\alpha _i ^2=2$), we have $a_i ^\vee = a_i$ and $h_\G = h_\G 
^\vee$. The dual Coxeter number equals the {\it quadratic Casimir} 
operator in the adjoint representation, $h_\G ^\vee = C_2 (\G)$.

\vfill\eject

\centerline{\epsfxsize 8.0 truein \epsfbox[-20 80 600 660]{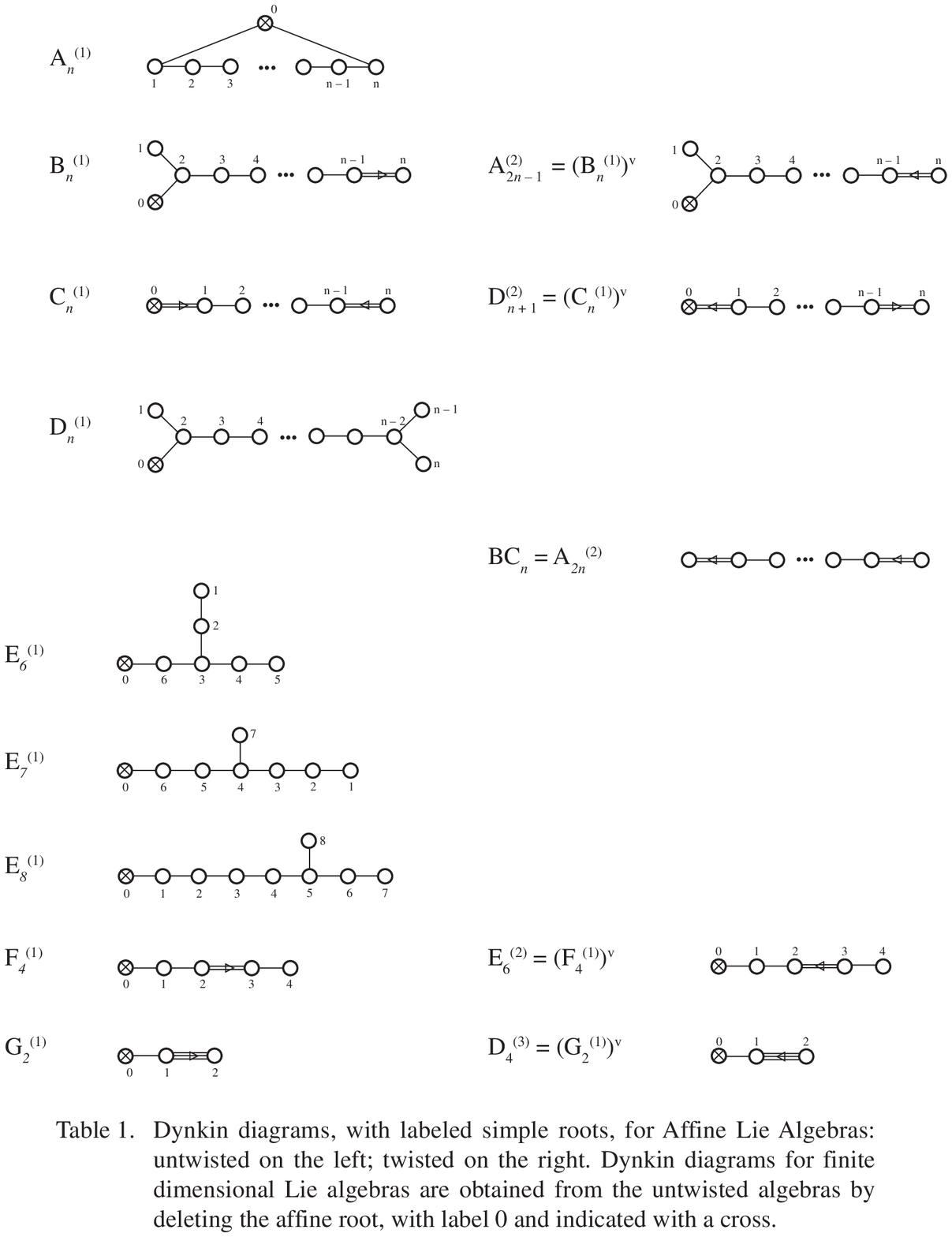}}

\vfill\eject

The highest weight vectors $\lambda _j$, $j=1, \cdots ,n$ of the {\it 
fundamental representations}  (also called fundamental weights) of $\G$ are
defined by
\be
\alpha _i ^\vee \cdot \lambda _j = \delta _{ij}.
\ee
The highest weight vector $\lambda$ of any finite dimensional representation
$\Lambda $ of $\G$ is then uniquely specified by positive or zero integers 
$q_i$, $i=1,\cdots ,n$, with
\be
\Lambda \equiv  (q_1, \cdots, q_n)
\qquad \qquad 
\lambda = \sum _{i=1} ^n q _i \lambda _i
\ee
The Weyl orbit of the highest weight vector of $(q_1, \cdots,
q_n)$ is denoted by $[q_1, \cdots , q_n]$.

\begin{table}[t]
\begin{center}
\begin{tabular}{|c||c|} \hline 
$\G$  & all  roots 
\\ \hline \hline
$A_n$      
   & $\pm (e_i - e_j), 1\leq i< j\leq n+1$ 
\\ \hline
$B_n$   
   & $\pm (e_i - e_j); ~ \pm (e_i + e_j), ~1\leq i<j\leq n; 
    {} ~\pm e_i, ~1\leq i \leq n$ 
\\ \hline
$C_n$       
   & $\pm (e_i - e_j); ~ \pm (e_i + e_j), ~1\leq i<j\leq n; 
    {} ~\pm 2e_i, ~ 1\leq i \leq n$ 
\\ \hline
$D_n$  
   & $\pm (e_i - e_j), ~ \pm (e_i + e_j), 1\leq i<j\leq n$ 
\\ \hline
$E_6$ 
   & $\pm (e_i - e_j), ~ \pm (e_i + e_j), ~1\leq i<j \leq 5;
    \pm \half (\sqrt 3 e_6 + \sum _{i=1} ^5 \epsilon _i e_i),~ \prod _i
    \epsilon _i =+1$
\\ \hline
$E_7$ 
   & $\pm (e_i - e_j), ~ \pm (e_i + e_j), ~1\leq i<j \leq 6;
    \pm \half (\sqrt 2 e_7 + \sum _{i=1} ^6 \epsilon _i e_i),~ \prod _i
    \epsilon _i =-1 $
\\ \hline
$E_8$
   & $\pm (e_i - e_j), ~ \pm (e_i + e_j), ~1\leq i<j \leq 8;
    \half  \sum _{i=1} ^8 \epsilon _i e_i,~ \prod _i
    \epsilon _i =+1$
\\ \hline 
$G_2$  
   & $\pm (e_i - e_j), ~ 1 \leq i<j\leq 3; ~\pm (e_i -{1 \over 3}
(e_1+e_2+e_3)),
    {}~ i=1,2,3$ 
\\ \hline
$F_4$
   & $\pm (e_i - e_j), ~ \pm (e_i + e_j), ~1\leq i<j \leq 4;~
    \pm e_i, ~1\leq i\leq 4;
    \pm \half  \sum _{i=1} ^4 \epsilon _i e_i$
\\ \hline \hline 
\end{tabular}
\end{center}
\caption{Root system of finite dimensional simple Lie algebras}
\label{table:10}
\end{table}

\begin{table}[t]
\begin{center}
\begin{tabular}{|c|| c|c|c|} \hline 
$\G$  & marks $(a_i)$ & comarks $(a_i ^\vee)$ & exponents $\gamma _i$
  \\ \hline
$A_n$ & (1,1,1,$\cdots$,1,1)  & (1,1,1,$\cdots$,1,1)&  1,2,3,$\cdots$ ,n   
  \\ \hline
$B_n$ & (1,2,2,$\cdots$,2,2)  & (1,2,2,$\cdots$,2,1)&  
1,3,5,$\cdots$,2n-1
  \\ \hline
$C_n$ & (2,2,2,$\cdots$,2,1)  & (1,1,1,$\cdots$,1,1)&  1,3,5,$\cdots$,2n-1 
  \\ \hline
$D_n$ & (1,2,$\cdots$,2,1,1)  & (1,2,$\cdots$,2,1,1)&
1,3,5,$\cdots$,2n-3,n-1
  \\ \hline
$E_6$ & (1,2,3,2,1,2)       & (1,2,3,2,1,2)       &  1,4,5,7,8,11 
  \\ \hline
$E_7$ & (2,3,4,3,2,1,2)     & (2,3,4,3,2,1,2)     &  1,5,7,9,11,13,17 
  \\ \hline 
$E_8$ & (2,3,4,5,6,4,2,3)   & (2,3,4,5,6,4,2,3)   &  
1,7,11,13,17,19,23,29
  \\ \hline
$G_2$ & (2,3)               & (2,1)               &  1,5  
  \\ \hline
$F_4$ & (2,3,4,2)           & (2,3,2,1)           &  1,5,7,11
\\ \hline \hline 
\end{tabular}
\end{center}
\caption{Marks, Co-marks and Exponents}
\label{table:11}
\end{table}

The {\it level} ~$l(\lambda)$ and the {\it co-level} ~$l^\vee (\alpha)$ are 
defined by
\bea
l(\lambda ) & = &  \lambda \cdot \rho ^\vee, 
\qquad  \qquad  
l(\alpha _i ) =  ~1, \qquad i=1,\cdots ,n; 
\nonumber \\
l^\vee (\lambda ) & = &  \lambda \cdot \rho,  
\qquad  \qquad  
l ^\vee (\alpha _i ^\vee ) =  ~1, \qquad i=1,\cdots ,n. 
\eea
Here, the {\it level vector} $\rho ^\vee$ is related to the {\it Weyl vector} 
$\rho$  by exchanging weights $\lambda _i$ and coweights $\lambda _i ^\vee =
2\lambda  _i / \alpha _i ^2$. Both are uniquely determined by the above
normalization,  and may be expressed in terms of the fundamental weights and
coweights by
\bea
\rho & = & \sum _{i=1} ^ n \lambda _i = \half \sum _{\alpha \in \R _+ (\G)} 
\alpha
\nonumber \\
\rho ^\vee & = & \sum _{i=1} ^ n \lambda _i ^\vee  = \half \sum _{\alpha ^\vee 
\in \R _+ (\G) ^\vee } \alpha ^\vee.
\eea
Here, we have provided the relation between the Weyl vector and 
the half sum of all positive roots, and its dual relation.
It is clear from (A.3), (A.4) and (A.8) that the Coxeter and dual Coxeter
numbers are related to the level of the highest root and the co-level of the 
highest coroot
\bea
h_\G & =   & ~1+ \alpha _0 \cdot \rho ^\vee = 1+ l(\alpha _0)
\nonumber \\
h_\G ^\vee  & =& ~1+ \alpha _0 ^\vee  \cdot \rho= 1+ l^\vee (\alpha _0 ^\vee).  
\eea
As $\alpha$ (resp. $\alpha ^\vee$) ranges through $\R (\G)$ (resp. $\R (\G) 
^\vee$), the maxima of $l(\alpha)$ and $l^\vee (\alpha ^\vee)$ are $h_\G -1$ 
and 
$h^\vee _\G -1$ respectively.

The {\it exponents} $\gamma _i $, $i=1,\cdots ,n$ are such that the 
numbers $\gamma _i+1$ are the degrees of the independent Casimir operators of 
the algebra $\G$. Their values are also listed in Table 4.

\vfill\eject

\setcounter{equation}{0}
\section{Appendix C : Elliptic Functions}

In this appendix we review some basic definitions and properties of elliptic
functions on an elliptic curve of periods $2\omega _1$ and $2\omega_2$ and
modulus $\tau = \omega _2 / \omega _1$, $\Im \tau >0$. The half periods are
$\omega _1$, $\omega _2$ and $\omega _3 = \omega _1 + \omega _2$. For a useful
source, see \cite{erdelyi}.

\subsection{ Basic Definitions and properties}

The Weierstrass function is defined by
\be
\wp(z; 2\omega _1, 2 \omega _2) \equiv 
{1 \over z^2} + \sum _{{(m_1,m_2) \atop \not=(0,0)}}
\left \{ {1 \over (z + 2 \omega _1 m_1 + 2 \omega _2 m_2) ^2}
- {1 \over (2 \omega _1 m_1 + 2 \omega _2 m_2) ^2} \right \}.
\ee
The function $\wp $ is related to the Weierstrass functions $\zeta$ and
$\sigma$ by
\be
\wp (z; 2 \omega _1, 2 \omega _2) = - {d \over dz} \zeta (z; 2 \omega _1, 2
\omega _2) = - {d^2 \over dz^2} \log \sigma (z; 2 \omega _1, 2 \omega _2).
\ee
These functions satisfy
\bea
\wp (-z) = \wp (z),
& \qquad &
\wp (z+2\omega _a) = \wp (z)
\quad\qquad \qquad a=1,2,3
\nonumber \\
\zeta (-z) = - \zeta (z),
& \qquad &
\zeta  (z+2\omega _a) = \zeta (z) + 2 \eta _a
\nonumber \\
\sigma (-z) = \sigma (z),
& \qquad &
\sigma (z+2 \omega _a) =-\sigma (z) e^{2 \eta _a (z + 2 \omega _a)},
\eea
where $\eta _a = \zeta (\omega _a)$ and 
\bea
\sigma(z) &= & z + {\cal O} (z^5) \nonumber \\
\zeta (z) &= & {1 \over z} + {\cal O} (z^3) \nonumber \\
\wp (z) & = & {1 \over z^2} + {\cal O}(z^2). 
\eea
The function $\sigma$ may be expressed in terms of the Jacobi
$\vartheta$-function
\be
\sigma (z; 2 \omega _1, 2 \omega _2)
=2 \omega _1 \exp \biggl ( {\eta _1 z^2 \over 2 \omega _1} \biggr )
{\vartheta _1 ({z \over 2 \omega _1} |\tau ) \over \vartheta _1 ' (0|\tau)},
\ee
which in turn is defined in terms of
\be
q = e^{2 \pi i\tau} 
\qquad \qquad
v= {z \over 2 \omega _1}
\ee
by
\be
\vartheta _1 (u|\tau)
=  2 q ^{1 \over 4} \sin \pi u \prod _{n=1} ^\infty
\bigl ( 1 - q ^n e^{2 \pi i u} \bigr )
\bigl ( 1 - q ^n e^{-2 \pi i u} \bigr )
\bigl ( 1 - q ^n  \bigr ).
\ee
An identity basic to the theory of $\theta$-functions is
\be \label{thetazero}
\vartheta_1'(0|\tau)
={2\pi\over 2\omega_1}q^{1/8}
\prod_{n=1}^{\infty}(1-q^n)^3
\ee
The function $\wp$ satisfies the differential equation
\be
\wp '(z) ^2 = 4 (\wp (z) - \wp (\omega _1) )(\wp (z) - \wp (\omega _2))
(\wp (z) - \wp (\omega _3)).
\ee
These and additional properties of elliptic functions may be found in 
\cite{erdelyi}. To evaluate scaling limits, the following formula for
$\wp(z;2\omega_1,2\omega_2)$ \cite{inozemtsev} is more useful
\be
\wp (z; 2\omega _1 , 2 \omega _2)
= - {\eta _1 \over \omega _1} + \bigl ( { \pi \over 2 \omega _1} \bigr ) ^2
\sum _{n=-\infty} ^ \infty {1 \over \sinh ^2 {i \pi \over 2 \omega _1}
(z-2n\omega _2) },
\ee
where 
\be
{\eta _1\over \omega _1} 
= - {1 \over 12} + \half \sum _{n=1} ^\infty
{1 \over \sinh ^2 i\pi n \tau}.
\ee
Since a proof of this formula is not given in \cite{inozemtsev},
we provide a proof here for the convenience of the reader.
First observe that 
\be
(1-q^ne^{2\pi iu})(1-q^ne^{-2\pi iu}) =
4q^{2n}{\rm sin}\,(\pi u+n\pi \tau){\rm sin}\,(\pi u-n\pi\tau)\, .
\ee
It follows that
\be
\vartheta_1(u|\tau)
=2q^{1/8}{\rm sin}\,(\pi u)\prod_{n=1}^{\infty}(1-q^n)^3
\prod_{n=1}^{\infty}(4q^2)^n
\prod_{n=1}^{\infty}{\rm sin}\,(nu+n\pi\tau)
\ {\rm sin}\,(nu-n\pi\tau)
\ee
However, 
\be
\ln \prod_{n=1}^{\infty}(4q^2)
=2\ln \,2\sum_{n=1}^{\infty}1+\ln q\sum_{n=1}^{\infty}n
=2\ln\,2\zeta(0)+\ln q\zeta(-1)\, ,
\ee 
where $\zeta(s)$ is the Riemann
zeta function. Since $\zeta(0)=-{1\over 2}$
and $\zeta(-1)=-{1\over 12}$, we find
$\ln\prod_{n=1}^{\infty}(4q^2)
={1\over 2}q^{-{1\over 12}}$.
In view of the identity (\ref{thetazero}), we arrive at
\be
{\vartheta_1(u|\tau)
\over
\vartheta_1'(0|\tau)}
=
{\omega_1\over\pi}q^{-{1\over 12}}\prod_{n=0}^{\infty}
{\rm sin}\,\pi(u-n\tau).
\ee
It follows that
\beq
\sigma(z)
=
{2\omega_1^2\over\pi}
{\rm exp}\,
(\eta_1{z^2\over 2\omega_1}-{i\pi\tau\over 6})
\prod_{n=-\infty}^{\infty}{\rm sin}\,\pi({z\over 2\omega_1}-n\tau)
\ee
Differentiating twice, we obtain
\bea
\zeta(z)&=&{\eta_1\over\omega_1}z
+{\pi\over 2\omega_1}\sum_{n=-\infty}^{\infty}
{\rm cotan}\,\pi({z\over 2\omega_1}-n\tau)\\
\wp(z)&=&-{\eta_1\over\omega_1}
+{\pi^2\over 4\omega_1^2}
\sum_{n=-\infty}^{\infty}{1\over{\rm sin}^2\pi({z\over 2\omega_1}-n\tau)}.
\eea
The constant $\eta_1/\omega_1$ can be easily determined. If we choose
$\omega_1=-i\pi$, ${\rm Re}\,\omega_2>0$,
and compare the expansion
(10.15) with the fact that $\wp(z)={1\over z^2}+{1\over 20}z^2g_2+O(z^4)$,
we find 
\be
{\eta_1\over\omega_1}
=
-{1\over 12}
+{1\over 2}
\sum_{n=1}^{\infty}{1\over {\rm sh}^2n\omega_1}.
\ee

\subsection{ Half and Third Period Functions}

Elliptic functions at half and third period (which are the only ones that we
shall need here, since the order of twisting is at most 3) are expressible in
terms of the original periods using Landen's transformations \cite{erdelyi}. It
is convenient to make a definite choice for the period that is to be twisted; we
shall choose this period to be $\omega _1$. It is straightforward to adapt
these formulas when an arbitrary period $2 \omega _a$, $a=1,\cdots ,3$ is
twisted.

\medskip

\noindent
{\it Formulas for Twists of Order 2 : Elliptic Functions with Half Periods}

\medskip

Henceforth, we shall reserve the notation $\wp (z)$, $\zeta (z)$ and $\sigma
(z)$ for the corresponding Weierstrass functions with periods $2 \omega _1$ 
and
$2 \omega _2$, as defined in \S B (a). The elliptic functions at half period
$\omega _1$ are given by
\bea
\wp _2 (z) = \wp (z; \omega _1 , 2 \omega _2) 
&=& \wp (z) + \wp (z+\omega _1) - \wp (\omega _1) 
\nonumber \\
&=& ~{1 \over \wp (\omega _1)} \big [ \wp (z) \wp (z+\omega _1) -(\wp (\omega 
_1) -\wp (\omega _2) ) (\wp (\omega _1 ) - \wp (\omega _3)) \big ] 
\nonumber \\
\zeta _2 (z) = \zeta (z;\omega _1, 2 \omega _2) 
&=& ~ \zeta (z) + \zeta (z+\omega _1) + z \wp (\omega _1) - \eta _1 
\nonumber \\
\sigma _2 (z) = \sigma (z; \omega _1 , 2 \omega _2) 
&=& ~ {\sigma (z) \sigma (z+ \omega _1) \over \sigma (\omega _1) }
e^{\half z^2 \wp (\omega _1) - z \eta _1} 
\eea
This gives rise to the duplication formula
\be
4\wp (2z) = \wp (z) + \wp (z+\omega _1) +\wp (z+\omega _2) + \wp (z+\omega _1 
+
\omega _2).
\ee

\medskip

\noindent
{\it Formulas for Twists of order 3 : Elliptic Functions at Third Periods}

\medskip

Similarly, we have the following formulas for the third period elliptic
functions
\bea
\wp _3 (z) = \wp (z; 2\omega _1/3 , 2 \omega _2) &= &
\ \wp (z) + \wp (z+2\omega _1/3) + \wp (z+4\omega _1/3) 
\nonumber \\ 
& & - \wp (2\omega _1/3) -\wp (4 \omega _1 /3)
\nonumber \\
\zeta _3 (z) = \zeta (z;2\omega _1/3 , 2 \omega _2) &=&
\ \zeta (z) + \zeta (z+2\omega _1/3) + \zeta (z +4 \omega _1 /3) 
\nonumber \\ 
& & + z \wp (2\omega _1/3) + z \wp (4 \omega _1 /3)  - \eta _1
\nonumber \\
\sigma _3 (z) = \sigma (z; 2\omega _1/3 , 2 \omega _2) &=&
\ {\sigma (z) \sigma (z + 2\omega _1/3 ) \sigma (z + 4 \omega _1 /3) 
\over \sigma (2\omega _1/3) \sigma (4 \omega _1 /3)} 
e^{\half z^2 \wp (\omega _1) - z \eta _1}
\eea
This gives rise to the triplication formula
\be
9\wp (3z) 
= 
\sum _{j,k=0} ^2 
\wp (z+ j {2 \omega _1 \over 3} + k{2 \omega _2 \over 3}). 
\ee
All of the above formulas may be established by identifying singularities in
$z$ and establishing that the remainder must be independent of $z$ by
Liouville's theorem.

\medskip

The functions at half and third periods, defined above are related to one
another in a way analogous to (B.4)
\be
\wp _\nu (z; 2 \omega _1, 2 \omega _2) 
= - {d \over dz} \zeta _\nu (z; 2 \omega _1, 2 \omega _2)
= - {d^2 \over dz^2} \log \sigma _\nu (z; 2 \omega _1, 2 \omega _2),
\ee
where $\nu =1,2,3$.

\subsection{The Function $\Phi$}

We define the function $\Phi$ by
\be
\Phi (x,z) = 
\Phi (x,z; 2 \omega _1 , 2 \omega _2) =
{ \sigma (z-x) \over \sigma (z) \sigma (x) } e ^{x \zeta (z)},
\ee
where $\sigma (z)$ and $\zeta (z)$ are the Weierstrass functions of (B.4) and 
(B.7).
As a function of $z$, $\Phi (x,z)$ is periodic with periods $2 \omega _1$ and 
$2 
\omega _2$, is holomorphic except for an essential singularity at $z=0$, and 
has 
a single zero at $z=x$. As a function of $x$, $\Phi (x,z)$ has multiplicative 
monodromy, given by
\be
\Phi (x+ 2 \omega _a,z ) = \Phi (x,z) e^{2 \omega _a \zeta (z) - 2 \eta _a 
z},
\ee
is holomorphic in $x$ except for a simple pole at $x=0$, and has a single 
zero 
at $x=z$.
Some useful asymptotics are given as follows
\bea
\Phi (x,z) & = & {1 \over x} - \half x \wp (z) + {\cal O}(x ^2 ) 
\nonumber \\
\Phi (x,z) & = & \bigl \{ -{1 \over z} + \zeta (x) + {\cal O}(z) \bigr \} e^{x 
\zeta (z)}
\qquad z\to 0.
\eea
Products of the function $\Phi (x_\alpha, z)$, with $\sum _{\alpha =1} ^n
x_\alpha =0$, are periodic in $z$, with periods $2 \omega _1$ and $2 \omega 
_2$, 
and meromorphic in $z$ since the essential singularities cancel. As a result, 
they satisfy
\be
\prod _{\alpha =1} ^n \Phi (x_\alpha , z) = P_n[\wp (z);x_\alpha ] + \wp '(z)
Q_n [\wp (z); x_ \alpha], 
\ee
where $P_n$ and $Q_n$ are polynomials of degrees $[{n\over 2}]$ and $[{n-3 
\over 2} ]$ in $\wp (z)$ respectively, with $x_\alpha$-dependent 
coefficients. 
The simplest case is 
\be\label{identity1}
\Phi (x,z) \Phi (-x,z) =  \wp (z) - \wp (x) 
\ee
In general, the polynomials $P$ and $Q$ may be determined by the fact that 
the
r.h.s. of (B.19) has a simple zero at each point $z=x_\alpha$, and that the 
pole highest order in $z$ has coefficient $(-1)^n$.

The function $\Phi (x,z)$ satisfies a fundamental differential equation,
\be \label{functional}
\Phi (x,z) \Phi ' (y,z) - \Phi (y,z) \Phi '(x,z) 
=  (\wp (x) - \wp (y) ) \Phi (x+y,z), 
\ee
where $\Phi '(x,z)$ denotes the derivative with respect to $x$ of $\Phi 
(x,z)$.

\subsection{ The Functions $\Lambda$, $\Phi _1$, $\Phi _2$}

The functions $\Lambda$ and $\Phi _2$ are defined by
\be
\Lambda (2x,z)  = \Phi _2 (x,z) =
{\Phi (x,z) \Phi (x+\omega _1,z) \over \Phi (\omega _1,z)}.
\ee
The essential singularity in $z$ and the monodromy in $x$ of $\Lambda (2x,z)$ 
coincide with those of $\Phi (x,z)$.
We shall need the following two basic differential equations,
\bea \label{functional1}
\Lambda  (2x,z) \Lambda ' (2y,z) - \Lambda '(2x,z) \Lambda (2y,z) 
& = & \half (\wp _2 (x) - \wp _2 (y) ) \Lambda (2x+2y,z), 
\nonumber \\
\Phi _2 (x,z) \Phi _2 ' (y,z) - \Phi _2(y,z) \Phi _2'(x,z) 
& = & (\wp _2 (x) - \wp _2 (y) ) \Phi _2(x+y,z), 
\eea
as well as differential equation that involves both $\Phi $ and $\Lambda$,
\bea \label{functional2}
\Lambda (2x,z) \Phi ' (-x-y,z) - \Lambda ' (2x,z) \Phi (-x-y,z) & & 
\qquad \quad \nonumber \\
-\Lambda (-2y,z) \Phi ' (x+y,z) + \Lambda ' (-2y,z) \Phi (x+y,z) & = & \half
(\wp _2 (x) - \wp _2 (y) ) \Phi (x-y,z). \qquad  
\eea
By letting $y\to x$ in (B.23), and taking into account the known zeros of 
$\Lambda $, we derive another useful formula
\be\label{functionalL}
\Lambda (2x,z) \Lambda (-2x,z) = \wp _2 (z) - \wp _2 (x).
\ee
Actually, $\Phi _2(x,z)$ may be viewed as the function $\Phi (x,z)$
associated with a torus of periods $\omega _1$ and $2 \omega _2$.

\medskip

The function $\Phi _1 (x,z)$ is defined by
\bea
\Phi _1(x,z) & = &\Phi  (x,z) +  f(z) \Phi  (x+\omega _1,z) 
\\
f(z) & = &-e^{\pi i \zeta (z) +  \eta _1 z}.
\eea
It obeys the monodromy relation $\Phi _1 (x+\omega _1, z)
= f(z) ^{-1} \Phi _1 (x,z)$, as well as the following differential equations
\bea
\Phi _1 (x,z) \Phi _1 ' (y,z) - \Phi _1 ' (x,z) \Phi _1(y,z)
&= & (\wp _2 (x) - \wp _2 (y) ) \Phi _1 (x+y,z) 
\\
\Phi _1 (x,z) \Phi  ' (y,z) - \Phi  (y,z) \Phi _1 '(x,z) 
&= & \ \{ \wp (x+\omega _1) - \wp (y) \} \Phi _1 (x+y,z) 
\nonumber \\
&& + \{ \wp (x) - \wp (x+\omega _1) \} \Phi  (x+y,z), 
\eea
and
\bea
\Phi (2x,z) \Phi _1 ' (-x-y,z) &-& \Phi  ' (2x,z) \Phi _1 (-x-y,z)  
-\Phi  (-2y,z) \Phi _1 ' (x+y,z) 
 \\
&+& \Phi  ' (-2y,z) \Phi _1 (x+y,z) = 
(\wp  (2x) - \wp (2y) )  \Phi _1 (x-y,z),  
\nonumber \\
 \Lambda (2x,z) \Phi _1 ' (-x-y,z) &-& \Lambda ' (2x,z) \Phi _1 (-x-y,z) 
 -\Lambda (-2y,z) \Phi  _1 ' (x+y,z) 
 \\
&+&  \Lambda ' (-2y,z) \Phi  _1(x+y,z) 
=  \half (\wp _2 (x) - \wp _2 (y) )  \Phi _1 (x-y,z).  
\nonumber   
\eea

\vfill\eject

\end{document}